\documentclass[12pt]{article}
\pdfoutput=1

\usepackage{booktabs}
\usepackage{tabulary}
\usepackage{multirow}
\usepackage{rotating}
\usepackage{epsfig}
\usepackage{psfrag}
\usepackage{latexsym}
\usepackage{indentfirst}
\usepackage{fancyhdr}
\usepackage{dsfont}
\usepackage{amsmath}
\usepackage{amssymb}
\usepackage{amsfonts}
\usepackage{mathrsfs}
\usepackage{amsthm}
\usepackage{pifont}
\usepackage{dsfont}
\usepackage{multirow}
\usepackage{array}
\usepackage{chngpage}
\usepackage{longtable}
\usepackage{cite}
\usepackage{bbold}
\usepackage{color}
\usepackage{braket}
\usepackage{colordvi}
\usepackage{fancybox}
\usepackage[footnotesize]{caption2}
\usepackage{graphicx}
\usepackage[center,footnotesize,hang]{subfigure}
\usepackage{bbm}
\usepackage{url}
\usepackage[colorlinks, linkcolor=red, anchorcolor=black, citecolor=green]{hyperref}
\usepackage{multirow}
\usepackage{harpoon}
\usepackage{array}
\usepackage{textcomp}
\usepackage{enumitem}
\usepackage{mathrsfs}
\newcommand{\PreserveBackslash}[1]{\let\temp=\\#1\let\\=\temp}
\newcolumntype{C}[1]{>{\PreserveBackslash\centering}p{#1}}
\newcolumntype{R}[1]{>{\PreserveBackslash\raggedleft}p{#1}}
\newcolumntype{L}[1]{>{\PreserveBackslash\raggedright}p{#1}}
\allowdisplaybreaks

\newcommand{\bq}{\begin{eqnarray}}
\newcommand{\nq}{\end{eqnarray}}
\newcommand*{\vect}[1]{\overrightharp{\ensuremath{#1}}}
\newcommand{\cleqn}{\setcounter{equation}{0}}

\begin{document}

\title{
\begin{flushright}
\hfill\mbox{\small USTC-ICTS-19-03} \\[5mm]
\begin{minipage}{0.2\linewidth}
\normalsize
\end{minipage}
\end{flushright}
{\Large \bf Neutrino Mass and Mixing with $A_5$ Modular Symmetry
\\[2mm]}}

\date{}

\author{
Gui-Jun~Ding$^{1}$\footnote{E-mail: {\tt
dinggj@ustc.edu.cn}},  \
Stephen~F.~King$^{2}$\footnote{E-mail: {\tt king@soton.ac.uk}}, \
Xiang-Gan Liu$^{1}$\footnote{E-mail: {\tt
hepliuxg@mail.ustc.edu.cn}}  \
\\*[20pt]
\centerline{
\begin{minipage}{\linewidth}
\begin{center}
$^1${\it \small
Interdisciplinary Center for Theoretical Study and  Department of Modern Physics,\\
University of Science and Technology of China, Hefei, Anhui 230026, China}\\[2mm]
$^2${\it \small
Physics and Astronomy,
University of Southampton,
Southampton, SO17 1BJ, U.K.}\\
\end{center}
\end{minipage}}
\\[10mm]}
\maketitle
\thispagestyle{empty}

\begin{abstract}
\noindent
We present a comprehensive analysis of neutrino mass and lepton mixing in theories with $A_5$ modular symmetry. We construct the weight 2, weight 4 and weight 6 modular forms of  level 5 in terms of Dedekind eta-functions and Klein forms, and their decomposition into irreducible representation of $A_5$. We construct all the simplest models based on $A_5$ modular symmetry, including scenarios of models with and without flavons in the charged lepton sectors. For each case, the neutrino masses can be generated through either the Weinberg operator or the type I seesaw mechanism. We perform an exhaustive numerical analysis, organising our results in an extensive set of figures and tables.
\end{abstract}
\newpage

\section{\label{sec:introduction}Introduction}

The flavour puzzle, in particular the origin of neutrino mass and lepton mixing, is a major unresolved problem of the Standard Model (SM).
The large mixing angles in the lepton sector can be explained using some discrete non-Abelian family symmetry \cite{King:2013eh,King:2017guk}.
Such a symmetry could either originate from a continuous non-Abelian gauge symmetry
\cite{Koide:2007sr,Banks:2010zn,Wu:2012ria,Merle:2011vy,Rachlin:2017rvm,Luhn:2011ip,King:2018fke},
or from extra dimensions \cite{Altarelli:2008bg,Burrows:2009pi,Burrows:2010wz,deAnda:2018oik,Adulpravitchai:2009id,Asaka:2001eh,Altarelli:2006kg,Adulpravitchai:2010na,Kobayashi:2006wq,deAnda:2018yfp,Kobayashi:2018rad,Baur:2019kwi}.
In the case of extra dimensions, the discrete non-Abelian family symmetry could either arise as an accidental symmetry
of the orbifold fixed points, or it could appear as a subgroup of the symmetry of the extra dimensional lattice vectors, commonly referred to as modular symmetry~\cite{Giveon:1988tt}.

Some time ago it was suggested that a finite subgroup of the modular symmetry group, when interpreted as a family symmetry, might help to provide
a possible explanation for the neutrino mass matrices
\cite{Altarelli:2005yx,deAdelhartToorop:2011re}.
Recently it has been suggested that neutrino masses might be modular forms
\cite{Feruglio:2017spp}, with constraints on the Yukawa couplings.
This has led to a revival of the idea that  modular symmetries are symmetries of the extra dimensional spacetime
with Yukawa couplings determined by their modular weights \cite{Criado:2018thu}.
The modular groups $\Gamma(2)$~\cite{Kobayashi:2018vbk,Kobayashi:2018wkl}, $\Gamma(3)$~\cite{Feruglio:2017spp,Criado:2018thu,Kobayashi:2018scp,Okada:2018yrn,Kobayashi:2018wkl,Novichkov:2018yse}, $\Gamma(4)$~\cite{Penedo:2018nmg,Novichkov:2018ovf} and $\Gamma(5)$~\cite{Novichkov:2018nkm}
and the phenomenological predictions for lepton mixing have all been discussed in the literature.

In the present work, we focus on the finite modular group
$\Gamma(5)\cong A_5$. In particular, we construct the weight 2, weight 4 and weight 6 modular forms of level 5 in terms of the Dedekind eta-function and Klein forms, and their decomposition into irreducible representation of $A_5$. This is complementary to~\cite{Novichkov:2018nkm} where only weight 2 and weight 4 modular forms are constructed using the Jacobi theta function. Since the results overlap, we have checked that the same $q-$expansions of the modular forms are obtained up to an overall irrelevant constant.
We construct all the simplest models based on $A_5$ modular symmetry, both
with and without flavons in the charged lepton sectors.
For each case, the neutrino mass is generated through both the Weinberg operator or the type I seesaw mechanism. The model building discussed here goes beyond that in~\cite{Novichkov:2018nkm} where only two models were constructed with diagonal charged lepton mass matrix, one of which is not viable.
For each model, we give an extensive numerical analysis, organising our results in a comprehensive
set of figures and tables.

The layout of the remainder of the paper is as follows.
In section~\ref{sec:modular_sym_gen} we review the basics of modular symmetry and modular forms of level $N=5$
corresponding to $A_5$. We also construct the modular space through an infinite product expansion, construct an eleven
dimensional basis for the lowest weight modular forms, and decompose this space into the two irreducible triplet and the quintuplet
representations of $A_5$, from which the higher weight representations may be obtained.
In section~\ref{sec:model} we systematically construct all the simplest models based on $\Gamma(5)\cong A_5$.
The charged lepton sector is analysed both with and without flavons.
The neutrino sector is analysed without flavons using both the Weinberg operator and the type I seesaw mechanism.
In section~\ref{sec:numerical} we perform a comprehensive numerical analysis for each of the models discussed in the previous section.
We give the best fit values of the parameters of each model and the corresponding predictions. We also give the possible regions of the parameters and ranges of the predictions in graphical form. In section~\ref{sec:conclusion} we summarize the main results of our paper. In Appendix~\ref{sec:App_CG} we present the group theory of $A_{5}$. In Appendix~\ref{sec:App_charged_lepton} we show that the charged lepton sector with flavons leads to diagonal
charged lepton Yukawa matrices when the flavons are aligned appropriately.

\section{\label{sec:modular_sym_gen}Modular symmetry and modular forms of level $N=5$}

\begin{figure}[t!]
\centering
\includegraphics[width=1.0\textwidth]{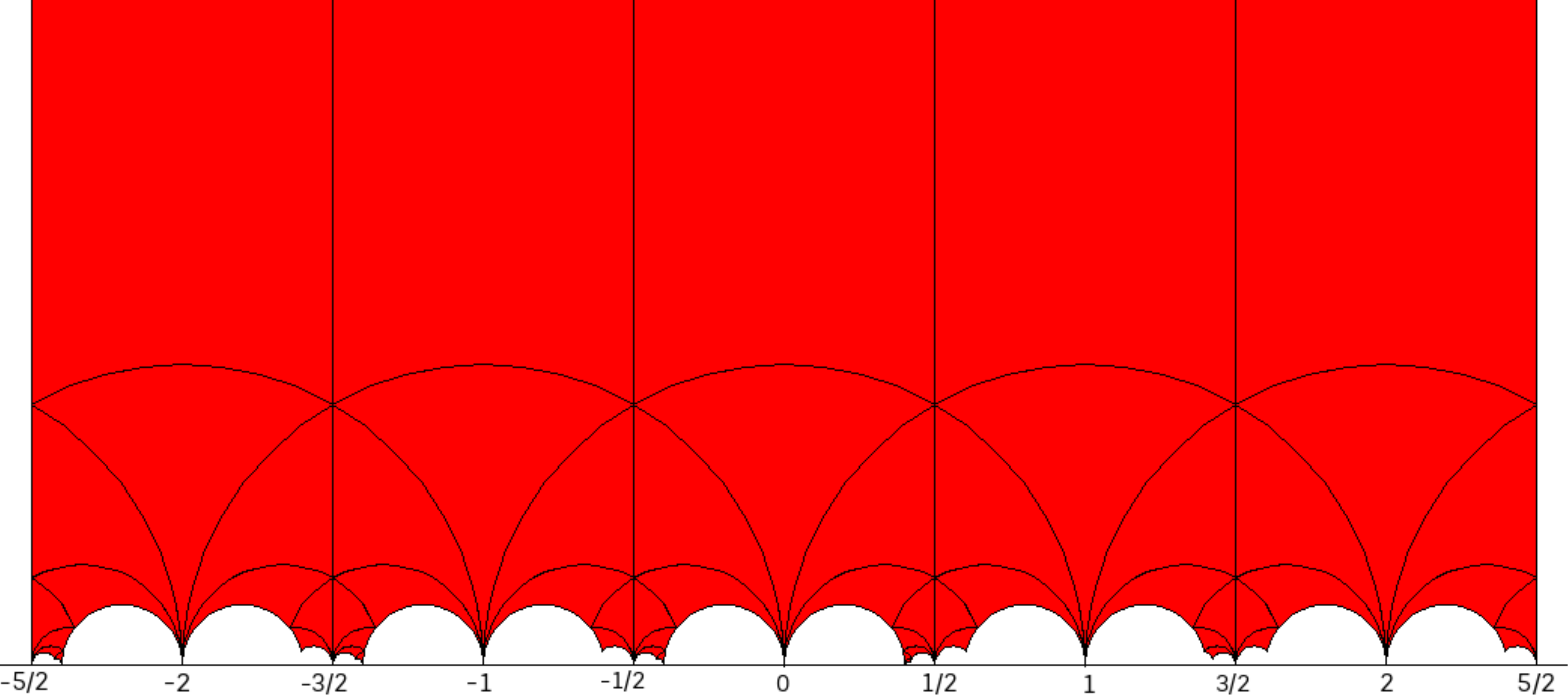}
\caption{\label{fig:Fund_domain}Fundamental domain for $\Gamma(5)$.}
\end{figure}

The \textit{full modular group} $SL(2,\mathbb{Z})$ is the group of 2-by-2 matrices with integer entries and determinant 1~\cite{Bruinier2008The,diamond2005first},\\
\begin{equation}
SL(2,\mathbb{Z})=\left\{\left(\begin{array}{cc}a&b\\c&d\end{array}\right)\bigg|a,b,c,d\in \mathbb{Z},ad-bc=1\right\}\,.
\end{equation}
The modular group can be regarded as the linear fraction transformations of the upper half complex plane ${\cal H}=\{\tau\in\mathbb{C}~|~\rm{Im}\,\tau >0\}$, and it has the following form
\begin{equation}
\tau \mapsto \gamma(\tau)=\frac{a\tau+b}{c\tau+d},\quad \gamma=\begin{pmatrix}
a  &  b  \\
c  &  d
\end{pmatrix}\in SL(2, \mathbb{Z})\,.
\end{equation}
It's easy to see that $\gamma$ and $-\gamma$ lead to the same linear fractional transformation. Therefore the modular group of transformations is is isomorphic to the projective special linear group $PSL(2, \mathbb{Z})=SL(2, \mathbb{Z})/\{\mathds{1}, -\mathds{1}\}$, which is the quotient of $SL(2, \mathbb{Z})$ over its center $\{\mathds{1}, -\mathds{1}\}$. The modular group can be generated by two elements $S$ and $T$~\cite{Bruinier2008The}
\begin{equation}
\begin{aligned}
&S: \tau \mapsto -\frac{1}{\tau},\qquad~~  S=\left(
\begin{array}{cc}
0 ~&~ 1\\
-1 ~&~ 0
\end{array}
\right),\\
&T: \tau \mapsto \tau+1,\qquad  T=\left(
\begin{array}{cc}
1 ~&~  1\\
0 ~&~ 1
\end{array}
\right)\,.
\end{aligned}
\end{equation}
We see that the generators $S$ and $T$ obey the relations:
\begin{equation}
\label{eq:muti_rules}S^2=(ST)^3=\mathds{1}\,.
\end{equation}
There are some important subgroups of the $SL_2(\mathbb{Z})$, i.e.
\begin{equation}
\Gamma(N)=\left\{\left(\begin{array}{cc}a&b\\c&d\end{array}\right)\in SL(2,\mathbb{Z}),~~ b\equiv c\equiv0, a\equiv d\equiv1~~({\tt mod}~N)\right\}\,,
\end{equation}
where $N$ is a positive integer, and $\Gamma(N)$ is usually called \textit{principal congruence subgroup of level N}. Obviously we have $\Gamma(1)\cong SL(2, \mathbb{Z})$, and $\Gamma(N>1)$ is the normal subgroup of $\Gamma(1)$. Taking the quotient of $\Gamma(N)$ over $\{\mathds{1}, -\mathds{1}\}$ or identifying $\gamma$ with $-\gamma$, we obtain the group $\overline{\Gamma}(N)=\Gamma(N)/\{\mathds{1}, -\mathds{1}\}$. To be more specific, $\overline{\Gamma}(1)=\Gamma(1)/\{\mathds{1}, -\mathds{1}\}$ and $\overline{\Gamma}(2)=\Gamma(2)/\{\mathds{1}, -\mathds{1}\}$, and $\overline{\Gamma}(N>2)=\Gamma(N)$ since the element $-\mathds{1}$ is not in $\Gamma(N)$ for $N>2$. The finite modular group $\Gamma_{N}=\overline{\Gamma}(1)/\overline{\Gamma}(N)$ has been used as flavor symmetry to explain quark and lepton flavor mixing in the past years~\cite{deAdelhartToorop:2011re}. For example, $\Gamma_2\cong S_3$, $\Gamma_3\cong A_4$, $\Gamma_4\cong S_4$ and $\Gamma_5\cong A_5$. The multiplication rules of the group $\Gamma_N$ can be obtain by extending Eq.~\eqref{eq:muti_rules} with the condition $T^{N}=1$, such that the generators $S$ and $T$ of $\Gamma_N$ satisfy
\begin{equation}
S^2=(ST)^3=T^{N}=\mathds{1}\,.
\end{equation}
The modular forms $f(\tau)$ of weight $2k$ and level $N$ are holomorphic functions of the complex variable $\tau$ with well-defined transformation properties under the group $\Gamma(N)$:
\begin{equation}
f(\frac{a\tau+b}{c\tau+d})=(c \tau+d)^{2k} f(\tau) \quad\text{for}\quad \forall~~\gamma=\begin{pmatrix}
a  &  b \\
c  &  d
\end{pmatrix}\in \Gamma(N)\,,
\end{equation}
where $k\ge 0$ is an integer. We shall only consider modular forms of even weight in this work. The function $f(\tau)$ is required to be holomorphic in ${\cal H}$ and at all the cusps. In particular, the element $T^{N}$ is in $\Gamma(N)$ and we have
\begin{equation}
f(\tau+N)=f(\tau)\,.
\end{equation}
In the present work, we shall study the $\Gamma(5)$ modular group and construct some models to explain neutrino masses and mixing. The quotient space ${\cal H}/\Gamma(5)$ can be described by a fundamental domain ${\cal F}_5$ for $\Gamma(5)$, that is a connected region of ${\cal H}$ such that
every $z\in\cal H$ can be mapped into ${\cal F}_5$ by a $\Gamma(5)$ transformation, but no two distinct points in the interior of ${\cal F}_5$ are related under $\Gamma(5)$. The space ${\cal H}/\Gamma(5)$ is simply ${\cal F}_5$ with certain boundary points identified. A fundamental domain for $\Gamma(5)$ is shown in figure~\ref{fig:Fund_domain}. It can be constructed by the fundamental domain ${\cal F}=\left\{\tau\in\mathcal{H}|-1/2\leq\rm{Re}\,\tau \leq1/2~\text{and}~|\tau|\geq1\right\}$ of $SL(2, \mathbb{Z})$~\cite{verrill2001algorithm}.
The cusps are $i\infty$, $-2$, $-1$, $-\frac{3}{2}$, $-\frac{1}{2}$, $0$, $\frac{2}{5}$, $\frac{1}{2}$, $1$, $\frac{3}{2}$, $2$ and $\frac{5}{2}$. ${\cal H}/\Gamma(5)$ can be made compact by adding these cusp points. The compactified space $\overline{{\cal H}/\Gamma(5)}$ has genus zero and can be thought of as an icosahedral whose vertices are the cusps. Indeed the cusps  are related by transformations of $\Gamma_5=\overline{\Gamma}(1)/\overline{\Gamma}(5)$, which is icosahedral group, given the isomorphism between $\Gamma_5$ and $A_5$.

The modular forms of weight $2k$ and level $N=5$ form a linear space $\mathcal{M}_{2k}(\Gamma(5))$, and its dimension turns out to be $10k+1$~\cite{Gunning1962,Feruglio:2017spp}. For the lowest nontrivial weight $2k=2$, the dimension is equal to $11$. The modular space $\mathcal{M}_{2k}(\Gamma(5))$ can be constructed from the Dedekind eta-function and the Klein form. In the following, we present the definition and basic properties of the eta-function and the Klein form which are the building blocks to construct modular forms of level 5.

For any complex number $\tau$ with $\rm{Im}\,\tau > 0$, the Dedekind eta-function $\eta(\tau)$ is defined to be the infinite product~\cite{diamond2005first,Bruinier2008The,lang2012introduction},
\begin{equation}
\eta(\tau)=q^{1/24}\prod_{n=1}^\infty \left(1-q^n \right),\qquad q\equiv e^{i 2 \pi\tau}\,.
\end{equation}
This function plays an important role in constructing various modular forms of integral or half-integral weight. It is well-known that $\eta(\tau)$ satisfies the following identities~\cite{diamond2005first,Bruinier2008The,lang2012introduction}
\begin{equation}
\eta(\tau+1)=e^{i \pi/12}\eta(\tau),\qquad \eta(-1/\tau)=\sqrt{-i \tau}~\eta(\tau)\,.
\end{equation}
Then we turn to the Klein form. Let $\Lambda=\mathbb{Z}\tau+\mathbb{Z}$ be a lattice in $\mathbb{C}$ and any pair $(r_1,r_2)\in\mathbb{Q}^2-\mathbb{Z}^2$ with $z=r_1\tau+r_2$, the Klein form $\mathfrak{k}_{(r_1,r_2)}(\tau)$ is defined by~\cite{K_lang1981,lang1987elliptic,lang2012introduction}
\begin{equation}
\label{eq:klein_form}\mathfrak{k}_{(r_1,r_2)}(\tau)=-2{\pi}ize^{{\pi}ir_1z}\prod_{\omega\in\Lambda'} (1+\frac{z}{\omega})e^{-\frac{z}{\omega}}\,,
\end{equation}
where $\Lambda'=\Lambda\setminus\{0\}$, and the product over $\omega=m\tau+n$ are performed over $n$ first and then $m$. Notice that there is an extra factor $i/2\pi$ in the original definition of~\cite{K_lang1981,lang1987elliptic,lang2012introduction}.
From the definition of Eq.~\eqref{eq:klein_form} we know that the Klein form $\mathfrak{k}_{(r_1,r_2)}(\tau)$ is a holomorphic function which
has no zeros and poles on $\mathcal{H}$. The Klein form has the following infinite product expansion~\cite{K_lang1981,lang1987elliptic,lang2012introduction,eum2011modularity}:
\begin{equation}\label{KleinForm}
\mathfrak{k}_{(r_1,r_2)}(\tau)=q^{(r_1-1)/2}_z(1-q_z)\prod_{n=1}^\infty(1-q^nq_z)(1-q^nq_z^{-1})(1-q^n)^{-2}\,,
\end{equation}
where $q_z=e^{2\pi iz}$. It is proved that the Klein form satisfies the following transformation formulas~\cite{K_lang1981,lang1987elliptic,lang2012introduction}
\begin{equation}
\label{eq:klein_pr2}
\begin{aligned}
\mathfrak{k}_{(-r_1, -r_2)}(\tau)&=-\mathfrak{k}_{(r_1, r_2)}(\tau)\,,\\
\mathfrak{k}_{(r_1, r_2)}(\gamma(\tau))&=(c\tau+d)^{-1}\mathfrak{k}_{(r_1, r_2)\cdot\gamma}(\tau)~~\mathrm{for}~~\gamma\in\overline{\Gamma}(1)\,,\\
\mathfrak{k}_{(r_1+s_1, r_2+s_2)}(\tau)&=\epsilon(\vect{r}, \vect{s})\mathfrak{k}_{(r_1, r_2)}(\tau)~~\mathrm{for}~~(s_1,s_2)\in\mathbb{Z}^2\,,
\end{aligned}
\end{equation}
with $\epsilon(\vect{r}, \vect{s})=(-1)^{s_1s_2+s_1+s_2}e^{-{\pi}i(s_1r_2-s_2r_1)}$. Using Eq.~\eqref{eq:klein_pr2}, we find the Klein form $\mathfrak{k}_{(r_1, r_2)}(\tau)$ fulfills
\begin{equation}
\mathfrak{k}_{(r_1, r_2)}(\tau+1)=\mathfrak{k}_{(r_1, r_1+r_2)}(\tau),\qquad \mathfrak{k}_{(r_1, r_2)}(-\frac{1}{\tau})=-\frac{1}{\tau}\mathfrak{k}_{(-r_2, r_1)}(\tau)\,.
\end{equation}
The modular space $\mathcal{M}_{2k}(\Gamma(5))$ has been explicitly constructed through $\eta$ function and Klein form as follow~\cite{schultz2015notes}
\begin{equation}
\mathcal{M}_{2k}(\Gamma(5))=\bigoplus_{a+b=10k,\,a,b\ge0} \mathbb{C} \frac{\eta^{15k}(5\tau)}{\eta^{3k}(\tau)} \mathfrak{k}^a_{\frac{1}{5},\frac{0}{5}}(5\tau)\mathfrak{k}^b_{\frac{2}{5},\frac{0}{5}}(5\tau)\,.
\end{equation}
Obviously the dimension of $\mathcal{M}_{2k}(\Gamma(5))$ is $10k+1$. For the lowest weight 2 modular forms, we could choose the basis vectors to be
\begin{eqnarray}
\nonumber&& \hat{e}_1(\tau)=\frac{\eta^{30}(5\tau)}{\eta^6(\tau)}\mathfrak{k}^{10}_{\frac{2}{5},\frac{0}{5}}(5\tau),\quad \qquad \hat{e}_2(\tau)=\frac{\eta^{30}(5\tau)}{\eta^6(\tau)}\mathfrak{k}_{\frac{1}{5},\frac{0}{5}}(5\tau)\mathfrak{k}^{9}_{\frac{2}{5},\frac{0}{5}}(5\tau),\\
\nonumber&& \hat{e}_3(\tau)=\frac{\eta^{30}(5\tau)}{\eta^6(\tau)}\mathfrak{k}^{2}_{\frac{1}{5},\frac{0}{5}}(5\tau)\mathfrak{k}^{8}_{\frac{2}{5},\frac{0}{5}}(5\tau),\quad\qquad \hat{e}_4(\tau)=\frac{\eta^{30}(5\tau)}{\eta^6(\tau)}\mathfrak{k}^{3}_{\frac{1}{5},\frac{0}{5}}(5\tau)\mathfrak{k}^{7}_{\frac{2}{5},\frac{0}{5}}(5\tau),\\
\nonumber&& \hat{e}_5(\tau)=\frac{\eta^{30}(5\tau)}{\eta^6(\tau)}\mathfrak{k}^{4}_{\frac{1}{5},\frac{0}{5}}(5\tau)\mathfrak{k}^{6}_{\frac{2}{5},\frac{0}{5}}(5\tau),\quad\qquad \hat{e}_6(\tau)=\frac{\eta^{30}(5\tau)}{\eta^6(\tau)}\mathfrak{k}^{5}_{\frac{1}{5},\frac{0}{5}}(5\tau)\mathfrak{k}^{5}_{\frac{2}{5},\frac{0}{5}}(5\tau),\\
\nonumber&& \hat{e}_7(\tau)=\frac{\eta^{30}(5\tau)}{\eta^6(\tau)}\mathfrak{k}^{6}_{\frac{1}{5},\frac{0}{5}}(5\tau)\mathfrak{k}^{4}_{\frac{2}{5},\frac{0}{5}}(5\tau),\quad\qquad \hat{e}_8(\tau)=\frac{\eta^{30}(5\tau)}{\eta^6(\tau)}\mathfrak{k}^{7}_{\frac{1}{5},\frac{0}{5}}(5\tau)\mathfrak{k}^{3}_{\frac{2}{5},\frac{0}{5}}(5\tau),\\
\nonumber&& \hat{e}_9(\tau)=\frac{\eta^{30}(5\tau)}{\eta^6(\tau)}\mathfrak{k}^{8}_{\frac{1}{5},\frac{0}{5}}(5\tau)\mathfrak{k}^{2}_{\frac{2}{5},\frac{0}{5}}(5\tau),\quad\qquad \hat{e}_{10}(\tau)=\frac{\eta^{30}(5\tau)}{\eta^6(\tau)}\mathfrak{k}^{9}_{\frac{1}{5},\frac{0}{5}}(5\tau)\mathfrak{k}_{\frac{2}{5},\frac{0}{5}}(5\tau),\\
&&\hat{e}_{11}(\tau)=\frac{\eta^{30}(5\tau)}{\eta^6(\tau)}\mathfrak{k}^{10}_{\frac{1}{5},\frac{0}{5}}(5\tau)\,.
\end{eqnarray}
The above basis vectors are linearly independent and they span the whole weight 2 modular space $\mathcal{M}_{2}(\Gamma(5))$. Any modular function of weight 2 and level 5 can be expressed as a linear combination of the basis elements $\hat{e}_i$ with $i=1, 2, \ldots, 11$. Under the action of the generator $T$, they transform as
\begin{eqnarray}
\nonumber&\hat{e}_1(\tau)\rightarrow\hat{e}_1(\tau),\qquad \hat{e}_2(\tau)\rightarrow e^{i\frac{2\pi}{5}}\hat{e}_2(\tau),\qquad \hat{e}_3(\tau)\rightarrow e^{i\frac{4\pi}{5}}\hat{e}_3(\tau),\\
\nonumber&\hat{e}_4(\tau)\rightarrow e^{i\frac{6\pi}{5}}\hat{e}_4(\tau),\qquad \hat{e}_5(\tau)\rightarrow e^{i\frac{8\pi}{5}}\hat{e}_5(\tau),\qquad \hat{e}_6(\tau)\rightarrow \hat{e}_6(\tau),\qquad \\
\nonumber&\hat{e}_7(\tau)\rightarrow e^{i\frac{2\pi}{5}}\hat{e}_7(\tau),\qquad \hat{e}_8(\tau)\rightarrow e^{i\frac{4\pi}{5}}\hat{e}_8(\tau),\qquad \hat{e}_9(\tau)\rightarrow
e^{i\frac{6\pi}{5}}\hat{e}_9(\tau),\qquad \\
&\hat{e}_{10}(\tau)\rightarrow e^{i\frac{8\pi}{5}}\hat{e}_{10}(\tau),\qquad
\hat{e}_{11}(\tau)\rightarrow\hat{e}_{11}(\tau)\,.
\end{eqnarray}
Furthermore, we find the following transformation properties under another generator $S$,
\begin{eqnarray}
\nonumber\hat{e}_1(\tau)&\rightarrow&\tau^2\Bigg\{-\left(\frac{1}{10}+\frac{11}{50\sqrt{5}}\right)\hat{e}_1(\tau)-\left(\frac{3}{5}+\frac{7}{5\sqrt{5}}\right)\hat{e}_2(\tau)
-\left(\frac{9}{5}+\frac{18}{5\sqrt{5}}\right)\hat{e}_3(\tau)\\
\nonumber&&-\left(\frac{12}{5}+\frac{36}{5\sqrt{5}}\right)\hat{e}_4(\tau)-\left(\frac{21}{5}+\frac{21}{5\sqrt{5}}\right)\hat{e}_5(\tau)-\frac{252}{25\sqrt{5}}\hat{e}_6(\tau)\\
\nonumber&&+\frac{21}{25}\left(-5+\sqrt{5}\right)\hat{e}_7(\tau)+\frac{12}{25}\left(5-3\sqrt{5}\right)\hat{e}_8(\tau)+\frac{9}{25}\left(-5+2\sqrt{5}\right)\hat{e}_9(\tau)\\
&&+\left(\frac{3}{5}-\frac{7}{5\sqrt{5}}\right)\hat{e}_{10}(\tau)+\frac{1}{250}\left(-25+11\sqrt{5}\right)\hat{e}_{11}(\tau)\Bigg\}\,.
\end{eqnarray}
\begin{eqnarray}
\nonumber\hat{e}_2(\tau)&\rightarrow&\tau^2\Bigg\{ -\left(\frac{3}{50}+\frac{7}{50\sqrt{5}}\right)\hat{e}_1(\tau)-\left(\frac{13}{50}+\frac{1}{2\sqrt{5}}\right)\hat{e}_2(\tau)-\left(\frac{9}{50}+\frac{9}{10\sqrt{5}}\right)\hat{e}_3(\tau)\\
\nonumber&&+\left(-\frac{6}{25}+\frac{6}{5\sqrt{5}}\right)\hat{e}_4(\tau)+\frac{42}{25}\hat{e}_5(\tau)+\frac{126}{25\sqrt{5}}\hat{e}_6(\tau)+\frac{42}{25}\hat{e}_7(\tau)+\frac{6}{25}\left(1+\sqrt{5}\right)\hat{e}_8(\tau)\\
&&+\frac{9}{50}\left(-1+\sqrt{5}\right)\hat{e}_9(\tau)+\frac{1}{50}\left(13-5\sqrt{5}\right)\hat{e}_{10}(\tau)+\frac{1}{250}\left(-15+7\sqrt{5}\right)\hat{e}_{11}(\tau)\Bigg\}\,.
\end{eqnarray}
\begin{eqnarray}
\nonumber\hat{e}_3(\tau)&\rightarrow&\tau^2\Bigg\{-\left(\frac{1}{25}+\frac{2}{25\sqrt{5}}\right)\hat{e}_1(\tau)-\left(\frac{1}{25}+\frac{1}{5\sqrt{5}}\right)\hat{e}_2(\tau)+\left(-\frac{1}{50}+\frac{1}{2\sqrt{5}}\right)\hat{e}_3(\tau)+\frac{16}{25}\hat{e}_4(\tau)\\
\nonumber&&+\left(-\frac{7}{25}+\frac{7}{5\sqrt{5}}\right)\hat{e}_5(\tau)-\frac{28}{25\sqrt{5}}\hat{e}_6(\tau)-\frac{7}{25}\left(1+\sqrt{5}\right)\hat{e}_7(\tau)-\frac{16}{25}\hat{e}_8(\tau)\\
&&+\frac{1}{50}\left(-1-5\sqrt{5}\right)\hat{e}_9(\tau)+\frac{1}{25}\left(1-\sqrt{5}\right)\hat{e}_{10}(\tau)+\frac{1}{125}\left(-5+2\sqrt{5}\right)\hat{e}_{11}(\tau)\Bigg\}\,.
\end{eqnarray}
\begin{eqnarray}
\nonumber\hat{e}_4(\tau)&\rightarrow&\tau^2\Bigg\{ -\left(\frac{1}{50}+\frac{3}{50\sqrt{5}}\right)\hat{e}_1(\tau)+\left(-\frac{1}{50}+\frac{1}{10\sqrt{5}}\right)\hat{e}_2(\tau)+\frac{6}{25}\hat{e}_3(\tau)+\left(-\frac{9}{50}+\frac{1}{2\sqrt{5}}\right)\hat{e}_4(\tau)\\
\nonumber&&-\left(\frac{7}{50}+\frac{7}{10\sqrt{5}}\right)\hat{e}_5(\tau)-\frac{21}{25\sqrt{5}}\hat{e}_6(\tau)+\frac{7}{50}\left(-1+\sqrt{5}\right)\hat{e}_7(\tau)+\frac{1}{50}\left(9+5\sqrt{5}\right)\hat{e}_8(\tau)\\
&&+\frac{6}{25}\hat{e}_9(\tau)+\frac{1}{50}\left(1+\sqrt{5}\right)\hat{e}_{10}(\tau)+\frac{1}{250}\left(-5+3\sqrt{5}\right)\hat{e}_{11}(\tau)\Bigg\}\,.
\end{eqnarray}
\begin{eqnarray}
\nonumber\hat{e}_5(\tau)&\rightarrow&\tau^2\Bigg\{ -\left(\frac{1}{50}+\frac{1}{50\sqrt{5}}\right)\hat{e}_1(\tau)+\frac{2}{25}\hat{e}_2(\tau)+\left(-\frac{3}{50}+\frac{3}{10\sqrt{5}}\right)\hat{e}_3(\tau)-\left(\frac{2}{25}+\frac{2}{5\sqrt{5}}\right)\hat{e}_4(\tau)\\
\nonumber&&+\left(\frac{3}{50}-\frac{1}{2\sqrt{5}}\right)\hat{e}_5(\tau)+\frac{18}{25\sqrt{5}}\hat{e}_6(\tau)+\frac{1}{50}\left(3+5\sqrt{5}\right)\hat{e}_7(\tau)-\frac{2}{25}\left(-1+\sqrt{5}\right)\hat{e}_8(\tau)\\
&&-\frac{3}{50}\left(1+\sqrt{5}\right)\hat{e}_9(\tau)-\frac{2}{25}\hat{e}_{10}(\tau)+\frac{1}{250}\left(-5+\sqrt{5}\right)\hat{e}_{11}(\tau)\Bigg\}\,.
\end{eqnarray}
\begin{eqnarray}
\nonumber\hat{e}_6(\tau)&\rightarrow&\tau^2\Bigg\{ -\frac{1}{25\sqrt{5}}\hat{e}_1(\tau)+\frac{1}{5\sqrt{5}}\hat{e}_2(\tau)-\frac{1}{5\sqrt{5}}\hat{e}_3(\tau)-\frac{2}{5\sqrt{5}}\hat{e}_4(\tau)+\frac{3}{5\sqrt{5}}\hat{e}_5(\tau)+\frac{11}{25\sqrt{5}}\hat{e}_6(\tau)\\
&&-\frac{3}{5\sqrt{5}}\hat{e}_7(\tau)-\frac{2}{5\sqrt{5}}\hat{e}_8(\tau)+\frac{1}{5\sqrt{5}}\hat{e}_9(\tau)+\frac{1}{5\sqrt{5}}\hat{e}_{10}(\tau)+\frac{1}{25\sqrt{5}}\hat{e}_{11}(\tau)\Bigg\}\,.
\end{eqnarray}
\begin{eqnarray}
\nonumber\hat{e}_7(\tau)&\rightarrow&\tau^2\Bigg\{ \frac{1}{250}\left(-5+\sqrt{5}\right)\hat{e}_1(\tau)+\frac{2}{25}\hat{e}_2(\tau)-\left(\frac{3}{50}+\frac{3}{10\sqrt{5}}\right)\hat{e}_3(\tau)+\left(-\frac{2}{25}+\frac{2}{5\sqrt{5}}\right)\hat{e}_4(\tau)\\
\nonumber&&+\left(\frac{3}{50}+\frac{1}{2\sqrt{5}}\right)\hat{e}_5(\tau)-\frac{18}{25\sqrt{5}}\hat{e}_6(\tau)+\frac{1}{50}\left(3-5\sqrt{5}\right)\hat{e}_7(\tau)+\frac{2}{25}\left(1+\sqrt{5}\right)\hat{e}_8(\tau)\\
&&\frac{3}{50}\left(-1+\sqrt{5}\right)\hat{e}_9(\tau)-\frac{2}{25}\hat{e}_{10}(\tau)-\frac{1}{250}\left(5+\sqrt{5}\right)\hat{e}_{11}(\tau)\Bigg\}\,.
\end{eqnarray}
\begin{eqnarray}
\nonumber\hat{e}_8(\tau)&\rightarrow&\tau^2\Bigg\{ \left(\frac{1}{50}-\frac{3}{50\sqrt{5}}\right)\hat{e}_1(\tau)+\left(\frac{1}{50}+\frac{1}{10\sqrt{5}}\right)\hat{e}_2(\tau)-\frac{6}{25}\hat{e}_3(\tau)+\left(\frac{9}{50}+\frac{1}{2\sqrt{5}}\right)\hat{e}_4(\tau)\\
\nonumber&&\left(\frac{7}{50}-\frac{7}{10\sqrt{5}}\right)\hat{e}_5(\tau)-\frac{21}{25\sqrt{5}}\hat{e}_6(\tau)+\frac{7}{50}\left(1+\sqrt{5}\right)\hat{e}_7(\tau)+\frac{1}{50}\left(-9+5\sqrt{5}\right)\hat{e}_8(\tau)\\
&&-\frac{6}{25}\hat{e}_9(\tau)+\frac{1}{50}\left(-1+\sqrt{5}\right)\hat{e}_{10}(\tau)+\frac{1}{250}\left(5+3\sqrt{5}\right)\hat{e}_{11}(\tau)\Bigg\}\,.
\end{eqnarray}
\begin{eqnarray}
\nonumber\hat{e}_9(\tau)&\rightarrow&\tau^2\Bigg\{\left(-\frac{1}{25}+\frac{2}{25\sqrt{5}}\right)\hat{e}_1(\tau)+\left(-\frac{1}{25}+\frac{1}{5\sqrt{5}}\right)\hat{e}_2(\tau)-\left(\frac{1}{50}+\frac{1}{2\sqrt{5}}\right)\hat{e}_3(\tau)+\frac{16}{25}\hat{e}_4(\tau)\\
\nonumber&&-\left(\frac{7}{25}+\frac{7}{5\sqrt{5}}\right)\hat{e}_5(\tau)+\frac{28}{25\sqrt{5}}\hat{e}_6(\tau)+\frac{7}{25}\left(-1+\sqrt{5}\right)\hat{e}_7(\tau)-\frac{16}{25}\hat{e}_8(\tau)\\
&&+\frac{1}{50}\left(-1+5\sqrt{5}\right)\hat{e}_9(\tau)+\frac{1}{25}\left(1+\sqrt{5}\right)\hat{e}_{10}(\tau)+\frac{1}{125}\left(-5-2\sqrt{5}\right)\hat{e}_{11}(\tau)\Bigg\}\,.
\end{eqnarray}
\begin{eqnarray}
\nonumber\hat{e}_{10}(\tau)&\rightarrow&\tau^2\Bigg\{ \left(\frac{3}{50}-\frac{7}{50\sqrt{5}}\right)\hat{e}_1(\tau)+\left(\frac{13}{50}-\frac{1}{2\sqrt{5}}\right)\hat{e}_2(\tau)+\left(\frac{9}{50}-\frac{9}{10\sqrt{5}}\right)\hat{e}_3(\tau)\\
\nonumber&&+\left(\frac{6}{25}+\frac{6}{5\sqrt{5}}\right)\hat{e}_4(\tau)-\frac{42}{25}\hat{e}_5(\tau)+\frac{126}{25\sqrt{5}}\hat{e}_6(\tau)-\frac{42}{25}\hat{e}_7(\tau)+\frac{6}{25}\left(-1+\sqrt{5}\right)\hat{e}_8(\tau)\\
&&+\frac{9}{50}\left(1+\sqrt{5}\right)\hat{e}_9(\tau)+\frac{1}{50}\left(-13-5\sqrt{5}\right)\hat{e}_{10}(\tau)+\frac{1}{250}\left(15+7\sqrt{5}\right)\hat{e}_{11}(\tau)\Bigg\}\,.
\end{eqnarray}
\begin{eqnarray}
\nonumber\hat{e}_{11}(\tau)&\rightarrow&\tau^2\Bigg\{\left(-\frac{1}{10}+\frac{11}{50\sqrt{5}}\right)\hat{e}_1(\tau)+\left(-\frac{3}{5}+\frac{7}{5\sqrt{5}}\right)\hat{e}_2(\tau)
+\left(-\frac{9}{5}+\frac{18}{5\sqrt{5}}\right)\hat{e}_3(\tau)\\
\nonumber&&+\left(-\frac{12}{5}+\frac{36}{5\sqrt{5}}\right)\hat{e}_4(\tau)+\left(-\frac{21}{5}+\frac{21}{5\sqrt{5}}\right)\hat{e}_5(\tau)+\frac{252}{25\sqrt{5}}\hat{e}_6(\tau)\\
\nonumber&&-\frac{21}{25}\left(5+\sqrt{5}\right)\hat{e}_7(\tau)+\frac{12}{25}\left(5+3\sqrt{5}\right)\hat{e}_8(\tau)-\frac{9}{25}\left(5+2\sqrt{5}\right)\hat{e}_9(\tau)\\
&&+\left(\frac{3}{5}+\frac{7}{5\sqrt{5}}\right)\hat{e}_{10}(\tau)-\frac{1}{250}\left(25+11\sqrt{5}\right)\hat{e}_{11}(\tau)\Bigg\}\,.
\end{eqnarray}
We see that the basis vectors $\hat{e}_i$ are closed under $S$ and $T$ up to multiplicative factors, and each element is exactly mapped into itself under the action of $S^2$, $(ST)^3$ and $T^5$. Therefore we conclude that $\hat{e}_i$ really span the whole modular space $\mathcal{M}_{2}(\Gamma(5))$. As shown in~\cite{Feruglio:2017spp}, the modular space of weight $2k$ and level $N$ can always be decomposed into different irreducible representations of $\Gamma_{N}$. As a consequence, there should exist modular forms $f_i(\tau)$ of weight $2$ and level $5$ which transform under $\Gamma_5\cong A_5$ as
\begin{equation}
\label{eq:MF_decomp}f_{i}(\gamma(\tau))=(c\tau+d)^{2}\rho(\gamma)_{ij}f_{j}(\tau)\,,
\end{equation}
where
\begin{equation}
\gamma=\begin{pmatrix}
a  ~&~ b  \\
c  ~&~ d
\end{pmatrix}\in \Gamma_5
\end{equation}
is a representative element of $\Gamma_{N}$ and $\rho(\gamma)$ is a unitary representation matrix of $\Gamma_{N}$. It is sufficient that Eq.~\eqref{eq:MF_decomp} is satisfied for the generators $S$ and $T$, i.e.
\begin{equation}
\label{eq:MF_decomp_ST}f_{i}(-1/\tau)=\tau^2\rho(S)_{ij}f_{j}(\tau),\qquad f_{i}(\tau+1)=\rho(T)_{ij}f_{j}(\tau)\,,
\end{equation}
where the representation matrix $\rho(S)$ and $\rho(T)$ are given in Eq.~\eqref{eq:irr_reps}. Solving the constraints in Eq.~\eqref{eq:MF_decomp_ST}, we can construct modular form $Y_{\mathbf{3}}$, $Y_{\mathbf{3}'}$ and $Y_{\mathbf{5}}$ which transform as $\mathbf{3}$, $\mathbf{3}'$ and $\mathbf{5}$ of $A_5$ respectively,
\begin{equation}
\label{eq:modular_space}Y_{\mathbf{3}}\equiv \begin{pmatrix} e_1(\tau) \\
e_2(\tau) \\
e_3(\tau)
\end{pmatrix},\quad Y_{\mathbf{3}'}\equiv \begin{pmatrix} e'_1(\tau) \\
e'_2(\tau) \\
e'_3(\tau)
\end{pmatrix},\quad Y_{\mathbf{5}}\equiv \begin{pmatrix} \widetilde{e}_1(\tau) \\
\widetilde{e}_2(\tau) \\
\widetilde{e}_3(\tau)  \\
\widetilde{e}_4(\tau) \\
\widetilde{e}_5(\tau)
\end{pmatrix}\,,
\end{equation}
with
\begin{eqnarray}
\nonumber&& e_1(\tau)=\hat{e}_1(\tau)-36\hat{e}_6(\tau)-\hat{e}_{11}(\tau),\\
\nonumber&& e_2(\tau)=5\sqrt{2}\,\hat{e}_2(\tau)-15\sqrt{2}\,\hat{e}_7(\tau),\\ \nonumber&&e_3(\tau)=15\sqrt{2}\,\hat{e}_5(\tau)+5\sqrt{2}\,\hat{e}_{10}(\tau),\\
\nonumber&& e'_1(\tau)=\hat{e}_1(\tau)+14\hat{e}_6(\tau)-\hat{e}_{11}(\tau),\\
\nonumber&& e'_2(\tau)=-5\sqrt{2}\,\hat{e}_3(\tau)-10\sqrt{2}\,\hat{e}_8(\tau),\\
\nonumber&&e'_3(\tau)=-10\sqrt{2}\,\hat{e}_4(\tau)+5\sqrt{2}\,\hat{e}_{9}(\tau),\\
\nonumber&& \widetilde{e}_1(\tau)=\hat{e}_1(\tau)+\hat{e}_{11}(\tau),\\
\nonumber&& \widetilde{e}_2(\tau)=-\sqrt{6}\,\hat{e}_2(\tau)-7\sqrt{6}\,\hat{e}_7(\tau),\\
\nonumber&&\widetilde{e}_3(\tau)=-3\sqrt{6}\,\hat{e}_3(\tau)+4\sqrt{6}\,\hat{e}_{8}(\tau),\\
\nonumber&&\widetilde{e}_4(\tau)=-4\sqrt{6}\,\hat{e}_4(\tau)-3\sqrt{6}\,\hat{e}_{9}(\tau),\\
&&\widetilde{e}_5(\tau)=-7\sqrt{6}\,\hat{e}_5(\tau)+\sqrt{6}\,\hat{e}_{10}(\tau)\,.
\end{eqnarray}
The $q$-expansion of the above linearly independent modular forms $e_i$, $e'_i$ and $\widetilde{e}_i$ are given by
\begin{eqnarray}
\nonumber e_1(\tau)&=&1-30q-20q^2-40q^3-90q^4+...\,,\\
\nonumber e_2(\tau)&=&5\sqrt{2}q^{\frac{1}{5}}(1+2q+12q^2+11q^3+12q^4+...)\,,\\
\nonumber e_3(\tau)&=&5\sqrt{2}q^{\frac{4}{5}}(3+7q+6q^2+20q^3+10q^4+...)\,, \\
\nonumber e'_1(\tau)&=&1+20q+30q^2+60^3+60q^4+...\,, \\
\nonumber e'_2(\tau)&=&-5\sqrt{2}q^{\frac{2}{5}}(1+6q+6q^2+16q^3+12q^4+...)\,,\\
\nonumber e'_3(\tau)&=&-5\sqrt{2}q^{\frac{3}{5}}(2+5q+12q^2+7q^3+22q^4+...) \,,\\
\nonumber \widetilde{e}_1(\tau)&=&1+6q+18q^2+24q^3+42q^4+... \,, \\
\nonumber\widetilde{e}_2(\tau)&=&-\sqrt{6}q^{\frac{1}{5}}(1+12q+12q^2+31q^3+32q^4+...) \,,\\
\nonumber\widetilde{e}_3(\tau)&=&-\sqrt{6}q^{\frac{2}{5}}(3+8q+28q^2+18q^3+36q^4+...) \,,\\
\nonumber\widetilde{e}_4(\tau)&=&-\sqrt{6}q^{\frac{3}{5}}(4+15q+14q^2+39q^3+24q^4+...)\,,\\
\widetilde{e}_5(\tau)&=&-\sqrt{6}q^{\frac{4}{5}}(7+13q+24q^2+20q^3+60q^4+...)\,.
\label{q_expansion}
\end{eqnarray}

\subsection{\label{subsec:higher_modular_forms}Weights 4 and 6 modular forms of level $N=5$ }

The weight 4 modular forms can be generated by the tensor products of weight 2 modular forms, and they can be arranged into different $A_5$ irreducible representations as well:
\begin{equation}
\begin{aligned}
&Y^{(4)}_{\mathbf{1}, I}=\left(Y_{\mathbf{3}}Y_{\mathbf{3}}\right)_{\mathbf{1}}=e_1^2+2 e_2 e_3, \qquad Y^{(4)}_{\mathbf{1},II}=\left(Y_{\mathbf{3}'}Y_{\mathbf{3}'}\right)_{\mathbf{1}}=e'^2_1+2 e_2'e_3',\\
&Y^{(4)}_{\mathbf{1}, III}=\left(Y_{\mathbf{5}}Y_{\mathbf{5}}\right)_{\mathbf{1}}=\widetilde{e}_1^2+2 \widetilde{e}_2\widetilde{e}_5+2 \widetilde{e}_3\widetilde{e}_4\,.
\end{aligned}
\end{equation}
\begin{eqnarray}
\nonumber&&Y^{(4)}_{\mathbf{3}, I}=\left(Y_{\mathbf{3}}Y_{\mathbf{5}}\right)_{\mathbf{3}}=\left(-2e_1\widetilde{e}_1+\sqrt{3}\,\left(e_2\widetilde{e}_5+e_3\widetilde{e}_2\right), \sqrt{3}\,e_1\widetilde{e}_2+e_2\widetilde{e}_1-\sqrt{6}\,e_3\widetilde{e}_3, \sqrt{3}\,e_1\widetilde{e}_5-\sqrt{6}\,e_2\widetilde{e}_4+e_3\widetilde{e}_1\right)^{T}\,,\\
&&Y^{(4)}_{\mathbf{3}, II}=\left(Y_{\mathbf{3}'}Y_{\mathbf{5}}\right)_{\mathbf{3}}=\left(\sqrt{3}\,e_1'\widetilde{e}_1+e_2'\widetilde{e}_4+e_3'\widetilde{e}_3, e_1'\widetilde{e}_2-\sqrt{2}\, \left(e_2'\widetilde{e}_5+e_3'\widetilde{e}_4\right), e_1'\widetilde{e}_5-\sqrt{2}\,\left(e_2'\widetilde{e}_3+e_3'\widetilde{e}_2\right)\right)^{T}\,.
\end{eqnarray}
\begin{eqnarray}
\nonumber&&Y^{(4)}_{\mathbf{3}', I}=\left(Y_{\mathbf{3}}Y_{\mathbf{5}}\right)_{\mathbf{3}'}=\left(\sqrt{3}\,e_1\widetilde{e}_1+e_2\widetilde{e}_5+e_3\widetilde{e}_2, e_1\widetilde{e}_3-\sqrt{2}\,(e_2\widetilde{e}_2+\,e_3\widetilde{e}_4),
e_1\widetilde{e}_4-\sqrt{2}\,(e_2\widetilde{e}_3+\,e_3\widetilde{e}_5)\right)^{T}\,,\\
\nonumber&&Y^{(4)}_{\mathbf{3}', II}=\left(Y_{\mathbf{3}'}Y_{\mathbf{5}}\right)_{\mathbf{3}'}= \Big(-2e_1'\widetilde{e}_1+\sqrt{3}\,\left(e_2'\widetilde{e}_4+e_3'\widetilde{e}_3\right), \sqrt{3}\,e_1'\widetilde{e}_3+e_2'\widetilde{e}_1-\sqrt{6}\,e_3'\widetilde{e}_5, \\
&&\qquad\qquad \qquad\qquad~~ \sqrt{3}\,e_1'\widetilde{e}_4-\sqrt{6}\,e_2'\widetilde{e}_2+e_3'\widetilde{e}_1\Big)^{T}\,.
\end{eqnarray}
\begin{eqnarray}
\nonumber&&Y^{(4)}_{\mathbf{4}, I}=\frac{1}{2\sqrt{3}}\left(Y_{\mathbf{5}}Y_{\mathbf{5}}\right)_{\mathbf{4}_S}=\Big(\sqrt{6}\,\widetilde{e}_1\widetilde{e}_2-\widetilde{e}_3 \widetilde{e}_5+2\widetilde{e}_4^2, \sqrt{6}\,\widetilde{e}_1\widetilde{e}_3+2\widetilde{e}_2^2-\widetilde{e}_4\widetilde{e}_5, \sqrt{6}\,\widetilde{e}_1\widetilde{e}_4-\widetilde{e}_2\widetilde{e}_3+2\widetilde{e}_5^2,\\
\nonumber&&\qquad\qquad\qquad\qquad\qquad~~ \sqrt{6}\,\widetilde{e}_1\widetilde{e}_5-\widetilde{e}_2\widetilde{e}_4+2\widetilde{e}_3^2\Big)^{T}\,, \\
\nonumber&&Y^{(4)}_{\mathbf{4}, II}=\left(Y_{\mathbf{3}}Y_{\mathbf{3}'}\right)_{\mathbf{4}}=\left(\sqrt{2}\,e_2 e_1'+e_3e_2', -\sqrt{2}\,e_1 e_2'-e_3e_3', -\sqrt{2}e_1e_3'-e_2 e_2', e_2e_3'+\sqrt{2}\,e_3e_1'\right)^{T}\,,\\
\nonumber&&Y^{(4)}_{\mathbf{4}, III}=\left(Y_{\mathbf{3}}Y_{\mathbf{5}}\right)_{\mathbf{4}}=\Big(2\sqrt{2}\,e_1\widetilde{e}_2-\sqrt{6}\,e_2\widetilde{e}_1+e_3\widetilde{e}_3, -\sqrt{2}\,e_1\widetilde{e}_3+2e_2\widetilde{e}_2-3e_3\widetilde{e}_4, \sqrt{2}\,e_1\widetilde{e}_4+3e_2\widetilde{e}_3 \\
\nonumber&&\qquad\qquad\qquad\qquad~~ -2e_3\widetilde{e}_5, -2\sqrt{2}\,e_1\widetilde{e}_5-e_2\widetilde{e}_4+\sqrt{6}\,e_3\widetilde{e}_1\Big)^{T}\,,\\
\nonumber&&Y^{(4)}_{\mathbf{4}, IV}=\left(Y_{\mathbf{3}'}Y_{\mathbf{5}}\right)_{\mathbf{4}}=
\Big(\sqrt{2}\,e_1'\widetilde{e}_2+3e_2'\widetilde{e}_5-2e_3'\widetilde{e}_4, 2\sqrt{2}\,e_1'\widetilde{e}_3-\sqrt{6}\,e_2'\widetilde{e}_1+e_3'\widetilde{e}_5, -2\sqrt{2}\,e_1'\widetilde{e}_4-e_2'\widetilde{e}_2\\
&&\qquad\qquad\qquad\qquad~~+\sqrt{6}\,e_3'\widetilde{e}_1,-\sqrt{2}\,e_1'\widetilde{e}_5+2e_2'\widetilde{e}_3-3e_3'\widetilde{e}_2\Big)^{T}\,.
\end{eqnarray}
\begin{eqnarray}
\nonumber&& Y^{(4)}_{\mathbf{5}, I}=\left(Y_{\mathbf{3}}Y_{\mathbf{3}}\right)_{\mathbf{5}}=\left(2 \left(e_1^2-e_2e_3\right),-2 \sqrt{3}\, e_1e_2, \sqrt{6}\, e_2^2,\sqrt{6}\,  e_3^2, -2 \sqrt{3}\, e_1 e_3\right)^{T}\,,\\
\nonumber&&Y^{(4)}_{\mathbf{5}, II}=\left(Y_{\mathbf{3}'}Y_{\mathbf{3}'}\right)_{\mathbf{5}}=\left(2\left(e'^{2}_1-e_2'e_3'\right),\sqrt{6}\,e'^{2}_3, -2\sqrt{3}\,e_1'e_2', -2\sqrt{3}\,e_1' e_3', \sqrt{6}\,e'^{2}_2\right)^{T},\\
\nonumber&&Y^{(4)}_{\mathbf{5}, III}=\frac{1}{2}\left(Y_{\mathbf{5}}Y_{\mathbf{5}}\right)_{\mathbf{5}_{S,1}}=\Big(\widetilde{e}_1^2+\widetilde{e}_2\widetilde{e}_5-2\widetilde{e}_3\widetilde{e}_4, \widetilde{e}_1\widetilde{e}_2+\sqrt{6}\,\widetilde{e}_3\widetilde{e}_5,\sqrt{\frac{3}{2}}\,\widetilde{e}_2^2-2   \widetilde{e}_1\widetilde{e}_3, \sqrt{\frac{3}{2}}\,\widetilde{e}_5^2-2\widetilde{e}_1\widetilde{e}_4, \\
\nonumber&&\qquad\qquad\qquad\qquad\qquad~~\widetilde{e}_1\widetilde{e}_5+\sqrt{6}\,\widetilde{e}_2\widetilde{e}_4\Big)^{T}\,,\\
\nonumber&&Y^{(4)}_{\mathbf{5}, IV}=\frac{1}{2}\left(Y_{\mathbf{5}}Y_{\mathbf{5}}\right)_{\mathbf{5}_{S,2}}=\Big(\widetilde{e}_1^2-2\widetilde{e}_2\widetilde{e}_5+\widetilde{e}_3\widetilde{e}_4, \sqrt{\frac{3}{2}}\,\widetilde{e}_4^2-2\widetilde{e}_1\widetilde{e}_2, \widetilde{e}_1\widetilde{e}_3+\sqrt{6}\,\widetilde{e}_4\widetilde{e}_5, \widetilde{e}_1\widetilde{e}_4+\sqrt{6}\,\widetilde{e}_2\widetilde{e}_3,\\
\nonumber&&\qquad\qquad\qquad\qquad\qquad
\sqrt{\frac{3}{2}}\,\widetilde{e}_3^2-2\widetilde{e}_1\widetilde{e}_5\Big)^{T}\,,\\
\nonumber&&Y^{(4)}_{\mathbf{5}, V}=\left(Y_{\mathbf{3}}Y_{\mathbf{3}'}\right)_{\mathbf{5}}=\left(\sqrt{3}\,e_1e_1', e_2e_1'-\sqrt{2}\,e_3 e_2', e_1e_2'-\sqrt{2}\,e_3 e_3', e_1e_3'-\sqrt{2}\,e_2 e_2', e_3e_1'-\sqrt{2}\, e_2 e_3'\right)^{T}\,, \\
\nonumber&&Y^{(4)}_{\mathbf{5}, VI}=\left(Y_{\mathbf{3}}Y_{\mathbf{5}}\right)_{\mathbf{5}}=\Big(\sqrt{3}\,\left(e_2\widetilde{e}_5-e_3\widetilde{e}_2\right), -e_1\widetilde{e}_2-\sqrt{3}\,e_2\widetilde{e}_1-\sqrt{2}\,e_3\widetilde{e}_3, -2e_1\widetilde{e}_3-\sqrt{2}\,e_2\widetilde{e}_2,\\
\nonumber&&\qquad\qquad\qquad\qquad~~ 2e_1\widetilde{e}_4+\sqrt{2}\,e_3\widetilde{e}_5, e_1\widetilde{e}_5+\sqrt{2}\,e_2\widetilde{e}_4+\sqrt{3}\,e_3\widetilde{e}_1\Big)^{T}\,,\\
\nonumber&&Y^{(4)}_{\mathbf{5}, VII}=\left(Y_{\mathbf{3}'}Y_{\mathbf{5}}\right)_{\mathbf{5}}=\Big(\sqrt{3}\,\left(e_2'\widetilde{e}_4-e_3'\widetilde{e}_3\right), 2e_1'\widetilde{e}_2+\sqrt{2}\,e_3'\widetilde{e}_4, -e_1'\widetilde{e}_3-\sqrt{3}\,e_2'\widetilde{e}_1-\sqrt{2}\,e_3'\widetilde{e}_5,\\
\label{eq:modularFunc_5plet}&&\qquad\qquad\qquad\qquad~~ e_1'\widetilde{e}_4+\sqrt{2}\,e_2'\widetilde{e}_2+\sqrt{3}\,e_3'\widetilde{e}_1, -2e_1'\widetilde{e}_5-\sqrt{2}\,e_2'\widetilde{e}_3\Big)^{T}\,,
\end{eqnarray}
where the upper index is the weight and the lower index refers to the representation under $A_5$. We notice that not all the above weight 4 modular functions are not linearly independent. From the expressions of the $q-$expansion of $e_{i}$, $e'_{i}$ and $\widetilde{e}_i$ given in Eq.~\eqref{q_expansion}, we find the following relations are satisfied,
\begin{eqnarray}
\label{eq:constraints_singlet}&&Y^{(4)}_{\mathbf{1}, I}=Y^{(4)}_{\mathbf{1},II}=Y^{(4)}_{\mathbf{1},III},\quad Y^{(4)}_{\mathbf{3}, II}=-\frac{\sqrt{3}}{2}Y^{(4)}_{\mathbf{3}, I},\quad Y^{(4)}_{\mathbf{3}', I}=-\frac{\sqrt{3}}{2}Y^{(4)}_{\mathbf{3}', II},\\
\label{eq:constraints_4plet}&&Y^{(4)}_{\mathbf{4}, I}=-\frac{3}{5}Y^{(4)}_{\mathbf{4}, II},\quad Y^{(4)}_{\mathbf{4}, III}=-\frac{7\sqrt{3}}{5}Y^{(4)}_{\mathbf{4}, II},\quad Y^{(4)}_{\mathbf{4}, IV}=-\frac{\sqrt{3}}{5}Y^{(4)}_{\mathbf{4}, II},\\
\label{eq:constraints_5plet1}&&\quad\qquad Y^{(4)}_{\mathbf{5}, III}=\frac{1}{10}Y^{(4)}_{\mathbf{5}, I}+\frac{2}{5}Y^{(4)}_{\mathbf{5}, II},\quad  Y^{(4)}_{\mathbf{5}, IV}=-\frac{1}{5}Y^{(4)}_{\mathbf{5}, I}+\frac{7}{10}Y^{(4)}_{\mathbf{5}, II}\,,\\
\label{eq:constraints_5plet2}&&Y^{(4)}_{\mathbf{5}, V}=-\frac{1}{2\sqrt{3}}Y^{(4)}_{\mathbf{5}, I}+\frac{2}{\sqrt{3}}Y^{(4)}_{\mathbf{5}, II},\quad Y^{(4)}_{\mathbf{5}, VI}=\frac{2}{5}Y^{(4)}_{\mathbf{5}, I}-\frac{2}{5}Y^{(4)}_{\mathbf{5}, II},\quad Y^{(4)}_{\mathbf{5}, VI}=2Y^{(4)}_{\mathbf{5}, VII}\,.
\end{eqnarray}
Therefore the singlet modular forms $Y^{(4)}_{\mathbf{1}, I}$, $Y^{(4)}_{\mathbf{1}, II}$ and $Y^{(4)}_{\mathbf{1},III}$ are exactly identical, the triplet modular functions $Y^{(4)}_{\mathbf{3}, II}$ and $Y^{(4)}_{\mathbf{3}', II}$ are parallel to $Y^{(4)}_{\mathbf{3}, I}$ and $Y^{(4)}_{\mathbf{3}', I}$ respectively. Analogously $Y^{(4)}_{\mathbf{4}, I}$, $Y^{(4)}_{\mathbf{4}, III}$ and $Y^{(4)}_{\mathbf{4}, IV}$ are proportional to $Y^{(4)}_{\mathbf{4}, II}$. Moreover, Eqs.~(\ref{eq:constraints_5plet1}, \ref{eq:constraints_5plet2}) imply that all the seven $\mathbf{5}$-plet modular functions in Eq.~\eqref{eq:modularFunc_5plet} can be expressed in terms of the linear combinations of $Y^{(4)}_{\mathbf{5}, I}$ and $Y^{(4)}_{\mathbf{5}, II}$. As a result, there are totally 21 independent modular functions which can be chosen as $Y^{(4)}_{\mathbf{1}, I}$, $Y^{(4)}_{\mathbf{3}, I}$, $Y^{(4)}_{\mathbf{3}', II}$, $Y^{(4)}_{\mathbf{4}, II}$, $Y^{(4)}_{\mathbf{5}, I}$ and $Y^{(4)}_{\mathbf{5}, II}$. Note the linear space of modular forms of weight $2k$ and level 5 is of dimension $10k+1$ which is equal to $21$ for $k=2$. Thus we have constructed an explicit basis for the weight 4 modular forms at level 5.

As regards the weight 6 modular forms for $k=3$, the dimension of the modular space is equal to $10\times3+1=31$. That is  to say, there should be 31 independent combination $Y_{i}Y_{j}Y_{k}$ where $i, j, k=\mathbf{3}, \mathbf{3}', \mathbf{5}$. In the same fashion as the $k=2$ case, all the basis vectors and the corresponding constraints can be straightforwardly found although the involved algebra is lengthy and tedious. Firstly we consider the weight 6 modular forms which are $A_5$ singlet. We can construct the following four singlet modular forms of weight 6 through the tensor products of weight 2 and weight 4 modular forms listed above,
\begin{equation}
\begin{aligned}
&Y^{(6)}_{\mathbf{1}, I}=\left(Y_{\mathbf{3}}Y^{(4)}_{\mathbf{3}, I}\right)_{\mathbf{1}}=\left[Y_{\mathbf{3}}\left(Y_{\mathbf{3}}Y_{\mathbf{5}}\right)_{\mathbf{3}}\right]_{\mathbf{1}},\quad
Y^{(6)}_{\mathbf{1}, II}=\left(Y_{\mathbf{3}'}Y^{(4)}_{\mathbf{3}', II}\right)_{\mathbf{1}}=\left[Y_{\mathbf{3}'}\left(Y_{\mathbf{3}'}Y_{\mathbf{5}}\right)_{\mathbf{3}'}\right]_{\mathbf{1}}\,,\\
&Y^{(6)}_{\mathbf{1}, III}=\left(Y_{\mathbf{5}}Y^{(4)}_{\mathbf{5}, I}\right)_{\mathbf{1}}=\left[Y_{\mathbf{5}}\left(Y_{\mathbf{3}}Y_{\mathbf{3}}\right)_{\mathbf{5}}\right]_{\mathbf{1}},\quad Y^{(6)}_{\mathbf{1}, IV}=\left(Y_{\mathbf{5}}Y^{(4)}_{\mathbf{5}, II}\right)_{\mathbf{1}}=\left[Y_{\mathbf{5}}\left(Y_{\mathbf{3}'}Y_{\mathbf{3}'}\right)_{\mathbf{5}}\right]_{\mathbf{1}}\,,
\end{aligned}
\end{equation}
which fulfill the relations
\begin{equation}
Y^{(6)}_{\mathbf{1}, II}=-Y^{(6)}_{\mathbf{1}, III}=-Y^{(6)}_{\mathbf{1}, IV}=Y^{(6)}_{\mathbf{1}, I}\,.
\end{equation}
Hence there is only one singlet modular form of weight 6 and it can be taken as
\begin{equation}
Y^{(6)}_{\mathbf{1}, I}=\left(Y_{\mathbf{3}}Y^{(4)}_{\mathbf{3}, I}\right)_{\mathbf{1}}=-2e_1^2\tilde{e}_1-\sqrt{6}\,e_2^2\tilde{e}_4-\sqrt{6}\,e_3^2\tilde{e}_3+2\sqrt{3}\,e_1e_2\tilde{e}_5+2\sqrt{3}\,e_1e_3\tilde{e}_2+2e_2e_3\tilde{e}_1\,.
\end{equation}
Analogously we can construct the following twelve weight 6 modular forms transforming as $\mathbf{3}$ under $A_5$,
\begin{eqnarray}
\nonumber&& Y_{\mathbf{3}}Y^{(4)}_{\mathbf{1}, I}=Y_{\mathbf{3}}\left(Y_{\mathbf{3}}Y_{\mathbf{3}}\right)_{\mathbf{1}},\quad \left(Y_{\mathbf{3}}Y^{(4)}_{\mathbf{3}, I}\right)_{\mathbf{3}}=\left[Y_{\mathbf{3}}\left(Y_{\mathbf{3}}Y_{\mathbf{5}}\right)_{\mathbf{3}}\right]_{\mathbf{3}},\quad
\left(Y_{\mathbf{3}}Y^{(4)}_{\mathbf{5}, I}\right)_{\mathbf{3}}=\left[Y_{\mathbf{3}}\left(Y_{\mathbf{3}}Y_{\mathbf{3}}\right)_{\mathbf{5}}\right]_{\mathbf{3}},\\
\nonumber&& \left(Y_{\mathbf{3}}Y^{(4)}_{\mathbf{5}, II}\right)_{\mathbf{3}}=\left[Y_{\mathbf{3}}\left(Y_{\mathbf{3}'}Y_{\mathbf{3}'}\right)_{\mathbf{5}}\right]_{\mathbf{3}},\quad \left(Y_{\mathbf{3}'}Y^{(4)}_{\mathbf{4}, II}\right)_{\mathbf{3}}=\left[Y_{\mathbf{3}'}\left(Y_{\mathbf{3}}Y_{\mathbf{3}'}\right)_{\mathbf{4}}\right]_{\mathbf{3}},
\quad
\left(Y_{\mathbf{3}'}Y^{(4)}_{\mathbf{5}, I}\right)_{\mathbf{3}}=\left[Y_{\mathbf{3}'}\left(Y_{\mathbf{3}}Y_{\mathbf{3}}\right)_{\mathbf{5}}\right]_{\mathbf{3}},\\
\nonumber&&\left(Y_{\mathbf{3}'}Y^{(4)}_{\mathbf{5}, II}\right)_{\mathbf{3}}=\left[Y_{\mathbf{3}'}\left(Y_{\mathbf{3}'}Y_{\mathbf{3}'}\right)_{\mathbf{5}}\right]_{\mathbf{3}},\quad
\left(Y_{\mathbf{5}}Y^{(4)}_{\mathbf{3}, I}\right)_{\mathbf{3}}=\left[Y_{\mathbf{5}}\left(Y_{\mathbf{3}}Y_{\mathbf{5}}\right)_{\mathbf{3}}\right]_{\mathbf{3}}, \quad \left(Y_{\mathbf{5}}Y^{(4)}_{\mathbf{3}', II}\right)_{\mathbf{3}}=\left[Y_{\mathbf{5}}\left(Y_{\mathbf{3}'}Y_{\mathbf{5}}\right)_{\mathbf{3}'}\right]_{\mathbf{3}},\\
&&\left(Y_{\mathbf{5}}Y^{(4)}_{\mathbf{4}, II}\right)_{\mathbf{3}}=\left[Y_{\mathbf{5}}\left(Y_{\mathbf{3}}Y_{\mathbf{3}'}\right)_{\mathbf{4}}\right]_{\mathbf{3}},\quad
\left(Y_{\mathbf{5}}Y^{(4)}_{\mathbf{5}, I}\right)_{\mathbf{3}}=\left[Y_{\mathbf{5}}\left(Y_{\mathbf{3}}Y_{\mathbf{3}}\right)_{\mathbf{5}}\right]_{\mathbf{3}},\quad \left(Y_{\mathbf{5}}Y^{(4)}_{\mathbf{5}, II}\right)_{\mathbf{3}}=\left[Y_{\mathbf{5}}\left(Y_{\mathbf{3}'}Y_{\mathbf{3}'}\right)_{\mathbf{5}}\right]_{\mathbf{3}}\,.
\end{eqnarray}
Furthermore, we find only two of them are linearly independent and they could be chosen to be
\begin{eqnarray}
\nonumber&&\qquad\qquad\qquad\qquad\qquad Y^{(6)}_{\mathbf{3}, I}=Y_{\mathbf{3}}Y^{(4)}_{\mathbf{1}, I}=(e_1^2+2 e_2 e_3)\left(e_1, e_2, e_3\right)^{T}\,,\\
&&Y^{(6)}_{\mathbf{3}, II}=\left(Y_{\mathbf{3}'}Y^{(4)}_{\mathbf{5}, II}\right)_{\mathbf{3}}=\Big(2\sqrt{3}\,e'^{3}_1-6\sqrt{3}\,e_1' e_2' e_3', 3\sqrt{6}\,e_1'e'^{2}_3-2\sqrt{3}\,e'^{3}_2, 3\sqrt{6}\,e_1'e'^{2}_2-2\sqrt{3}\,e'^{3}_3\Big)^{T}\,.
\end{eqnarray}
In a similar manner, we can construct the following twelve $A_5$ triplet $\mathbf{3}'$ modular forms of weight 6,
\begin{eqnarray}
\nonumber&& \left(Y_{\mathbf{3}}Y^{(4)}_{\mathbf{4}, II}\right)_{\mathbf{3}'}=\left[Y_{\mathbf{3}}\left(Y_{\mathbf{3}}Y_{\mathbf{3}'}\right)_{\mathbf{4}}\right]_{\mathbf{3}'},
\quad
\left(Y_{\mathbf{3}}Y^{(4)}_{\mathbf{5}, I}\right)_{\mathbf{3}'}=\left[Y_{\mathbf{3}}\left(Y_{\mathbf{3}}Y_{\mathbf{3}}\right)_{\mathbf{5}}\right]_{\mathbf{3}'},\quad
\left(Y_{\mathbf{3}}Y^{(4)}_{\mathbf{5}, II}\right)_{\mathbf{3}'}=\left[Y_{\mathbf{3}}\left(Y_{\mathbf{3}'}Y_{\mathbf{3}'}\right)_{\mathbf{5}}\right]_{\mathbf{3}'},
\\
\nonumber&&
Y_{\mathbf{3}'}Y^{(4)}_{\mathbf{1}, I}=Y_{\mathbf{3}'}\left(Y_{\mathbf{3}}Y_{\mathbf{3}}\right)_{\mathbf{1}}, \quad \left(Y_{\mathbf{3}'}Y^{(4)}_{\mathbf{3}', II}\right)_{\mathbf{3}}=\left[Y_{\mathbf{3}'}\left(Y_{\mathbf{3}'}Y_{\mathbf{5}}\right)_{\mathbf{3}'}\right]_{\mathbf{3}'},\quad \left(Y_{\mathbf{3}'}Y^{(4)}_{\mathbf{5}, I}\right)_{\mathbf{3}'}=\left[Y_{\mathbf{3}'}\left(Y_{\mathbf{3}}Y_{\mathbf{3}}\right)_{\mathbf{5}}\right]_{\mathbf{3}'}, \\
\nonumber&&\left(Y_{\mathbf{3}'}Y^{(4)}_{\mathbf{5}, II}\right)_{\mathbf{3}'}=\left[Y_{\mathbf{3}'}\left(Y_{\mathbf{3}'}Y_{\mathbf{3}'}\right)_{\mathbf{5}}\right]_{\mathbf{3}'},\quad
\left(Y_{\mathbf{5}}Y^{(4)}_{\mathbf{3}, I}\right)_{\mathbf{3}'}=\left[Y_{\mathbf{5}}\left(Y_{\mathbf{3}}Y_{\mathbf{5}}\right)_{\mathbf{3}}\right]_{\mathbf{3}'}, \quad \left(Y_{\mathbf{5}}Y^{(4)}_{\mathbf{3}', II}\right)_{\mathbf{3}'}=\left[Y_{\mathbf{5}}\left(Y_{\mathbf{3}'}Y_{\mathbf{5}}\right)_{\mathbf{3}'}\right]_{\mathbf{3}'},\\
&&\left(Y_{\mathbf{5}}Y^{(4)}_{\mathbf{4}, II}\right)_{\mathbf{3}'}=\left[Y_{\mathbf{5}}\left(Y_{\mathbf{3}}Y_{\mathbf{3}'}\right)_{\mathbf{4}}\right]_{\mathbf{3}'},\quad
\left(Y_{\mathbf{5}}Y^{(4)}_{\mathbf{5}, I}\right)_{\mathbf{3}'}=\left[Y_{\mathbf{5}}\left(Y_{\mathbf{3}}Y_{\mathbf{3}}\right)_{\mathbf{5}}\right]_{\mathbf{3}'},\quad \left(Y_{\mathbf{5}}Y^{(4)}_{\mathbf{5}, II}\right)_{\mathbf{3}'}=\left[Y_{\mathbf{5}}\left(Y_{\mathbf{3}'}Y_{\mathbf{3}'}\right)_{\mathbf{5}}\right]_{\mathbf{3}'}\,.
\end{eqnarray}
Only two out of the above twelve modular forms are linearly independent and they can be taken as
\begin{eqnarray}
\nonumber&&\qquad\qquad\qquad\qquad\quad  Y^{(6)}_{\mathbf{3}', I}=Y_{\mathbf{3}'}Y^{(4)}_{\mathbf{1}, I}=(e_1^2+2 e_2 e_3)\left(e'_1, e'_2, e'_3\right)^{T}\,,\\
&&Y^{(6)}_{\mathbf{3}', II}=
\left(Y_{\mathbf{3}}Y^{(4)}_{\mathbf{5}, I}\right)_{\mathbf{3}'}=
\Big(2\sqrt{3}\,e^{3}_1-6\sqrt{3}\,e_1 e_2 e_3, 3\sqrt{6}\,e_1e^{2}_2-2\sqrt{3}\,e^{3}_3, 3\sqrt{6}\,e_1e^{2}_3-2\sqrt{3}\,e^{3}_2\Big)^{T}\,.
\end{eqnarray}
Eventually we find that there are two linearly independent four-dimensional and five-dimensional modular forms of weight 6
\begin{eqnarray}
\nonumber&& Y^{(6)}_{\mathbf{4}, I}=\left(Y_{\mathbf{3}}Y^{(4)}_{\mathbf{5}, I}\right)_{\mathbf{4}}/6\sqrt{6}=\left(-2e_1^2e_2+e_2^2e_3,-e_3^3-\sqrt{2}e_1e_2^2, e_2^3+\sqrt{2}e_1e_3^2, 2e_1^2e_3-e_2e_3^2\right)^{T}\,,\\
\nonumber&&Y^{(6)}_{\mathbf{4}, II}=\left(Y_{\mathbf{3}'}Y^{(4)}_{\mathbf{5}, II}\right)_{\mathbf{4}}/6\sqrt{6}=\left(e_2'^3+\sqrt{2} e_1'e_3'^2, -2e_1'^2e_2'+e_2'^2e_3', 2e_1'^2e_3'-e_2'e_3'^2, -e_3'^3-\sqrt{2}e_1'e_2'^2\right)^T\,,\\
\nonumber&& Y^{(6)}_{\mathbf{5}, I}=Y_{\mathbf{5}}Y^{(4)}_{\mathbf{1}, I}=(e_1^2+2 e_2 e_3)\left(\widetilde{e}_1, \widetilde{e}_2, \widetilde{e}_3, \widetilde{e}_4, \widetilde{e}_5\right)^{T}\,,\\
\nonumber&&Y^{(6)}_{\mathbf{5}, II}=\left(Y_{\mathbf{3}}Y^{(4)}_{\mathbf{4}, II}\right)_{\mathbf{5}}=\Big(\sqrt{6} e_2^2 e_3'-\sqrt{6} e_3^2 e_2', 4 e_1 e_2 e_1'-2 e_3^2 e_3', 2e_1^2 e_2'+\sqrt{2} e_2^2 e_1'-2\sqrt{2} e_1e_3 e_3' \\
&&\qquad\qquad -2 e_2e_3e_2', -2e_1^2 e_3'-\sqrt{2}e_3^2e_1'+2 \sqrt{2}e_1e_2 e_2'+2 e_2 e_3 e_3', 2e_2^2 e_2'-4 e_1 e_3 e_1'\Big)^{T}\,.
\end{eqnarray}
In short, a linearly independent basis of weight 6 modular forms for $\Gamma(5)$ can be chosen to be $Y^{(6)}_{\mathbf{1}, I}$, $Y^{(6)}_{\mathbf{3}, I}$, $Y^{(6)}_{\mathbf{3}, II}$, $Y^{(6)}_{\mathbf{3}', I}$, $Y^{(6)}_{\mathbf{3}', II}$, $Y^{(6)}_{\mathbf{4}, I}$, $Y^{(6)}_{\mathbf{4}, II}$, $Y^{(6)}_{\mathbf{5}, I}$ and $Y^{(6)}_{\mathbf{5}, II}$.

\section{\label{sec:model}Models based on $\Gamma_5$}

In this section,we shall construct some minimal modular-invariant supersymmetric model based on $\Gamma_5$. Adopting the $N=1$ global supersymmetry, the most general form of the action is
\begin{equation}
\mathcal{S}=\int d^4 x d^2\theta d^2\bar \theta~ K(\Phi_I,\bar{\Phi}_I; \tau,\bar{\tau})+\int d^4 x d^2\theta~ W(\Phi_I,\tau)+\mathrm{h.c.}\,,
\end{equation}
where the K$\ddot{\mathrm{a}}$hler potential $K(\Phi_I,\bar{\Phi}_I; \tau,\bar{\tau})$ is a real gauge invariant function of the set of chiral superfields $\Phi_I$. The action is required to be invariant under the finite modular group $\Gamma_{N}$. The supermultiplets $\Phi_I$ are assumed to transform in a representation $\rho_{I}$ of the quotient group $\Gamma_{N}$ with a weight $-k_{I}$,
\begin{equation}
\label{eq:modular_transform}\left\{\begin{array}{l}
\tau\to \gamma(\tau) = \dfrac{a \tau+b}{c \tau+d}\,,\\ \\
\Phi_I\to (c\tau+d)^{-k_I}\rho_I(\gamma)\Phi_I\,,
\end{array} \right.\quad \text{with}\quad \gamma=\begin{pmatrix}
a  &  b  \\
c  &  d
\end{pmatrix}\in\Gamma_{N}\,,
\end{equation}
where $\rho_I(\gamma)$ is the representation matrix of the element $\gamma$ and $k_{I}$ are integers. Under the modular transformation of Eq.~\eqref{eq:modular_transform}, the K$\ddot{\mathrm{a}}$hler potential should be invariant
up to K$\ddot{\mathrm{a}}$hler transformations. Thus one can determine the  K$\ddot{\mathrm{a}}$hler potential as
\begin{equation}
K(\Phi_I,\bar{\Phi}_I; \tau,\bar{\tau}) =-h \log(-i\tau+i\bar\tau)+ \sum_I (-i\tau+i\bar\tau)^{-k_I} |\Phi_I|^2~~~,
\end{equation}
where the constant $h>0$. After the modulus $\tau$ develops a vacuum expectation value (VEV), the above K$\ddot{\mathrm{a}}$hler potential leads to the following kinetic term for the scalar components of the supermultiplets $\Phi_I$ and $\tau$,
\begin{equation}
\frac{h}{\langle-i\tau+i\bar\tau\rangle^2}\partial_\mu \bar\tau\partial^\mu \tau+\sum_I \frac{\partial_\mu  \bar{\phi}_I \partial^\mu \phi_I}{\langle-i\tau+i\bar\tau\rangle^{k_I}}\,.
\end{equation}
For each given value of the VEV of $\tau$, the kinetic term of $\phi_I$ can be made into canonical form by rescaling the fields $\phi_I$. This effect amounts to a redefinition of the superpotential parameters in a concrete model.

One the other hand, the superpotential $W(\Phi_I,\tau)$ should be invariant under the group $\Gamma_{N}$, and the total weight of $W(\Phi_I,\tau)$ should be vanishing. In general, the modular invariant superpotential $W(\Phi_I,\tau)$ can be expanded in power series of the supermultiplets $\Phi_I$,
\begin{equation}
W(\Phi_I,\tau) =\sum_n Y_{I_1...I_n}(\tau)~ \Phi_{I_1}... \Phi_{I_n}\,.
\end{equation}
For the transformation rule of the field $\Phi_{I}$ in Eq.~\eqref{eq:modular_transform}, the functions $Y_{I_1...I_n}$ should be modular forms of weight $k_Y$ and transform in the presentation $\rho_{Y}$ of $\Gamma_{N}$,
\begin{equation}
\left\{\begin{array}{l}
\tau\to \gamma(\tau) =\dfrac{a\tau+b}{c\tau+d}\,,\\ \\
Y(\tau)\to Y(\gamma(\tau))=(c\tau+d)^{k_Y}\rho_{Y}(\gamma)Y(\tau)\,.
\end{array}
\right.
\end{equation}
The modular invariance requires
\begin{equation}
k_Y=k_{I_1}+...+k_{I_n}\,,\qquad \rho_Y\otimes\rho_{I_1}\otimes...\otimes\rho_{I_n}\ni\mathbf{1}\,.
\end{equation}
In the following, we shall construct some simple models of lepton masses and mixing based on the $\Gamma_5$ modular symmetry. The neutrinos are assumed to be Majorana particles. We shall formulate our models in the framework of supersymmetry, i.e. the minimal supersymmetric standard model (MSSM) and its extension with right-handed neutrinos. In order to be as simple as possible, we suppose the Higgs doublets $H_{u,d}$ are singlets of $A_5$ and their modular weights are zero.
There are freedoms for the assignments of irreducible representation and modular weight to the matter fields. We assume the three left-handed lepton doublets transform as an irreducible three-dimensional representation $\mathbf{3}$ or $\mathbf{3}'$ of the $\Gamma_5$ flavor group. The generic assignments of representations and modular weights to the MSSM fields are listed in table~\ref{tab:generalmod}. In the following, we proceed to discuss the possible structures of the models in the charged lepton and neutrino sectors.
\begin{table}[t!]
\centering
\begin{tabular}{|c|c|c|c|c|c|}
\hline\hline
 &$N^c$& $E^c_{1,2,3}$& $L$& $H_d$& $H_u$\\
\hline
$SU(2)_L\times U(1)_Y$&$(1,0)$& $(1,1)$& $(2,-1/2)$& $(2,-1/2)$& $(2,+1/2)$\\
\hline
$\Gamma_5$&$\rho_N$& $\rho_{E_{1,2,3}}$& $\rho_L$& $\mathbf{1}$ & $\mathbf{1}$ \\
\hline
$k_I$&$k_N$& $k_{E_{1,2,3}}$& $k_L$& $0$& $0$\\
\hline\hline
\end{tabular}
\caption{\label{tab:generalmod}The transformation properties of the MSSM chiral superfields under the SM gauge group $SU(2)_L\times U(1)_Y$ and the $\Gamma_5$ modular symmetry, where $-k_{I}$ denotes the modular weight of the fields.}
\end{table}

\subsection{\label{subsec:charged_lepton}Charged lepton sector}

The most general superpotential for the charged lepton masses can be written as
\begin{equation}
\label{eq:We}W_e =\alpha \big(E^c L H_d f_E(Y,\varphi)\big)_\mathbf{1}\,,
\end{equation}
where $\varphi$ denotes a generic flavon field which is a SM singlet, and $f_E(Y,\varphi)$ is the most general function of the modular forms $Y(\tau)$ and the flavon field $\varphi$. Notice that the total weight of each term of $W_e$ should be vanishing. In the present work, for simplicity we shall consider the case that $f_E(Y,\varphi)$ only depends on either the flavon field $\varphi$ or the modular forms $Y$. For the scenario that only $\varphi$ is involved in $W_e$, we can choose the modular weights $k_{I}$ to forbid a dependence of $W_e$ on the modulus $\tau$. As shown in Appendix~\ref{sec:App_charged_lepton}, the flavon multiplets can develop the VEVs along certain directions such that the charged lepton mass matrix is diagonal. Then the lepton flavor mixing completely arises from the neutrino sector.

In the second scenario that the function $f_E(Y,\varphi)$ only depends on
the modulus $\tau$, the three generations of left-handed lepton doublets are assigned to the triplet representation $\mathbf{3}$ or $\mathbf{3}'$ while all the right-handed charged leptons $E^{c}_i$ transform as $\mathbf{1}$ under $\Gamma_5$. In order to avoid a charged lepton mass matrix with rank less than $3$, we assume that $E^{c}_1$, $E^{c}_2$ and $E^{c}_3$ have different modular weights such that modular forms of weights $2,\,4,\,6$ (i.e. $Y$, $Y^{(4)}$, $Y^{(6)}$) are involved. Thus the superpotential for the charged lepton masses is of the following form,
\begin{equation}
\label{eq:WE1} W_e=\alpha \big(E^c_1\, L\, Y\,\big)_\mathbf{1}H_d\, +\,\beta \big(E^c_2\, L\, Y^{(4)}\,\big)_\mathbf{1}H_d\, +\, \gamma \big(E^c_3\, L\, Y^{(6)}\,\big)_\mathbf{1}H_d\,,
\end{equation}
where the condition of weight cancellation requires
\begin{equation}
\begin{cases}
2\,=\,k_{E_1}\,+\,k_L\,,\\
4\,=\,k_{E_2}\,+\,k_L\,, \\
6\,=\,k_{E_3}\,+\,k_L\,,
\end{cases}
\Rightarrow\quad
k_{E_3}\,=\,6\,-\,k_L\,=\,k_{E_2}\,+2\,=\,k_{E_1}\,+4\,.
\label{weightconstr1}
\end{equation}\\
Notice that $Y$, $Y^{(4)}$ and $Y^{(6)}$ should transform in the same way as $L$ under $A_5$ in order to form $A_5$ singlet. All the linearly independent modular forms of weights 2, 4 and 6 have been found out in section~\ref{sec:modular_sym_gen}. The explicit form of the superpotential $W_e$ is different for $\rho_L\sim\mathbf{3} $ and $\rho_L\sim\mathbf{3}'$. To be more specific, using the Clebsch-Gordan coefficients given in Appendix~\ref{sec:App_CG}, we can expand $W_e$ in Eq.~\eqref{eq:WE1} as follow,
\begin{itemize}[labelindent=-0.8em, leftmargin=1.3em]
\item {$(\rho_{E^c_{1,2,3}},\,\rho_L) \,=\, (\mathbf{1},\mathbf{3})$}
\begin{align}
\nonumber
W_e\,&\,
=\alpha\,E^c_1\big(\,L\,Y_\mathbf{3}\big)_\mathbf{1}\,H_d
+\beta\,E^c_2\big(\,L\,Y^{(4)}_{\mathbf{3},I}\big)_\mathbf{1}\,H_d + \gamma_1\,E^c_3\,\big(\,L\,Y^{(6)}_{\mathbf{3},I}\big)_\mathbf{1}\,H_d + \gamma_2\,E^c_3\,\big(\,L\,Y^{(6)}_{\mathbf{3},II}\big)_\mathbf{1}\,H_d\,\\
\nonumber
&\,=\alpha\,E^c_1\left(L_1\,e_1+L_2\,e_3+L_3\,e_2\right)H_d\,+\beta\,E^c_2\big[L_1(-2e_1\widetilde{e}_1+\sqrt{3}e_2\widetilde{e}_5+\sqrt{3}e_3\widetilde{e}_2)\,\\
\nonumber
&\,\quad+\,L_2(\sqrt{3}e_1\widetilde{e}_5-\sqrt{6}e_2\widetilde{e}_4+e_3\widetilde{e}_1)\,+\,L_3(\sqrt{3}e_1\widetilde{e}_2+e_2\widetilde{e}_1-\sqrt{6}e_3\widetilde{e}_3)\big]H_d\,\\
\nonumber
&\,\quad+\gamma_1\,E^c_3\left[L_1(e^2_1+2e_2e_3)e_1\,+\,L_2(e^2_1+2e_2e_3)e_3\,+\,L_3(e^2_1+2e_2e_3)e_2\right]H_d\,\\
\nonumber
&\,\quad+\gamma_2\,E^c_3\big[L_1(2\sqrt{3}e'^3_1-6\sqrt{3}e'_1e'_2e'_3)\,+\,L_2(3\sqrt{6}e'_1e'^2_2-2\sqrt{3}e'^3_3)\,\\
\label{eq:We_1}&\,\quad+L_3(3\sqrt{6}e'_1e'^2_3-2\sqrt{3}e'^3_2)\big]H_d\,.
\end{align}

\item {$(\rho_{E^c_{1,2,3}},\rho_L)\,=\,(\mathbf{1},\mathbf{3'})$}
\begin{align}
\nonumber
W_e\,&\,=\alpha\,E^c_1\big(\,L\,Y_\mathbf{3'}\big)_\mathbf{1}\,H_d
+\beta\,E^c_2\big(\,L\,Y^{(4)}_{\mathbf{3'}, II}\big)_\mathbf{1}\,H_d + \gamma_1\,E^c_3\,\big(\,L\,Y^{(6)}_{\mathbf{3'},I}\big)_\mathbf{1}\,H_d + \gamma_2\,E^c_3\,\big(\,L\,Y^{(6)}_{\mathbf{3'},II}\big)_\mathbf{1}\,H_d\,\\
\nonumber
&\,=\alpha\,E^c_1\left(L_1\,e'_1+L_2\,e'_3+L_3\,e'_2\right)H_d\,+\beta\,E^c_2\big[L_1(-2e'_1\widetilde{e}_1+\sqrt{3}e'_2\widetilde{e}_4+\sqrt{3}e'_3\widetilde{e}_3)\,\\
\nonumber
&\,\quad+\,L_2(\sqrt{3}e'_1\widetilde{e}_4-\sqrt{6}e'_2\widetilde{e}_2+e'_3\widetilde{e}_1)\,+\,L_3(\sqrt{3}e'_1\widetilde{e}_3+e'_2\widetilde{e}_1-\sqrt{6}e'_3\widetilde{e}_5)\big]H_d\,\\
\nonumber
&\,\quad+\gamma_1\,E^c_3\left[L_1(e^2_1+2e_2e_3)e'_1\,+\,L_2(e^2_1+2e_2e_3)e'_3\,+\,L_3(e^2_1+2e_2e_3)e'_2\right]H_d\,\\
\nonumber
&\,\quad+\gamma_2\,E^c_3\big[L_1(2\sqrt{3}e^3_1-6\sqrt{3}e_1e_2e_3)\,+\,L_2(3\sqrt{6}e_1e^2_3-2\sqrt{3}e^3_2)\,\\
\label{eq:We_2}&\,\quad+\,L_3(3\sqrt{6}e_1e^2_2-2\sqrt{3}e^3_3)\big]H_d\,.
\end{align}
\end{itemize}
After electroweak symmetry breaking, we can read out the charged lepton mass matrices which are summarized in the table~\ref{tab:charged lepton}.
\begin{table}[h!]
\centering
\begin{tabular}{|c|c|c|} \hline\hline
\multicolumn{2}{|c|}{\texttt{Models}}  &\multirow{2}{*}{\texttt{Mass Matrices} $M_e$}\\ \cline{1-2}
\texttt{cases}	& $(\rho_{E^c},\,\rho_{L})$ &	\\ \hline
C1  & $(\mathbf{1}\,,\mathbf{3})$  &  {\scriptsize $ \begin{pmatrix}
~~&~~     ~~&~~    ~~&~~ \\[-0.01in]
 \alpha\,e_1 ~~&~~ \alpha\,e_3 ~~&~~ \alpha\,e_2 \\
 ~~&~~     ~~&~~    ~~&~~ \\[-0.03in]
\beta(-2e_1\widetilde{e}_1+\sqrt{3}e_2\widetilde{e}_5+\sqrt{3}e_3\widetilde{e}_2) ~&~ \beta(\sqrt{3}e_1\widetilde{e}_5-\sqrt{6}e_2\widetilde{e}_4+e_3\widetilde{e}_1)  ~&~ \beta(\sqrt{3}e_1\widetilde{e}_2+e_2\widetilde{e}_1-\sqrt{6}e_3\widetilde{e}_3) \\
 ~&~        ~&~      ~&~  \\[-0.03in]
\gamma_1(e^2_1+2e_2e_3)e_1 ~&~ \gamma_1(e^2_1+2e_2e_3)e_3 ~&~\gamma_1(e^2_1+2e_2e_3)e_2 \\
+\gamma_2(2\sqrt{3}e'^3_1-6\sqrt{3}e'_1e'_2e'_3) &    +\gamma_2(3\sqrt{6}e'_1e'^2_2-2\sqrt{3}e'^3_3)   &
+\gamma_2(3\sqrt{6}e'_1e'^2_3-2\sqrt{3}e'^3_2)\\
 ~~&~~     ~~&~~    ~~&~~ \\[-0.01in]
 \end{pmatrix}v_d $}\\ \hline
 C2  & $(\mathbf{1}\,,\mathbf{3'})$  &  {\scriptsize $\begin{pmatrix}
     ~~&~~     ~~&~~    ~~&~~ \\[-0.01in]
 \alpha\,e'_1 ~~&~~ \alpha\,e'_3 ~~&~~ \alpha\,e'_2 \\
 ~~&~~     ~~&~~    ~~&~~ \\[-0.03in]
\beta(-2e'_1\widetilde{e}_1+\sqrt{3}e'_2\widetilde{e}_4+\sqrt{3}e'_3\widetilde{e}_3) ~&~ \beta(\sqrt{3}e'_1\widetilde{e}_4-\sqrt{6}e'_2\widetilde{e}_2+e'_3\widetilde{e}_1)  ~&~ \beta(\sqrt{3}e'_1\widetilde{e}_3+e'_2\widetilde{e}_1-\sqrt{6}e'_3\widetilde{e}_5) \\
 ~&~        ~&~      ~&~  \\[-0.03in]
\gamma_1(e^2_1+2e_2e_3)e'_1 ~&~ \gamma_1(e^2_1+2e_2e_3)e'_3 ~&~\gamma_1(e^2_1+2e_2e_3)e'_2 \\
+\gamma_2(2\sqrt{3}e^3_1-6\sqrt{3}e_1e_2e_3) &    +\gamma_2(3\sqrt{6}e_1e^2_3-2\sqrt{3}e^3_2)   &
+\gamma_2(3\sqrt{6}e_1e^2_2-2\sqrt{3}e'^3_3)\\
 ~~&~~     ~~&~~    ~~&~~ \\[-0.01in]
\end{pmatrix}v_d $}\\ \hline\hline			 		
\end{tabular}
\caption{\label{tab:charged lepton}The prediction for the charged lepton mass matrix in the models without flavon fields, where the charged lepton mass matrix is given in the right-left basis $E^cM_eL$ with $v_d \equiv \langle H^0_d \rangle$.}
\end{table}

\subsection{\label{subsec:neutrino}Neutrino sector}

We shall assume neutrinos are Majorana particles and the left-handed leptons are assigned to a triplet representation $\mathbf{3}$ or $\mathbf{3}'$ of the modular group $\Gamma_5$. We shall consider two different scenarios: the neutrinos are described by the effective Weinberg operator or are generated by the type-I seesaw mechanism. If the neutrino masses originate from the Weinberg operator, the most general form of the superpotential responsible for neutrino masses is,
\begin{equation}
\label{eq:Wnu1}W_\nu = \frac{1}{\Lambda} \big(H_u\, H_u\, L\, L\, f_W\left(Y\right)\big)_\mathbf{1}\,,
\end{equation}
where $f_W(Y)$ is a generic function of the modular form $Y$. For the sake of simplicity, we are concerned with the case that $f_W(Y)$ is the modular form of lowest weight 2. Thus the neutrino superpotential reads
\begin{equation}
\label{eq:Wnu1_simp}W_\nu=\frac{1}{\Lambda} \big(H_u\, H_u\, L\, L\, Y\big)_\mathbf{1}\,.
\end{equation}
The modular invariance requires the weight of lepton doublet should be equal to 1, i.e.
\begin{equation}
\label{eq:weightconstr2}k_L=1\,.
\end{equation}
Since the linearly independent modular forms of weight 2 decompose into three $\Gamma_5\cong A_5$ irreducible representations $Y_{\mathbf{3}}, Y_{\mathbf{3}'}, Y_{\mathbf{5}}$, from the Kronecker products $\mathbf{3}\otimes\mathbf{3}=\mathbf{1}_S\oplus\mathbf{3}_A \oplus\mathbf{5}_S$ and $\mathbf{3}'\otimes\mathbf{3}'=\mathbf{1}_S\oplus\mathbf{3}'_A \oplus\mathbf{5}_S$ we know that only the quintuplet modular form $Y_{\mathbf{5}}$ contributes to $W_\nu$ in Eq.~\eqref{eq:Wnu1_simp}. The concrete form of $W_{\nu}$ depends on the transformation property of $L$ under $\Gamma_5$.
\begin{itemize}[labelindent=-0.8em, leftmargin=1.3em]
\item {$\rho_L\,=\,\mathbf{3}$ }
\begin{align}
\nonumber
W_\nu\,& =\,\frac{1}{\Lambda} \big(\,L\, L\, Y\big)_\mathbf{1}\,H^2_u\,=\,\frac{1}{\Lambda} \big((\,L\, L\,)_{\mathbf{5}} Y_{\mathbf{5}}\big)_\mathbf{1}\,H^2_u\,\\
\nonumber
&\,=\,\frac{1}{\Lambda}\,\big[\,2(\,L_1\,L_1-L_2\,L_3)\,\widetilde{e}_1-2\sqrt{3}\,L_1L_2\,\widetilde{e}_5\,+\,\sqrt{6}\,L_2\,L_2\,\widetilde{e}_4\,\\
\label{eq:Ww1}&\qquad+\sqrt{6}\,L_3\,L_3\,\widetilde{e}_3-2\sqrt{3}\,L_1L_3\widetilde{e}_2\big]\,H^2_u\,,
\end{align}
which gives rise to the following light neutrino mass matrix
\begin{equation}
\label{eq:mnu_WO1}M_\nu=\begin{pmatrix}
2\widetilde{e}_1 ~&~ -\sqrt{3}\,\widetilde{e}_5 ~&~ -\sqrt{3}\,\widetilde{e}_2 \\
-\sqrt{3}\,\widetilde{e}_5 ~&~ \sqrt{6}\,\widetilde{e}_4  ~&~ -\widetilde{e}_1  \\
-\sqrt{3}\,\widetilde{e}_2 ~&~ -\widetilde{e}_1 ~&~\sqrt{6}\,\widetilde{e}_3 \end{pmatrix}\frac{v^2_u}{\Lambda}\,,
\end{equation}
with $v_u=\langle H_u\rangle$.

\item {$\rho_L\,=\,\mathbf{3}'$ }
\begin{align}
\nonumber
W_\nu\,& =\,\frac{1}{\Lambda} \big(\,L\, L\, Y\big)_\mathbf{1}\,H^2_u=\frac{1}{\Lambda} \big((\,L\, L\,)_{\mathbf{5}} Y_{\mathbf{5}}\big)_\mathbf{1}\,H^2_u\,\\
\nonumber
&\,=\,\frac{1}{\Lambda}\,\big[2(L_1L_1-L_2L_3)\,\widetilde{e}_1\,+\,\sqrt{6}\,L_3L_3\widetilde{e}_5-2\sqrt{3}\,L_1L_2\,\widetilde{e}_4\,+\\
\label{eq:Ww2}&\qquad-2\sqrt{3}\,L_1L_3\widetilde{e}_3\,+\sqrt{6}\,L_2L_2\widetilde{e}_2\big]H^2_u\,.
\end{align}
Then the neutrino mass matrix reads
\begin{equation}
\label{eq:mnu_WO2}M_\nu =\begin{pmatrix}
2\widetilde{e}_1 ~&~ -\sqrt{3}\,\widetilde{e}_4 ~&~ -\sqrt{3}\,\widetilde{e}_3 \\
-\sqrt{3}\,\widetilde{e}_4 ~&~ \sqrt{6}\,\widetilde{e}_2  ~&~ -\widetilde{e}_1  \\
-\sqrt{3}\,\widetilde{e}_3 ~&~ -\widetilde{e}_1 ~&~\sqrt{6}\,\widetilde{e}_5 \end{pmatrix}\frac{v^2_u}{\Lambda}\,.
\end{equation}
\end{itemize}
In the second scenario, the neutrino masses are generated via the type-I seesaw mechanism. Three right-handed neutrinos $N^{c}=(N^c_1, N^{c}_2, N^c_3)^{T}$ are introduced and they are assigned to transform as a $\Gamma_5$ triplet $\mathbf{3}$ or $\mathbf{3}'$. The superpotential of the neutrino sector can be generally written as
\begin{equation}
\label{eq:Wnu2}W_\nu = g \left(N^c L H_uf_N\left(Y\right)\right)_\mathbf{1}
+ \Lambda \left(N^c N^cf_M\left(Y\right)\right)_\mathbf{1}\,,
\end{equation}
where $f_N(Y)$ and $f_M(Y)$ are generic functions of the modular form $Y$. In order to build models with minimal number of parameters, we consider the case that both $f_N(Y)$ and $f_M(Y)$ are of weight 2 yet only one of them is involved. If the modulus parameter $\tau$ enters into the neutrino masses through the Yukawa coupling, the neutrino superpotential would be
\begin{equation}
\label{eq:Wnu2_1}W^I_\nu = g \left(N^cL H_u Y\right)_\mathbf{1}
+ \Lambda \left(N^c N^c\right)_\mathbf{1}\,.
\end{equation}
In this case the weights of $N^c$ and $L$ should be
\begin{equation}
\label{eq:weightconstr3} k_N=0,\quad k_L=2\,.
\end{equation}
If the complex modulus $\tau$ only appears in the right-handed neutrino mass term, the neutrino superpotential would be
\begin{equation}
W^{II}_\nu = g \label{eq:Wnu2_2}\left(N^cLH_u\right)_\mathbf{1}+\Lambda\left(N^cN^cY\right)_\mathbf{1}\,,
\end{equation}
with the weights
\begin{equation}
\label{eq:weightconstr4}k_N=1\,,\quad k_L=-1\,.
\end{equation}
The explicit form of $W_{\nu}$ depends on the assignments of $\rho_{L}$ and $\rho_{N^c}$.

\begin{itemize}[labelindent=-0.8em, leftmargin=1.3em]
\item {$\left(\rho_L, \rho_{N^c}\right)~$=$~(\mathbf{3}\,,\mathbf{3})$ }
\begin{align}
\nonumber
W^I_\nu&=g_1 \left((N^cL)_\mathbf{3}Y_{\mathbf{3}}\right)_\mathbf{1}H_u+g_2 \left((N^cL)_\mathbf{5}Y_{\mathbf{5}}\right)_\mathbf{1}H_u+\Lambda \left(N^cN^c\right)_\mathbf{1}\,,\\
\nonumber
&=g_1\big[(N^c_2L_3-N^c_3L_2)e_1+(N^c_1L_2-N^c_2L_1)e_3\,+(N^c_3L_1-N^c_1L_3)e_2\big]H_u\,,\\
\nonumber
&\quad+g_2\big[(2N^c_1L_1-N^c_2L_3-N^c_3L_2)\widetilde{e}_1-\sqrt{3}(N^c_1L_2+N^c_2L_1)\widetilde{e}_5+\sqrt{6}N^c_2L_2\widetilde{e}_4\,\\
&\quad+\sqrt{6}N^c_3L_3\widetilde{e}_3-\sqrt{3}(N^c_1L_3+N^c_3L_1)\widetilde{e}_2\big]H_u+\Lambda\left(N^c_1N^c_1+2N^c_2N^c_3\right)\,,
\end{align}
and
\begin{align}
\nonumber
W^{II}_\nu&=g \left(N^cL\right)_\mathbf{1}H_u+\Lambda \left((N^cN^c)_{\mathbf{5}}Y_{\mathbf{5}}\right)_\mathbf{1}\,,\\
\nonumber
&=g\left(N^c_1L_1+N^c_2L_3+N^c_3L_2\right)H_u+\Lambda\big[2(N^c_1N^c_1-N^c_2N^c_3)\widetilde{e}_1-2\sqrt{3}N^c_1N^c_2\widetilde{e}_5\\
&\quad+\sqrt{6}N^c_2N^c_2\widetilde{e}_4+\sqrt{6}N^c_3N^c_3\widetilde{e}_3-2\sqrt{3}N^c_1N^c_3\widetilde{e}_2\big]\,.
\end{align}

\item{$\left(\rho_L,\rho_{N^c}\right)~$=$~(\mathbf{3}'\,,\mathbf{3})$}

From the multiplication rule $\mathbf{3}\otimes\mathbf{3}'=\mathbf{4}\oplus\mathbf{5}$ which doesn't contain singlet, we know that the Yukawa coupling $N^cLH_u$ is not invariant under the $\Gamma_5$ modular symmetry. Consequently the superpotential $W^{II}_\nu$ is absent in this case, and we have
\begin{align}
\nonumber
W^I_\nu&=g \left((N^cL)_\mathbf{5}Y_{\mathbf{5}}\right)_\mathbf{1}H_u+\Lambda \left(N^cN^c\right)_\mathbf{1}\,,\\
\nonumber
&=g\big[\sqrt{3}N^c_1L_1\widetilde{e}_1+(N^c_2L_1-\sqrt{2}N^c_3L_2)\widetilde{e}_5+(N^c_1L_2-\sqrt{2}N^c_3L_3)\widetilde{e}_4\,\\
&\quad+(N^c_1L_3-\sqrt{2}N^c_2L_2)\widetilde{e}_3+(N^c_3L_1-\sqrt{2}N^c_2L_3)\widetilde{e}_2\big]H_u+
\Lambda(N^c_1N^c_1+2N^c_2N^c_3)\,.
\end{align}

\item{$\left(\rho_L, \rho_{N^c}\right)~$=$~(\mathbf{3}\,,\mathbf{3}')$}

Similar to previous case, $W^{II}_\nu$ is not allowed by the $\Gamma_{5}\cong A_5$ flavor symmetry, and the neutrino superpotential is
\begin{align}
\nonumber
W^I_\nu &=g \left((N^cL)_\mathbf{5}Y_{\mathbf{5}}\right)_\mathbf{1}H_u+\Lambda \left(N^cN^c\right)_\mathbf{1}\,,\\
\nonumber
&=g\big[\sqrt{3}N^c_1L_1\widetilde{e}_1+(N^c_1L_2-\sqrt{2}N^c_2L_3)\widetilde{e}_5+(N^c_2L_1-\sqrt{2}N^c_3L_3)\widetilde{e}_4\\
&\quad+(N^c_3L_1-\sqrt{2}N^c_2L_2)\widetilde{e}_3+(N^c_1L_3-\sqrt{2}N^c_3L_2)\widetilde{e}_2\big]H_u+\Lambda\left(N^c_1N^c_1+2N^c_2N^c_3\right)\,.
\end{align}

\item{$\left(\rho_L,\rho_{N^c}\right)~$=$~(\mathbf{3}'\,, \mathbf{3}')$}

Analogous to the case of $\left(\rho_L, \rho_{N^c}\right)~$=$~(\mathbf{3}\,,\mathbf{3})$, the neutrino superpotential reads
\begin{align}
\nonumber
W^I_\nu&=g_1 \left((N^cL)_\mathbf{3}'Y_{\mathbf{3}'}\right)_\mathbf{1}H_u+g_2\left((N^cL)_\mathbf{5}Y_{\mathbf{5}}\right)_\mathbf{1}H_u
+\Lambda\left(N^cN^c\right)_\mathbf{1}\,,\\
\nonumber
&=g_1\big[(N^c_2L_3-N^c_3L_2)e'_1+(N^c_1L_2-N^c_2L_1)e'_3+(N^c_3L_1-N^c_1L_3)e'_2\big]H_u\,\\
\nonumber
&+g_2\big[(2N^c_1L_1-N^c_2L_3-N^c_3L_2)\widetilde{e}_1+\sqrt{6}N^c_3L_3\widetilde{e}_5-\sqrt{3}(N^c_1L_2+N^c_2L_1)\widetilde{e}_4\\
&\quad-\sqrt{3}(N^c_1L_3+N^c_3L_1)\widetilde{e}_3+\sqrt{6}N^c_2L_2\widetilde{e}_2\big]H_u+\Lambda\left(N^c_1N^c_1+2N^c_2N^c_3\right)\,,
\end{align}
and
\begin{align}
\nonumber
W^{II}_\nu&=g\left(N^c L\right)_\mathbf{1}H_u+\Lambda \left((N^cN^c)_{\mathbf{5}}Y_{\mathbf{5}}\right)_\mathbf{1}\,,\\
\nonumber&=g (N^c_1L_1+N^c_2L_3+N^c_3L_2)\,H_u+\Lambda\big[2(N^c_1N^c_1-N^c_2N^c_3)\widetilde{e}_1+\sqrt{6}N^c_3N^c_3\widetilde{e}_5\\
&\quad-2\sqrt{3}N^c_1N^c_2\widetilde{e}_4-2\sqrt{3}N^c_1N^c_3\widetilde{e}_3+\sqrt{6}N^c_2N^c_2\widetilde{e}_2\big]\,.
\end{align}

\end{itemize}
Subsequently we can straightforwardly read out the modular invariant Dirac neutrino mass matrix $M_D$ and the right-handed Majorana neutrino mass matrix $M_{N}$ for each case above, and the results are summarized in table~\ref{tab:neutrino}. After integrating out the heavy neutrinos $N^c$, we can obtain the effective neutrino mass matrix given by the well-known seesaw formula,
\begin{equation}
\label{eq:mnufactor}M_\nu\,=\,-M_D^TM^{-1}_NM_D\,.
\end{equation}
Since there are a few constructions in the charged lepton and neutrino sectors listed in above, we can obtain plenty of possible models with $\Gamma_5$ modular symmetry, as summarized in table~\ref{tab:models}. We see sixteen models are constructed, eight of them named as $\mathcal{A}i\,(i=1,\ldots,8)$ involve flavon fields in the charged lepton mass terms while the remaining eight models $\mathcal{B}i\,(i=1,\ldots,8)$ only depend on the modulus $\tau$.

\begin{table}[t!]
\centering
\begin{tabular}{|c|c|c|c|} \hline\hline
\multicolumn{2}{|c|}{\texttt{Models}}  &\multirow{2}{*}{\texttt{Neutrino mass matrices}} & \multirow{2}{*}{\texttt{cases}}\\ \cline{1-2}
	& $(\rho_L,\,\rho_{N^c})$ &	 & \\ \hline

 &   &  &     \\ [-0.16in]

	\multirow{2}{*}{\texttt{Weinberg}} & $(\mathbf{3}\,,-)$  &  {\scriptsize $ M_\nu =\begin{pmatrix}
 2\widetilde{e}_1 ~&~ -\sqrt{3}\widetilde{e}_5 ~&~ -\sqrt{3}\widetilde{e}_2 \\
-\sqrt{3}\widetilde{e}_5 ~&~ \sqrt{6}\widetilde{e}_4  ~&~ -\widetilde{e}_1  \\
-\sqrt{3}\widetilde{e}_2 ~&~ -\widetilde{e}_1 ~&~\sqrt{6}\widetilde{e}_3 \end{pmatrix}\frac{v^2_u}{\Lambda}$} & W1\\
  &   & &     \\ [-0.16in]\cline{2-4}
 &   &  &     \\ [-0.16in]

 \texttt{operator} & $(\mathbf{3'}\,,-)$  &  {\scriptsize $ M_\nu =\begin{pmatrix}
 2\widetilde{e}_1 ~&~ -\sqrt{3}\widetilde{e}_4 ~&~ -\sqrt{3}\widetilde{e}_3 \\
-\sqrt{3}\widetilde{e}_4 ~&~ \sqrt{6}\widetilde{e}_2  ~&~ -\widetilde{e}_1  \\
-\sqrt{3}\widetilde{e}_3 ~&~ -\widetilde{e}_1 ~&~\sqrt{6}\widetilde{e}_5 \end{pmatrix}\frac{v^2_u}{\Lambda} $} & W2\\

 &   & &     \\ [-0.16in]\hline
 &   &  &     \\ [-0.16in]

 \multirow{6}{*}{} & \multirow{2}{*}{$(\mathbf{3}\,,\mathbf{3})$} &  {\scriptsize $ M_D=\begin{pmatrix}
2g_2\widetilde{e}_1~&~ g_1e_3-\sqrt{3}g_2\widetilde{e}_5 ~&~ -g_1e_2-\sqrt{3}g_2\widetilde{e}_2 \\
-g_1e_3-\sqrt{3}g_2\widetilde{e}_5 ~&~ \sqrt{6}g_2\widetilde{e}_4 ~&~ g_1e_1-g_2\widetilde{e}_1  \\
g_1e_2-\sqrt{3}g_2\widetilde{e}_2 ~&~ -g_1e_1-g_2\widetilde{e}_1 ~&~ \sqrt{6}g_2\widetilde{e}_3 \end{pmatrix}v_{u} $},~~{\scriptsize	$M_N=\begin{pmatrix}
			1 & 0 & 0 \\
			0 & 0 & 1 \\
			0 & 1 & 0 \end{pmatrix}\Lambda$} & S1 \\

 &   & &     \\ [-0.16in]\cline{3-4}
 &   &  &     \\ [-0.16in]

  &  & {\scriptsize $ M_D=g\begin{pmatrix}
			1 & 0 & 0 \\
			0 & 0 & 1 \\
			0 & 1 & 0\end{pmatrix}v_u $},~~ {\scriptsize $M_N= \begin{pmatrix}
2\widetilde{e}_1 ~&~ -\sqrt{3}\widetilde{e}_5 ~&~ -\sqrt{3}\widetilde{e}_2 \\
-\sqrt{3}\widetilde{e}_5 ~&~ \sqrt{6}\widetilde{e}_4  ~&~ -\widetilde{e}_1  \\
-\sqrt{3}\widetilde{e}_2 ~&~ -\widetilde{e}_1 ~&~\sqrt{6}\widetilde{e}_3\end{pmatrix}\Lambda$} & S2 \\		

  &   & &     \\ [-0.16in]\cline{3-4}
 &   &  &     \\ [-0.16in]

\texttt{Type I}  & $(\mathbf{3'}\,,\mathbf{3})$ &  {\scriptsize $ M_D= g\begin{pmatrix}
\sqrt{3}\widetilde{e}_1~&~ \widetilde{e}_4 ~&~ \widetilde{e}_3 \\
\widetilde{e}_5 ~&~ -\sqrt{2}\widetilde{e}_3 ~&~ -\sqrt{2}\widetilde{e}_2  \\
\widetilde{e}_2 ~&~ -\sqrt{2}\widetilde{e}_5~&~ -\sqrt{2}\widetilde{e}_4\end{pmatrix}v_u $},~~ {\scriptsize	$M_N=\begin{pmatrix}
			1 & 0 & 0 \\
			0 & 0 & 1 \\
			0 & 1 & 0\end{pmatrix}\Lambda$} & S3 \\

&   & &     \\ [-0.16in]\cline{2-4}
 &   &  &     \\ [-0.16in]

\texttt{see-saw}  & $(\mathbf{3}\,,\mathbf{3'})$ &   {\scriptsize $ M_D= g\begin{pmatrix}
\sqrt{3}\widetilde{e}_1~&~ \widetilde{e}_5 ~&~ \widetilde{e}_2 \\
\widetilde{e}_4 ~&~ -\sqrt{2}\widetilde{e}_3 ~&~ -\sqrt{2}\widetilde{e}_5  \\
\widetilde{e}_3 ~&~ -\sqrt{2}\widetilde{e}_2 ~&~ -\sqrt{2}\widetilde{e}_4\end{pmatrix}v_u $},~~ {\scriptsize	$M_N=\begin{pmatrix}
			1 & 0 & 0 \\
			0 & 0 & 1 \\
			0 & 1 & 0\end{pmatrix}\Lambda$} & S4\\

 &   & &     \\ [-0.16in]\cline{2-4}
 &   &  &     \\ [-0.16in]

& \multirow{2}{*}{$(\mathbf{3'}\,,\mathbf{3'})$} &  {\scriptsize $ M_D=\begin{pmatrix}
2g_2\widetilde{e}_1~&~ g_1e'_3-\sqrt{3}g_2\widetilde{e}_4 ~&~ -g_1e'_2-\sqrt{3}g_2\widetilde{e}_3 \\
-g_1e'_3-\sqrt{3}g_2\widetilde{e}_4 ~&~ \sqrt{6}g_2\widetilde{e}_2 ~&~ g_1e'_1-g_2\widetilde{e}_1  \\
g_1e'_2-\sqrt{3}g_2\widetilde{e}_3 ~&~ -g_1e'_1-g_2\widetilde{e}_1 ~&~ \sqrt{6}g_2\widetilde{e}_5 \end{pmatrix}v_u $},~~ {\scriptsize	$M_N=\begin{pmatrix}
			1 & 0 & 0 \\
			0 & 0 & 1 \\
			0 & 1 & 0 \end{pmatrix}\Lambda$} & S5\\

  &   & &     \\ [-0.16in]\cline{3-4}
 &   &  &     \\ [-0.16in]

&  & {\scriptsize $ M_D=g\begin{pmatrix}
			1 & 0 & 0 \\
			0 & 0 & 1 \\
			0 & 1 & 0\end{pmatrix}v_u $},~~ {\scriptsize $M_N= \begin{pmatrix}
2\widetilde{e}_1 ~&~ -\sqrt{3}\widetilde{e}_4 ~&~ -\sqrt{3}\widetilde{e}_3 \\
-\sqrt{3}\widetilde{e}_4 ~&~ \sqrt{6}\widetilde{e}_2  ~&~ -\widetilde{e}_1  \\
-\sqrt{3}\widetilde{e}_3 ~&~ -\widetilde{e}_1 ~&~\sqrt{6}\widetilde{e}_5 \end{pmatrix}\Lambda$} & S6\\
\hline \hline			 		
\end{tabular}
\caption{\label{tab:neutrino}The classification of neutrino mass matrices for neutrino mass arising from Weinberg operator or the type-I seesaw. The last column is the name of each possible case. }
\end{table}

\begin{table}[hptb!]
\centering
\begin{tabular}{|c|c|c|c|c|c|c|} \hline\hline
 \multicolumn{2}{|c|}{\multirow{2}{*}{\texttt{Models}}} & \multirow{2}{*}{\texttt{mass matrices}} & \texttt{assignment} & \multicolumn{3}{|c|}{\texttt{weight}} \\ \cline{4-7}
 \multicolumn{2}{|c|}{}  & & $(\,\rho_{E^c}\,,\rho_L\,,\rho_{N^c}\,)$ & $\,k_{E_{1,2,3}}\,$ &$\,k_L\,$ & $\,k_{N^c}\,$ \\ \hline
 & $\mathcal{A}1$ & $W1$ & $(\mathbf{1},\mathbf{3}\,,-)$ & $-$ & $1$ & $-$\\ \cline{2-7}	
 & $\mathcal{A}2$ & $W2$ & $(\mathbf{1},\mathbf{3'}\,,-)$ & $-$ & $1$ & $-$\\ \cline{2-7}	
 & $\mathcal{A}3$ & $S1$ & $(\mathbf{1},\mathbf{3}\,,\mathbf{3})$ & $-$ & $2$ & $0$\\ \cline{2-7}	
 \texttt{With}& $\mathcal{A}4$ & $S2$ & $(\mathbf{1},\mathbf{3}\,,\mathbf{3})$ & $-$ & $-1$ & $1$\\ \cline{2-7}	
\texttt{flavons} & $\mathcal{A}5$ & $S3$ & $(\mathbf{1},\mathbf{3'}\,,\mathbf{3})$ & $-$ & $2$ & $0$\\ \cline{2-7}	 	
 & $\mathcal{A}6$ & $S4$ & $(\mathbf{1},\mathbf{3}\,,\mathbf{3'})$ & $-$ & $2$ & $0$\\ \cline{2-7}	
 & $\mathcal{A}7$ & $S5$ & $(\mathbf{1},\mathbf{3'}\,,\mathbf{3'})$ & $-$ & $2$ & $0$\\ \cline{2-7}	
 & $\mathcal{A}8$ & $S6$ & $(\mathbf{1},\mathbf{3'}\,,\mathbf{3'})$ & $-$ & $-1$ & $1$\\ \hline\hline

 & $\mathcal{B}1$ & $C1\,~,~W1\,$ & $(\mathbf{1}\,,\mathbf{3}\,,-)$ & $1\,,3\,,5$ & $1$ & $-$\\ \cline{2-7}			
 & $\mathcal{B}2$ & $C2\,~,~W2\,$ & $(\mathbf{1}\,,\mathbf{3'}\,,-)$ & $1\,,3\,,5$ & $1$ & $-$\\ \cline{2-7}	
 & $\mathcal{B}3$& $C1\,~,~S1\,$ & $(\mathbf{1}\,,\mathbf{3}\,,\mathbf{3})$ & $0\,,2\,,4$ & $2$ & $0$\\ \cline{2-7}	
\texttt{Without}& $\mathcal{B}4$ & $C1\,~,~S2\,$ & $(\mathbf{1}\,,\mathbf{3}\,,\mathbf{3})$ & $3\,,5\,,7$ & $-1$ & $1$\\ \cline{2-7}	
\texttt{flavons} & $\mathcal{B}5$ & $C2\,~,~S3\,$ & $(\mathbf{1}\,,\mathbf{3'}\,,\mathbf{3})$ & $0\,,2\,,4$ & $2$ & $0$\\ \cline{2-7}	 	
& $\mathcal{B}6$ & $C1\,~,~S4\,$ & $(\mathbf{1}\,,\mathbf{3}\,,\mathbf{3'})$ & $0\,,2\,,4$ & $2$ & $0$\\ \cline{2-7}	
& $\mathcal{B}7$ & $C2\,~,~S5\,$ & $(\mathbf{1}\,,\mathbf{3'}\,,\mathbf{3'})$ & $0\,,2\,,4$ & $2$ & $0$\\ \cline{2-7}	
& $\mathcal{B}8$& $C2\,~,~S6\,$ & $(\mathbf{1}\,,\mathbf{3'}\,,\mathbf{3'})$ & $3\,,5\,,7$ & $-1$ & $1$\\ \hline\hline	      	      	
\end{tabular}
\caption{\label{tab:models}The summary of models and the corresponding predictions for neutrino and charged lepton mass matrices. For the models with flavons in the charged lepton sector, the weights of the right-handed charged leptons should satisfy the constraint in Eq.~\eqref{eq:weight_ch1}, i.e. $k_{E_1}=5k_{E_3}+4k_L$ and $k_{E_2}=4k_{E_3}+3k_L$.  	}
\end{table}

\section{\label{sec:numerical}Numerical Analysis}
In this section we shall perform a comprehensive numerical analysis for each possible model listed in table~\ref{tab:models}. Since some phases can be absorbed through field redefinition, some coupling constants of the models can be taken to be real. We first count the number of independent real free parameters of each model. If flavon fields are used, the resulting charged lepton mass matrix would be diagonal in our models and the observed charged lepton masses can be reproduced for certain values of the coupling constants and the flavon VEVs. If the models make no use of any flavon field other than the complex modulus $\tau$, the charged lepton mass matrix is summarized in table~\ref{tab:charged lepton}. We can rephase the charged lepton superfields $E^c_1,\,E^c_2,\,E^c_3$ to make the parameters $\alpha, \beta, \gamma_1$ real while the phase of $\gamma_2$ can not be removed. Thus the charged lepton mass matrix depends on four independent real parameters ${\beta/\alpha, \gamma_1/\alpha, |\gamma_2/\alpha|, \text{Arg}(\gamma_2/\alpha)}$ except the overall scale factor $\alpha v_d$. If the neutrino masses originate from the Weinberg operator, the effective neutrino mass matrix would be expressed in terms of modular forms as functions of the modulus $\tau$ besides the overall factor $v^2_u/\Lambda$. If the neutrino masses are generated through the seesaw mechanism, the light neutrino mass matrix has two independent real parameters ${|g_1/g_2|, \text{Arg}(g_1/g_2)}$ and the overall scale factor is $g^2_2v^2_u/\Lambda$ for the models $S1$ and $S5$. For the cases of $S2, S3, S4$ and $S6$, the neutrino mass matrix uniquely depends on the modulus parameter $\tau$ and $g^2v^2_u/\Lambda$ controls the absolute scale of neutrino masses, as can seen from table~\ref{tab:neutrino}. In this way, we can easily read out the independent real input parameters of our models and the results are collected in table~\ref{tab:freeparams}.
\begin{table}[hptb!]
\centering
\begin{tabular}{|c||c||c||c|} \hline\hline
 \multicolumn{2}{|c||}{\texttt{Models}} & \texttt{free input parameters} $p_i$ & \texttt{overall factors}  \\ \hline
\multirow{3}{*}{} & $\mathcal{A}1,\,\mathcal{A}2$& $\{\rm{Re}\,\tau,\,\rm{Im} \,\tau\}$ & $v_u^2/\Lambda$  \\ \cline{2-4}
  \texttt{With} & $\mathcal{A}4,\,\mathcal{A}5,\,\mathcal{A}6,\,\mathcal{A}8$ & $\{\rm{Re}\,\tau,\,\rm{Im} \,\tau\}$ & $g^2v_u^2/\Lambda$ \\ \cline{2-4}			
 \texttt{flavons} & $\mathcal{A}3,\,\mathcal{A}7$ & $\{\rm{Re}\,\tau,\,\rm{Im} \,\tau,$ $\,|g_1/g_2|,\,\text{Arg}(g_1/g_2)\}$ & $g_2^2v_u^2/\Lambda$ \\ \hline\hline
\multirow{3}{*}{} & $\mathcal{B}1,\,\mathcal{B}2$& $\{\rm{Re}\,\tau,\,\rm{Im} \,\tau,\,\beta/\alpha,\,\gamma_1/\alpha,\,|\gamma_2/\alpha|,$ $\,\text{Arg}(\gamma_2/\alpha)\}$ & $\alpha v_d,\,v_u^2/\Lambda$   \\ \cline{2-4}
\texttt{Without} & $\mathcal{B}4,\,\mathcal{B}5,\,\mathcal{B}6,\,\mathcal{B}8$& $\{\rm{Re}\,\tau,\,\rm{Im} \,\tau,\,\beta/\alpha,\,\gamma_1/\alpha,\,|\gamma_2/\alpha|,$ $\,\text{Arg}(\gamma_2/\alpha)\}$ & $\alpha v_d,\,g^2v_u^2/\Lambda$ \\ \cline{2-4}			
\texttt{flavons}  & $\mathcal{B}3,\,\mathcal{B}7$ &
{$\! \begin{aligned}
&\{\rm{Re}\,\tau,\,\rm{Im} \,\tau,\beta/\alpha,\,\gamma_1/\alpha,\,|\gamma_2/\alpha|,\\[-0.04in]
&\,\text{Arg}(\gamma_2/\alpha),\,|g_1/g_2|,\,\text{Arg}(g_1/g_2)\}
\end{aligned} $} & $\alpha v_d,\,g_2^2v_u^2/\Lambda$\\ \hline\hline    	
\end{tabular}
\caption{\label{tab:freeparams}The input parameters of each model,
where the freedom of field redefinition has been used to absorb the physically irrelevant phases. Notice that the values of the input parameters are real. }
\end{table}

If the modulus $\tau$ is changed to $-\tau^{\star}$, from the $q-$expansion of the modular forms we see that each modular form would become its complex conjugate. If we also set all the coupling constants to be their complex conjugate, i.e.
\begin{equation}
\label{eq:conjParams}\tau \rightarrow -\tau^{\star},\quad g_i \to g_i^{\star}\,,
\end{equation}
the charged lepton and neutrino mass matrices as well as the lepton mixing matrix will become complex conjugate. As a consequence, this transformation leaves lepton masses and mixing angles unchanged while the signs of both Dirac and Majorana CP phases would be flipped. Hence it is sufficient to limit in the range $\mathrm{Re}\tau>0$ during the numerical analysis. We call two sets of input parameters are conjugate if they are related through the transformation in Eq.~\eqref{eq:conjParams}.

Moreover we notice that the number of free parameters of each model is less than the number of low energy observables including three charged lepton masses $m_{e, \mu, \tau}$, three light neutrino masses $m_{1, 2, 3}$ and three lepton mixing angles $\theta_{12}$, $\theta_{13}$, $\theta_{23}$, one Dirac CP phase $\delta_{CP}$ and two Majorana phases $\alpha_{21}$ and $\alpha_{31}$. In order to quantitatively measure how well the models can describe the experimental data, we define the $\chi^2$ function to serve as a test-statistic for the goodness-of-fit,
\begin{equation}
{\chi}^2(p_i) = \sum_j \left( \frac{Q_j(p_i) - Q_{j, \text{best-fit}}}{\sigma_j} \right)^2\,,
\end{equation}
where $p_i$ denote the input free parameters of a model and they are listed in table~\ref{tab:freeparams}, and $Q_i\in\{\sin^2\theta_{12}, \sin^2\theta_{13}, \sin^2\theta_{23}, r \equiv \Delta m^2_{21}/|\Delta m^2_{3\ell}|, m_e/m_\mu, m_\mu/m_\tau\}$ are observable quantities derived from the neutrino and charged lepton mass matrices as complex nonlinear functions of the free parameters of the models\footnote{For the models with flavons, the mass ratios $m_e/m_\mu$ and $m_\mu/m_\tau$ are not included in the $\chi^2$ function since their measured values can be obtained exactly for particular values of the coupling constants and flavon VEVs.}. The parameters $Q_{j, \text{best-fit}}$ and $\sigma_j$ refer to the current central values
and $1\sigma$ deviations respectively of the corresponding observable quantities listed in the table~\ref{tab:globalFit}. Since the indication of a preferred value of the Dirac CP violation phase $\delta_{CP}$ coming from global data analyses is rather weak~\cite{Esteban:2018azc}, we do not include any information on $\delta_{CP}$ in the $\chi^2$ function. We consider both the normal ordering (NO) neutrino mass spectrum $m_1<m_2<m_3$ and the inverted ordering (IO) neutrino masses $m_3<m_1<m_2$,
where $m_{1,2,3}$ denote three light neutrino masses. We define a quantity $N\sigma\equiv \sqrt{\chi^2_{\text{min}}}$ where $\chi^2_{\text{min}}$ is the global minimum of function $\chi^2$. The free parameters of the models are scanned over the following ranges,
\begin{equation}
\begin{aligned}
&\beta/\alpha,\,\gamma_1/\alpha,\,|\gamma_2/\alpha|,\,|g_1/g_2|\,\in [0,10^4],\\
&\,\text{Arg}(\gamma_2/\alpha),\,\text{Arg}(g_1/g_2)\,\in [0,2\pi]\,.
\end{aligned}
\end{equation}
The modulus $\tau$ is taken from the fundamental domain ${\cal F}_5$ shown in figure~\ref{fig:Fund_domain}, and we restrict the parameter search in $\rm{Re}\tau\in [0, 0.5]$. The predictions of the mixing parameters in the conjugate region $\rm{Re}\tau\in [-0.5, 0]$ can be easily obtained by only shifting the overall signs of the Dirac as well as Majorana CP phases. Hence all the numerical results given in the following come in pair with opposite CP violation phases, and we show only one of them for the sake of readability. The function $\chi^2$ is numerically minimized by using the minimization algorithms incorporated in the package MINUIT developed by CERN to determine the best fit values of the input parameters~\cite{minuit}. Requiring all the three mixing angles $\theta_{12}$, $\theta_{13}$, $\theta_{23}$ and $r$ in the $3\sigma$ ranges of global data analysis~\cite{Esteban:2018azc} and the mass ratios $m_e/m_{\mu}$, $m_{\mu}/m_{\tau}$ in the experimentally favored intervals, we can obtain the allowed regions of the mixing angles, CP violation phases and light neutrino masses. The overall scale factors $\alpha v_d$ and $v_u^2/\Lambda$ (or $g^2v_u^2/\Lambda$, $g^2_2v_u^2/\Lambda$) of the charged lepton and neutrino mass matrices can fixed the measured values of the electron mass $m_e$ and the squared mass difference $\Delta m^2_{21}$.
\begin{table}[!t]
\centering
\renewcommand{\arraystretch}{1.2}
\begin{tabular}{|l|cc|} \hline\hline

\texttt{Observables}  & \multicolumn{2}{c|}{\texttt{Best-fit value and $1\sigma$ range} } \\
\hline
$m_e / m_\mu$ & \multicolumn{2}{c|}{$0.0048 \pm 0.0002$} \\
$m_\mu / m_\tau$ & \multicolumn{2}{c|}{$0.0565 \pm 0.0045$} \\
\hline
& Normal Ordering ~~&~~ Inverted Ordering \\ \hline

$\sin^2\theta_{12}$ & $0.310^{+0.013}_{-0.012}$ & $0.310^{+0.013}_{-0.012}$ \\
$\sin^2\theta_{13}$ & $0.02241^{+0.00065}_{-0.00065}$ & $0.02264^{+0.00066}_{-0.00066}$ \\
$\sin^2\theta_{23}$ & $0.580^{+0.017}_{-0.021}$ & $0.584^{+0.016}_{-0.020}$\\
$\delta_{CP}/\pi$ & $1.194^{+0.222}_{-0.161}$  & $1.578^{+0.150}_{-0.161}$ \\
$\Delta m^2_{21}/(10^{-5}\text{ eV}^2)$ & $7.39^{+0.21}_{-0.20}$ & $7.39^{+0.21}_{-0.20}$ \\
$|\Delta m^2_{3\ell}|/(10^{-3}\text{ eV}^2)$ & $2.525^{+0.033}_{-0.032}$ & $2.512^{+0.034}_{-0.032}$ \\
$r \equiv \Delta m^2_{21}/|\Delta m^2_{3\ell}|$ & $ 0.02927^{+0.000895}_{-0.000895}$ & $ 0.02941^{+0.00125}_{-0.00116}$\\ \hline\hline
\end{tabular}
\caption{\label{tab:globalFit}The best-fit values and 1$\sigma$ errors of the observable quantities used in the $\chi^2$ analysis. The values of the neutrino oscillation parameters are adapted from NuFIT 4.0 for normal ordering and inverted ordering without SK atmospheric data~\cite{Esteban:2018azc}, where $\Delta m^2_{21} \equiv m_2^2- m_1^2$, $\Delta m^2_{3\ell} \equiv m_3^2-m_1^2 >0 $ for normal ordering and $\Delta m^2_{3\ell} \equiv m_3^2 - m_2^2 < 0$ for inverted ordering. The charged lepton mass ratios $m_e/m_{\mu}$ and $m_{\mu}/m_{\tau}$ given at the scale $2\times10^{16}\,GeV$ are taken from Refs.~\cite{Feruglio:2014jla,Ross:2007az}. }
\end{table}

\subsection{\label{subsec:numerical_W_flavons}Numerical results of the models with flavons}

We have extensively scanned over the parameter space of each model. As shown in table~\ref{tab:freeparams}, the models $\mathcal{A}1$, $\mathcal{A}2$, $\mathcal{A}4$, $\mathcal{A}5$, $\mathcal{A}6$ and $\mathcal{A}8$ only depend on the complex modulus $\tau$ besides the overall mass scale. These models are too constrained to give a realistic description of lepton mixing angles as well as neutrino masses. However, the model $\mathcal{A}1$ can approach the experimental data better. We find the best fit values of the neutrino mixing parameters and neutrino masses are
\begin{equation}
\begin{gathered}
\sin^2\theta_{12}=0.2937\,,\quad
\sin^2\theta_{13}=0.0346\,,\quad
\sin^2\theta_{23}=0.7954\,,\\
\delta_{CP}/\pi=1.3963\,,\quad
\alpha_{21}/\pi=1.6286\,,\quad
\alpha_{31}/\pi=0.1542\,,\\
m_1=0.04903 \text{ eV}\,,\quad
m_2=0.04978 \text{ eV}\,,\quad
m_3=0.00106 \text{ eV}\,.
\end{gathered}
\end{equation}
We see that the observed quantities $\sin^2\theta_{12}$, $r\equiv \Delta m^2_{21}/|\Delta m^2_{3\ell}|$ and $\delta_{CP}$ can be reproduced very well, particularly the Dirac phase is close to $3\pi/2$. However, both reactor mixing angle $\theta_{13}$ and atmospheric angle $\theta_{23}$ are slightly larger than the $3\sigma$ upper bounds, and their pulls dominate the minimum of $\chi^2$. Similar to Ref.~\cite{Criado:2018thu}, we expect that the discrepancy could be resolved if there is small deviation from the leading order charged lepton flavon alignment such that the charged lepton mass matrix is not diagonal.

For the remaining two models $\mathcal{A}3$ and $\mathcal{A}7$, the neutrino mass matrix $m_{\nu}$ only involves the modulus $\tau$ and the parameter $g_1/g_2$ apart from the overall scale factor $g^2_2v^2_u/\Lambda$. The model $\mathcal{A}3$ can give realistic values of the observables except $\theta_{23}$ in certain parameter space for NO neutrino mass spectrum. We find that a best fit to the experimental data can be obtained at the point
\begin{equation}
\begin{gathered}
\langle \tau \rangle = 0.2497+0.1935i\,,\quad ~|g_1/g_2|=2.2913\,,\\
~\text{Arg}(g_1/g_2)=0.00456\,,\quad ~g_2^2v_u^2/\Lambda=0.00021 \text{ eV}\,.
\end{gathered}
\end{equation}
The neutrino masses and mixing parameters are determined to be
\begin{equation}
\label{eq:mixpara_A3NO}
\begin{gathered}
\sin^2\theta_{12}=0.3109\,,\quad
\sin^2\theta_{13}=0.0234\,,\quad
\sin^2\theta_{23}=0.6875,\\
\delta_{CP}/\pi=1.6784,\quad
\alpha_{21}/\pi=0.6628,\quad
\alpha_{31}/\pi = 0.5088\,,\\
m_1=0.08291 \text{ eV},\quad
m_2=0.08336 \text{ eV},\quad
m_3=0.09697 \text{ eV}\,.
\end{gathered}
\end{equation}
We can see that both $\sin^2 \theta_{12}$ and $r\equiv \Delta m^2_{21}/|\Delta m^2_{3\ell}|$ are in the $1\sigma$ region~\cite{Esteban:2018azc}, while $\sin^2 \theta_{23}$ is a bit larger than its 3$\sigma$ upper limit. Moreover, from Eq.~\eqref{eq:mixpara_A3NO} we know the sum of neutrino masses is
\begin{equation}
\sum_i m_i=0.2632 \text{ eV}\,.
\end{equation}
Under the assumption of $\Lambda$CDM cosmology, the latest Planck result on the neutrino mass sum is~\cite{Aghanim:2018eyx}.
\begin{equation}
\label{eq:cons_mnu:plus}\sum_{i}m_{i}<0.12 \text{ eV}-0.54 \text{ eV}\,.
\end{equation}
Since the upper bound of the neutrino mass sum depends on the cosmological model and whether other experimental data such as baryon acoustic oscillation, gravitational lensing of galaxies and the high multipole TT, TE and EE polarization spectra~\cite{Aghanim:2018eyx} are considered, we shall not seriously take the constraint in Eq.~\eqref{eq:cons_mnu:plus}.

The neutrinoless double beta ($0\nu\beta\beta$) decay is the unique probe for the Majorana nature of neutrinos, and it depends on the values of the Majorana CP violation phases. The $0\nu\beta\beta$ decay experiments can provide valuable information on the neutrino mass spectrum and constrain the Majorana phases. The $0\nu\beta\beta$ decay rate is proportional to the effective Majorana mass $|m_{ee}|$ defined as~\cite{Tanabashi:2018oca},
\begin{equation}
|m_{ee}|=|m_1\cos^2\theta_{12}\cos^2\theta_{13}+m_2\sin^2\theta_{12}\cos^2\theta_{13}e^{i\alpha_{21}}+m_3\sin^2\theta_{13}e^{i(\alpha_{31}-2\delta_{CP})}|\,.
\end{equation}
From Eq.~\eqref{eq:mixpara_A3NO} we can extract the following predicted value for $|m_{ee}|$ in the $\mathcal{A}3$ model with NO,
\begin{equation}
|m_{ee}|=0.04643 \text{ eV}\,,
\end{equation}
which is within the reach of future $0\nu\beta\beta$ experiments.

\begin{figure}[hptb]
\centering
\includegraphics[width=\textwidth]{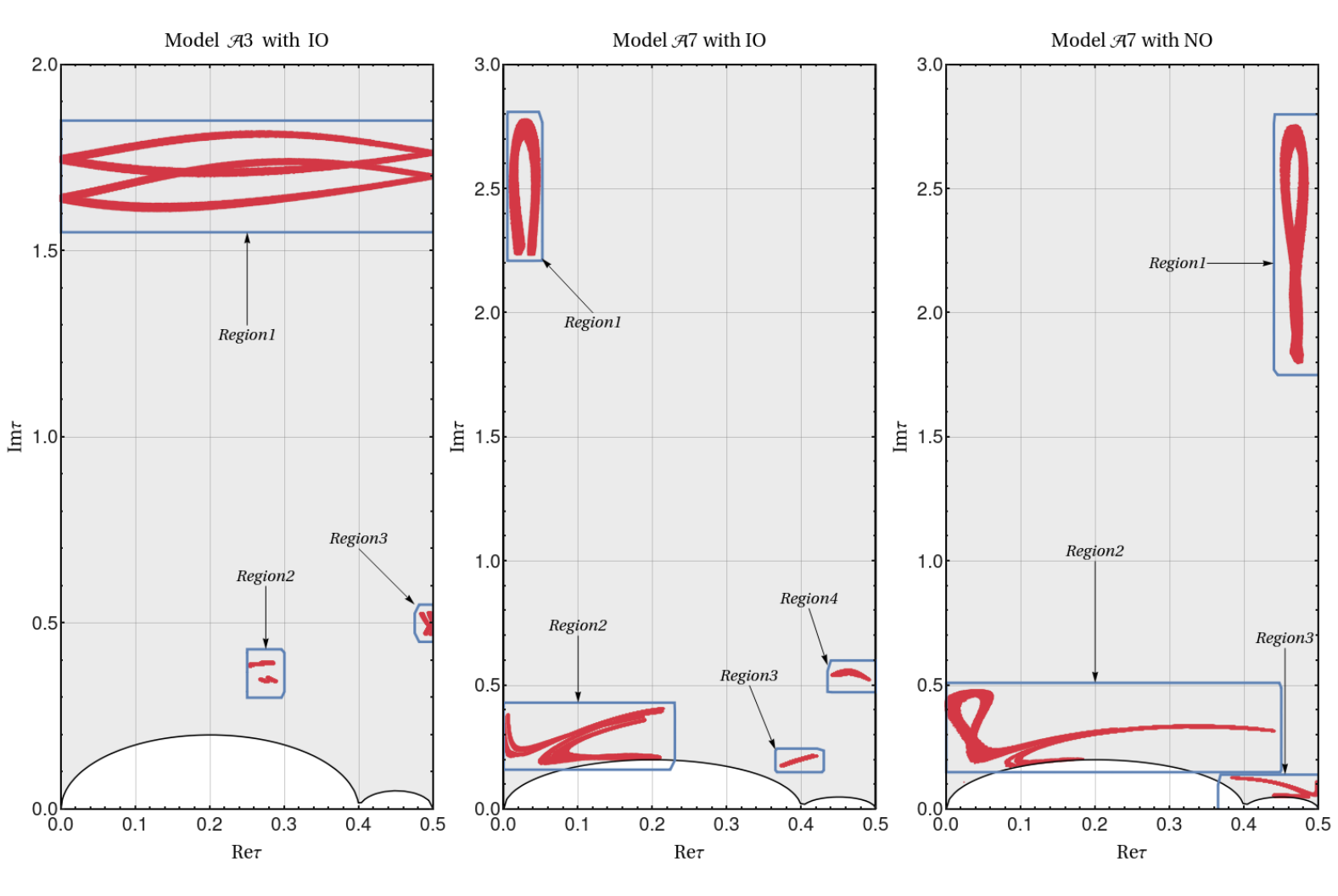}
\caption{The phenomenologically viable regions of the modular parameter $\tau$ in the fundamental domain ${\cal F}_5$ for the models $\mathcal{A}3$ and $\mathcal{A}7$. }
\label{fig:region_viable}
\end{figure}

For the model $\mathcal{A}3$ with IO and model $\mathcal{A}7$ with either NO or IO, there exist parameter spaces to be completely consistent with the
experimental data on neutrino masses and mixing angles. We find that the allowed ranges of the coupling constants and theoretical predictions crucially depend on the values of the complex modulus $\tau$. In each case, we find several distinct local minima of $\chi^2$ in the fundamental domain of $\tau$. We denote the phenomenologically viable regions of $\tau$ as region1, region2, region3 and so on, as shown in figure~\ref{fig:region_viable}, where we have focused on $\text{Re}\tau>0$ and the conjugate regions are not shown for simplicity. We perform an extensive numerical scan over the parameter space, and the results of the numerical analysis are summarized in tables~\ref{tab:modela3_IH1and2}-\ref{tab:modela7_NH3}. The allowed regions of the input parameters and observables are determined by
requiring all the lepton mixing angles and the squared mass splittings $\Delta m^2_{21}$ and $\Delta m^2_{3\ell}$ within the $3\sigma$ intervals~\cite{Esteban:2018azc}.

\begin{figure}[hptb]
\centering
\includegraphics[width=0.94\textwidth]{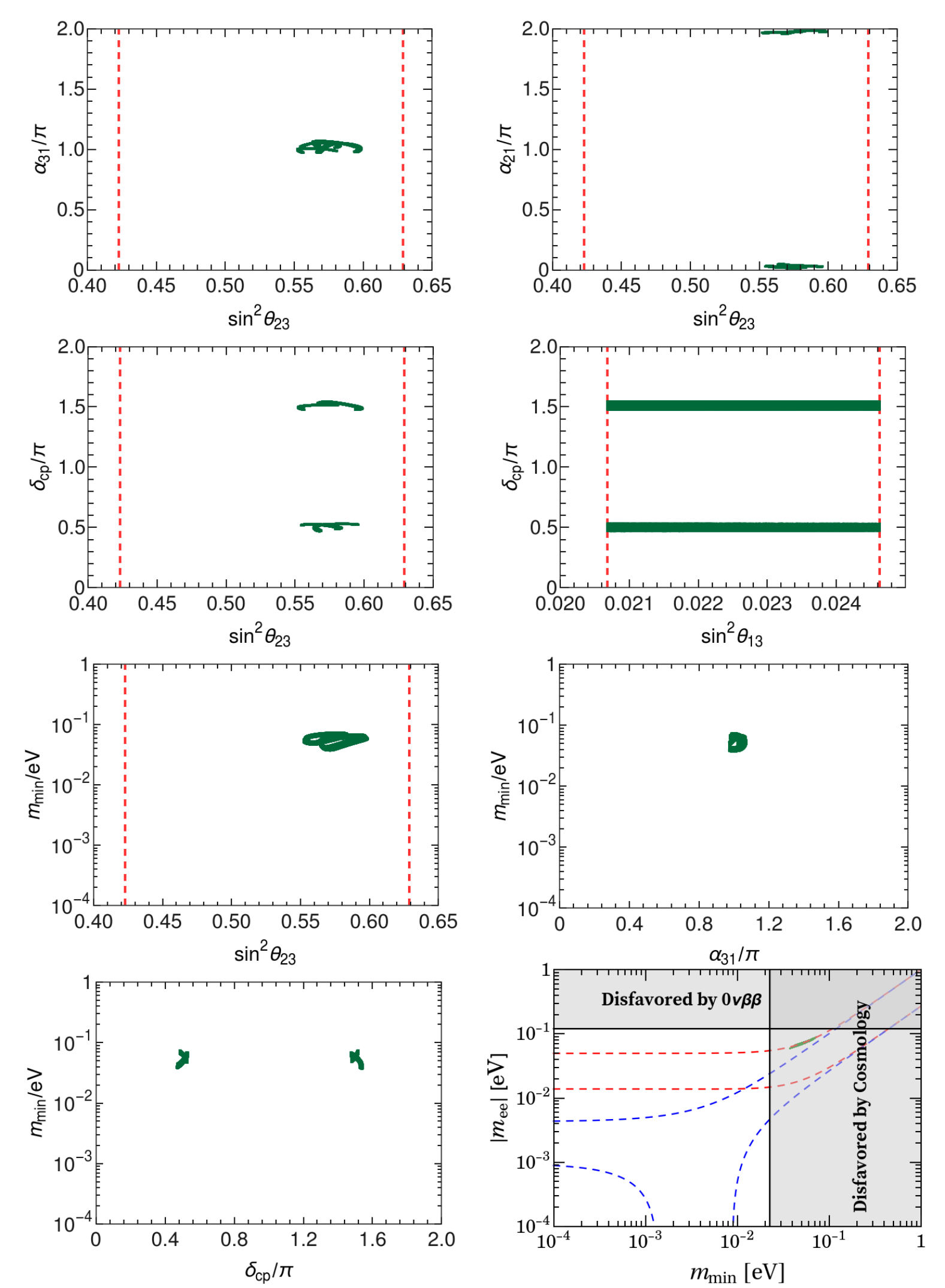}
\caption{The predictions for the correlations among the
neutrino mixing angles, CP violation phases and neutrino masses in the model $\mathcal{A}3$ with IO in region$1$. The $3\sigma$ bounds of the mixing angles are shown by vertical red dashed lines. In the last panel of $|m_{ee}|$ versus $m_{\text{min}}$, the red (blue) dashed lines denote the most general allowed regions for IO (NO) neutrino mass spectrum obtained by varying the mixing parameters over their $3\sigma$ ranges~\cite{Esteban:2018azc}. The present most stringent upper limits $|m_{ee}|<0.120$ eV from EXO-200~\cite{Auger:2012ar,Albert:2014awa} and KamLAND-ZEN~\cite{Gando:2012zm} is represented by horizontal grey band. The vertical grey exclusion band denotes the current bound from the cosmological data of $\sum_i m_i<0.130$ eV at $95\%$ confidence level obtained by the Planck collaboration~\cite{Aghanim:2018eyx}.
Notice that the bound on $\sum_i m_i$ sensitively depends on the cosmological model and whether other experimental data such as BAO and gravitational lensing etc are included~\cite{Aghanim:2018eyx}. }
\label{fig:a3IH1}
\end{figure}

\begin{figure}[hptb]
\centering
\includegraphics[width=\textwidth]{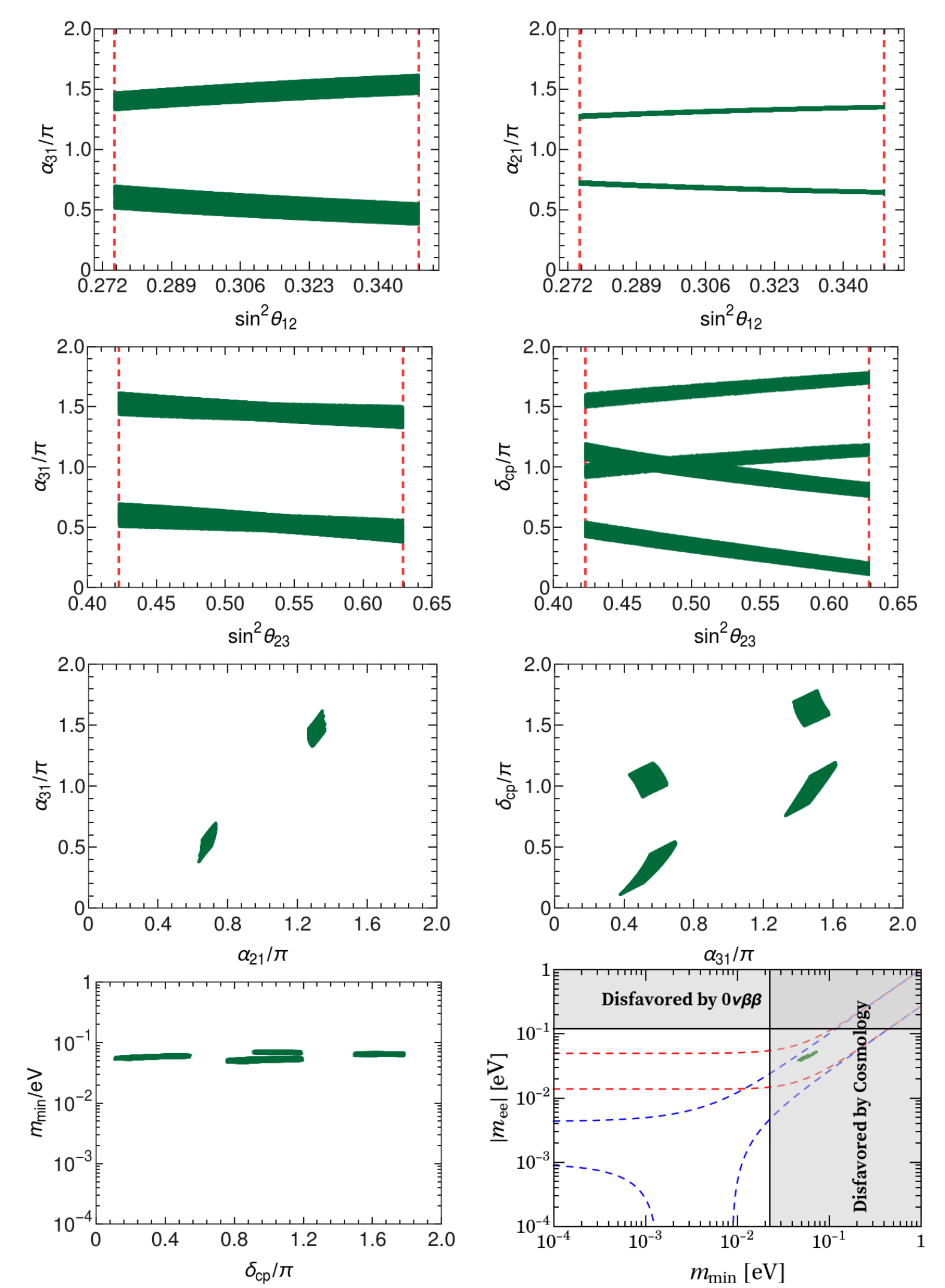}
\caption{The predictions for the correlations among the
neutrino mixing angles, CP violation phases and neutrino masses in the model $\mathcal{A}3$ with IO in region$2$. Here we adopt the same conventions as figure~\ref{fig:a3IH1}. }
\label{fig:a3IH2}
\end{figure}

\begin{figure}[hptb]
\centering
\includegraphics[width=\textwidth]{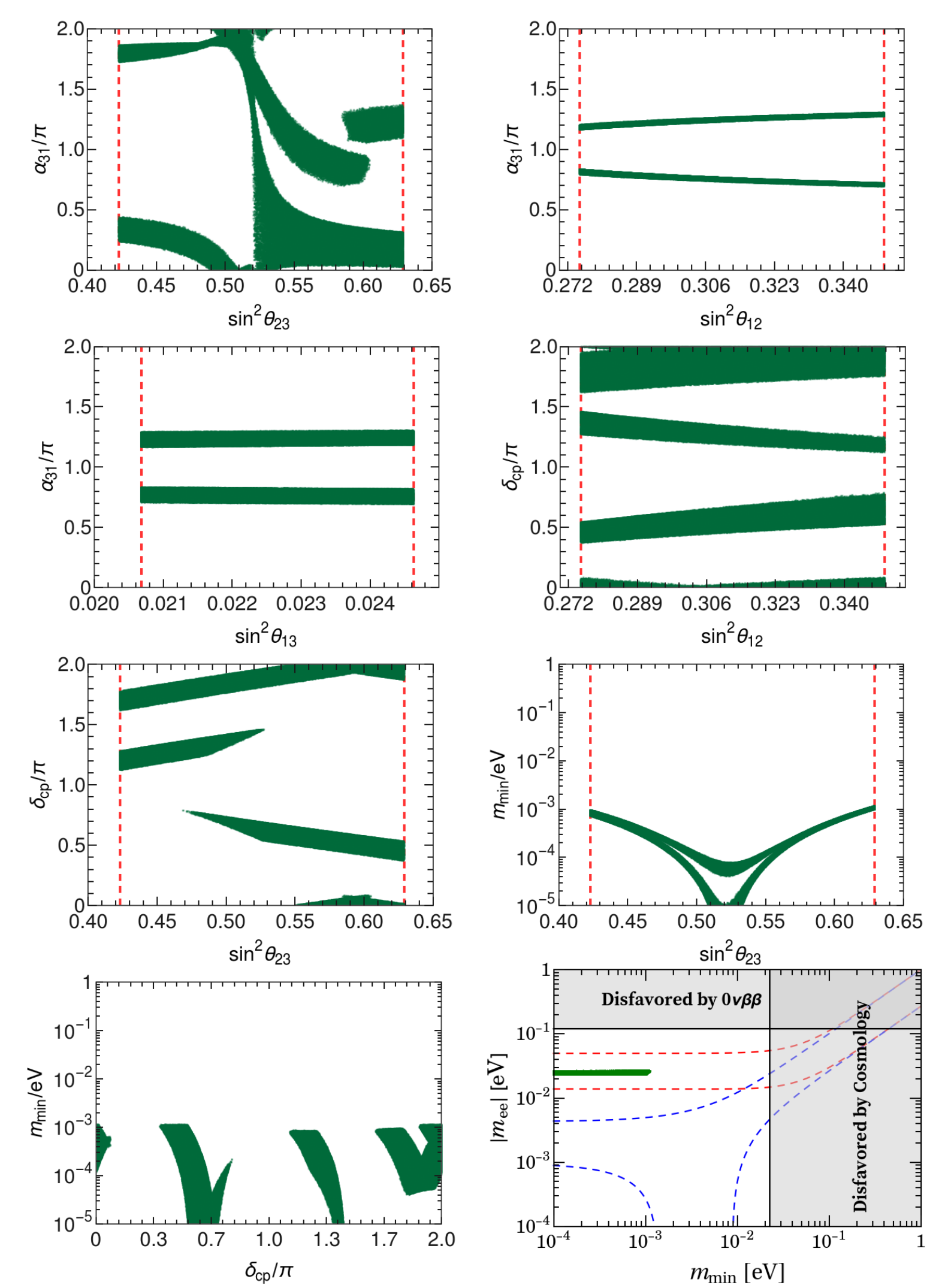}
\caption{The predictions for the correlations among the
neutrino mixing angles, CP violation phases and neutrino masses in the model $\mathcal{A}3$ with IO in region$3$.  Here we adopt the same conventions as figure~\ref{fig:a3IH1}. }
\label{fig:a3IH3}
\end{figure}

\begin{figure}[hptb]
\centering
\includegraphics[width=\textwidth]{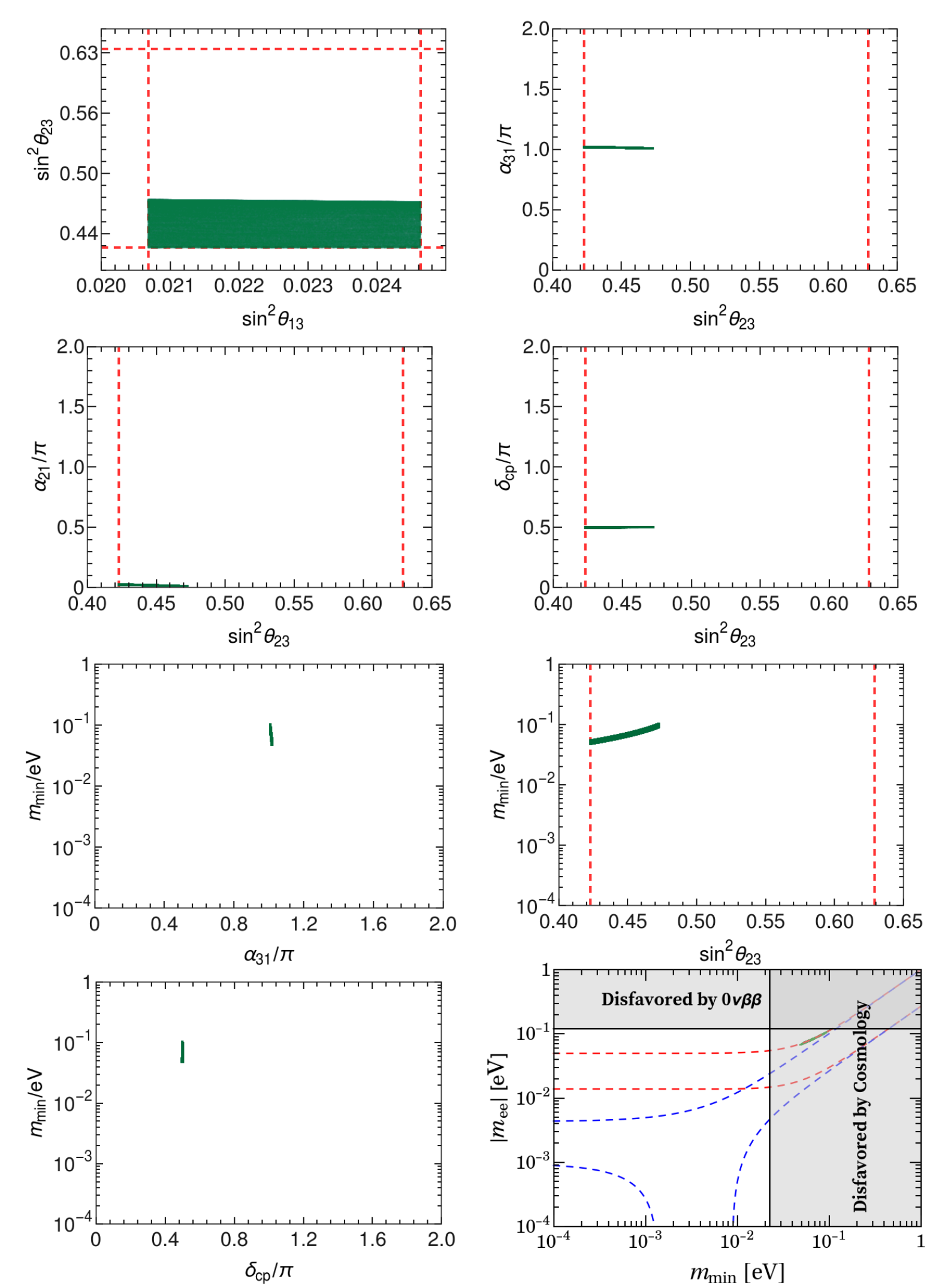}
\caption{The predictions for the correlations among the
neutrino mixing angles, CP violation phases and neutrino masses in the model $\mathcal{A}7$ with IO in region$1$.  Here we adopt the same conventions as figure~\ref{fig:a3IH1}. }
\label{fig:a7IH1}
\end{figure}

\begin{figure}[hptb]
\centering
\includegraphics[width=\textwidth]{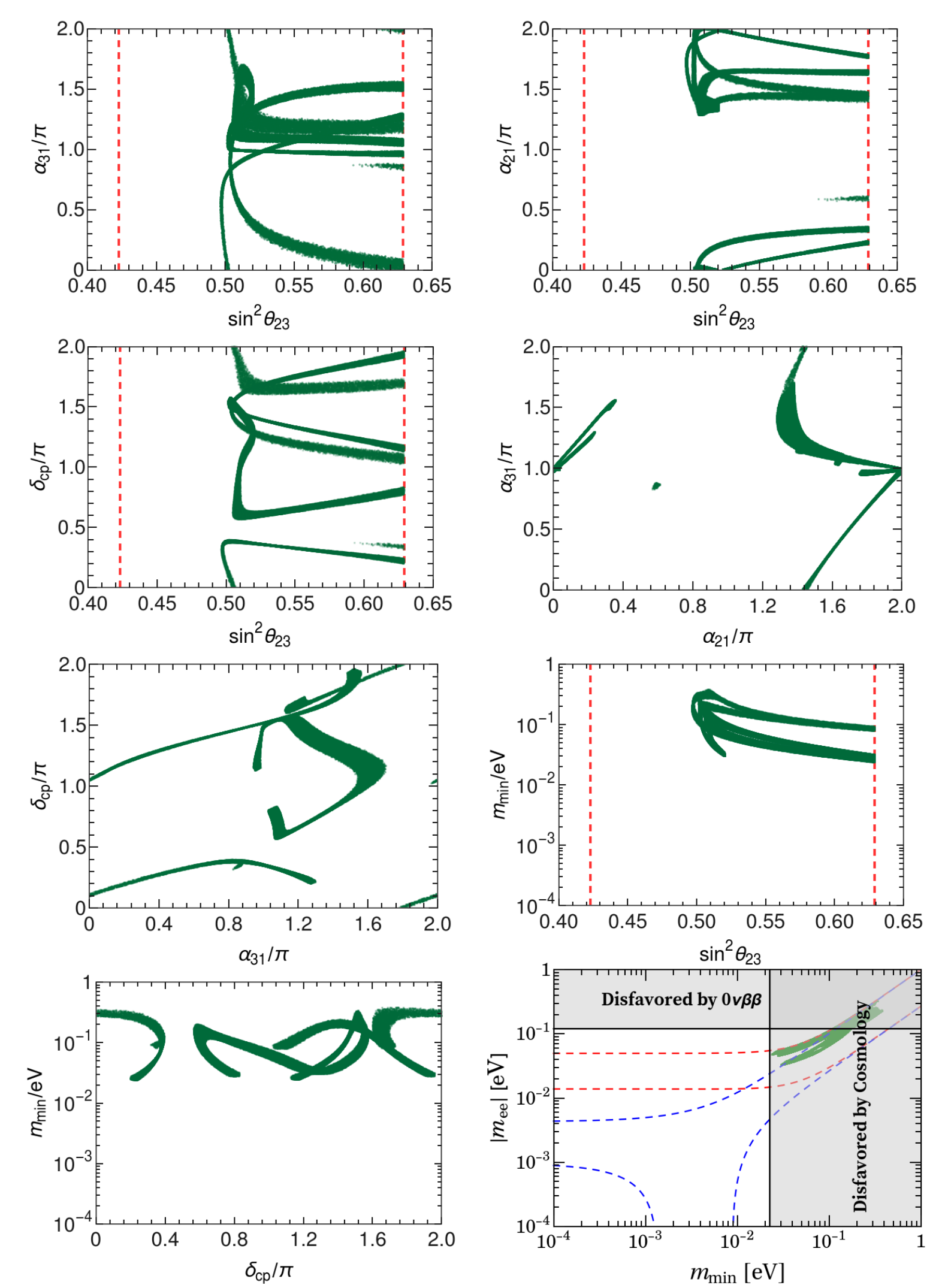}
\caption{The predictions for the correlations among the
neutrino mixing angles, CP violation phases and neutrino masses in the model $\mathcal{A}7$ with IO in region$2$.  Here we adopt the same conventions as figure~\ref{fig:a3IH1}. }
\label{fig:a7IH2}
\end{figure}

\begin{figure}[hptb]
\centering
\includegraphics[width=\textwidth]{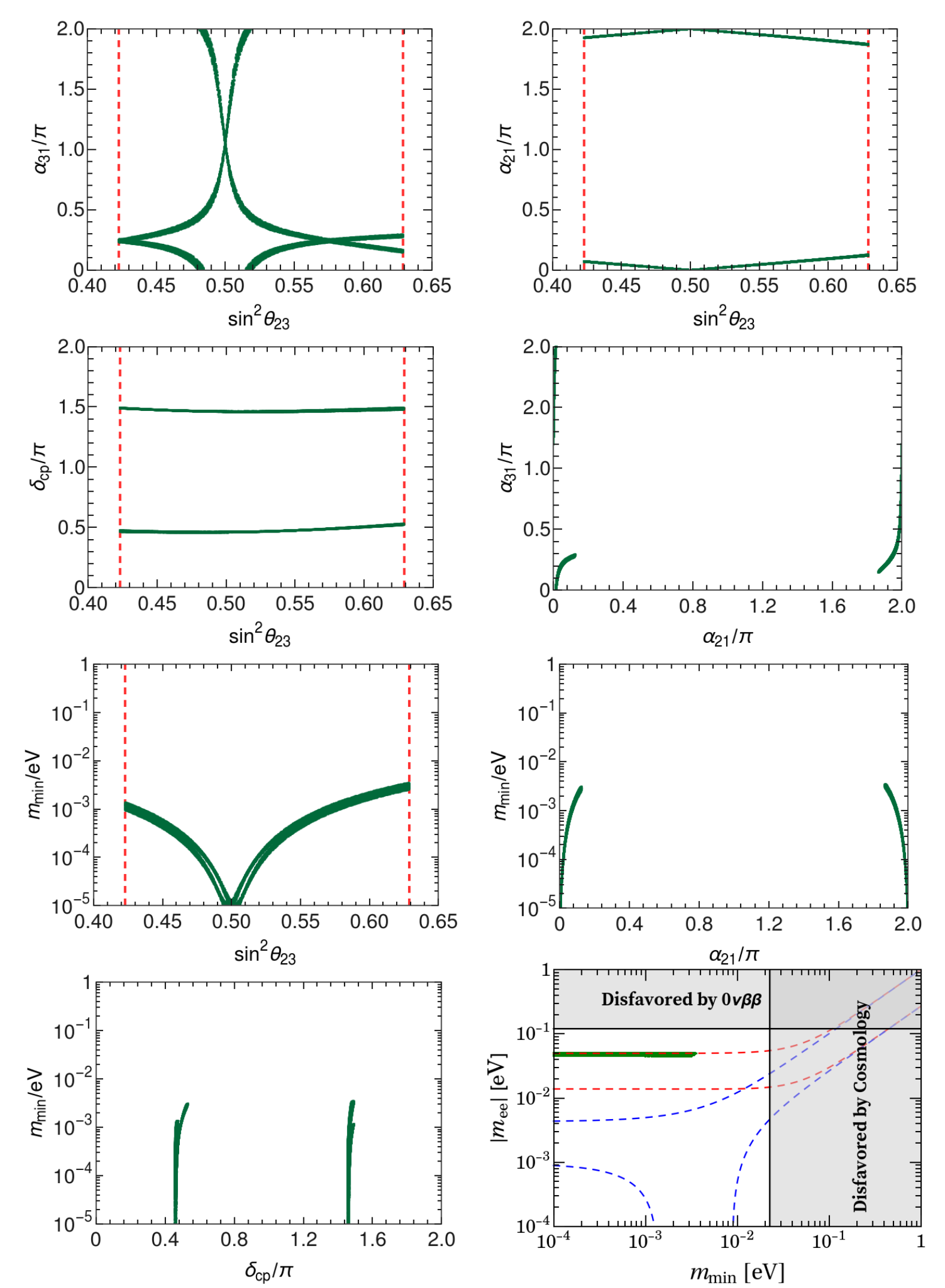}
\caption{The predictions for the correlations among the
neutrino mixing angles, CP violation phases and neutrino masses in the model $\mathcal{A}7$ with IO in region$3$.  Here we adopt the same conventions as figure~\ref{fig:a3IH1}. }
\label{fig:a7IH3}
\end{figure}

\begin{figure}[hptb]
\centering
\includegraphics[width=\textwidth]{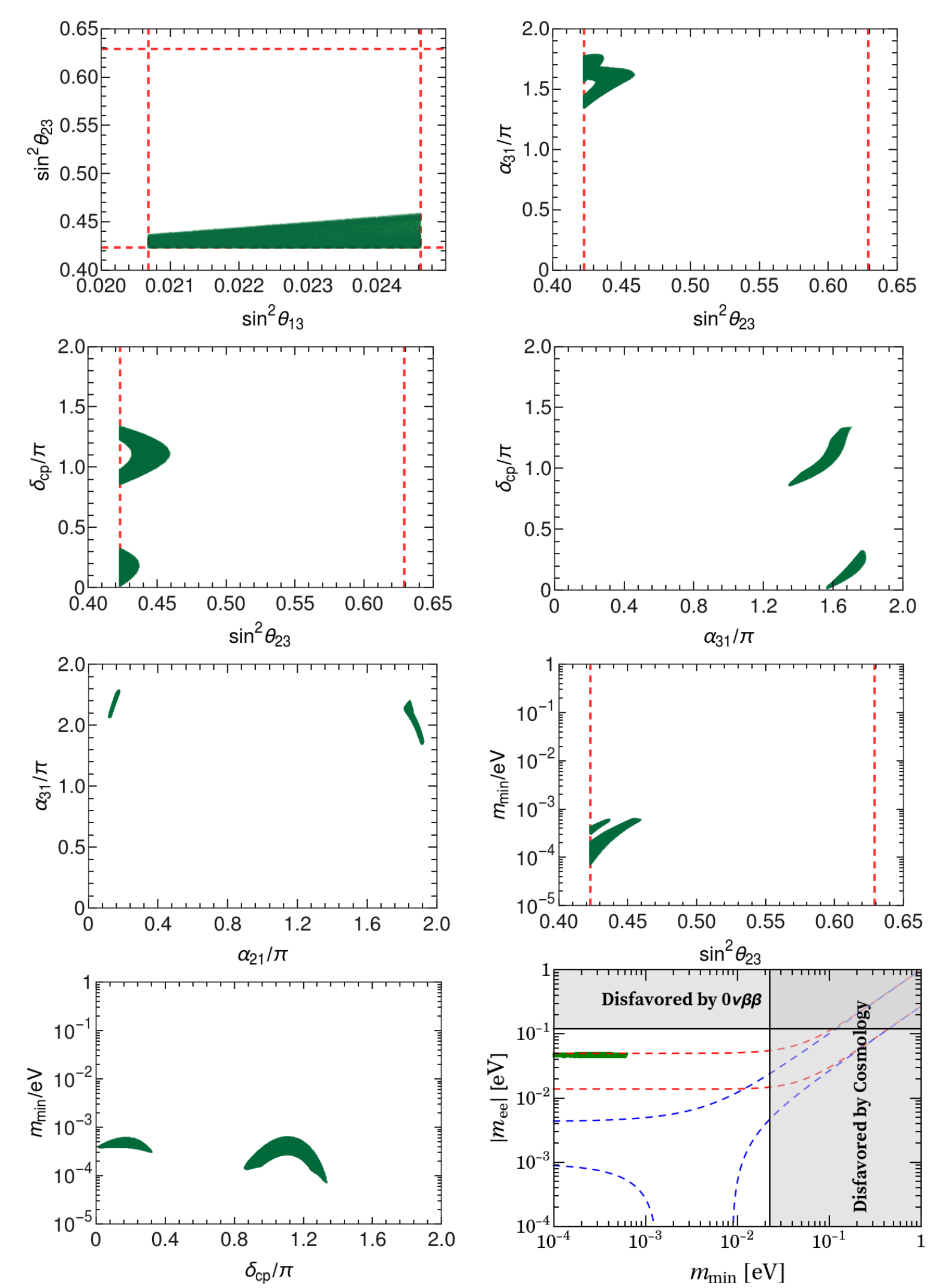}
\caption{The predictions for the correlations among the
neutrino mixing angles, CP violation phases and neutrino masses in the model $\mathcal{A}7$ with IO in region$4$.  Here we adopt the same conventions as figure~\ref{fig:a3IH1}. }
\label{fig:a7IH4}
\end{figure}

\begin{figure}[hptb]
\centering
\includegraphics[width=\textwidth]{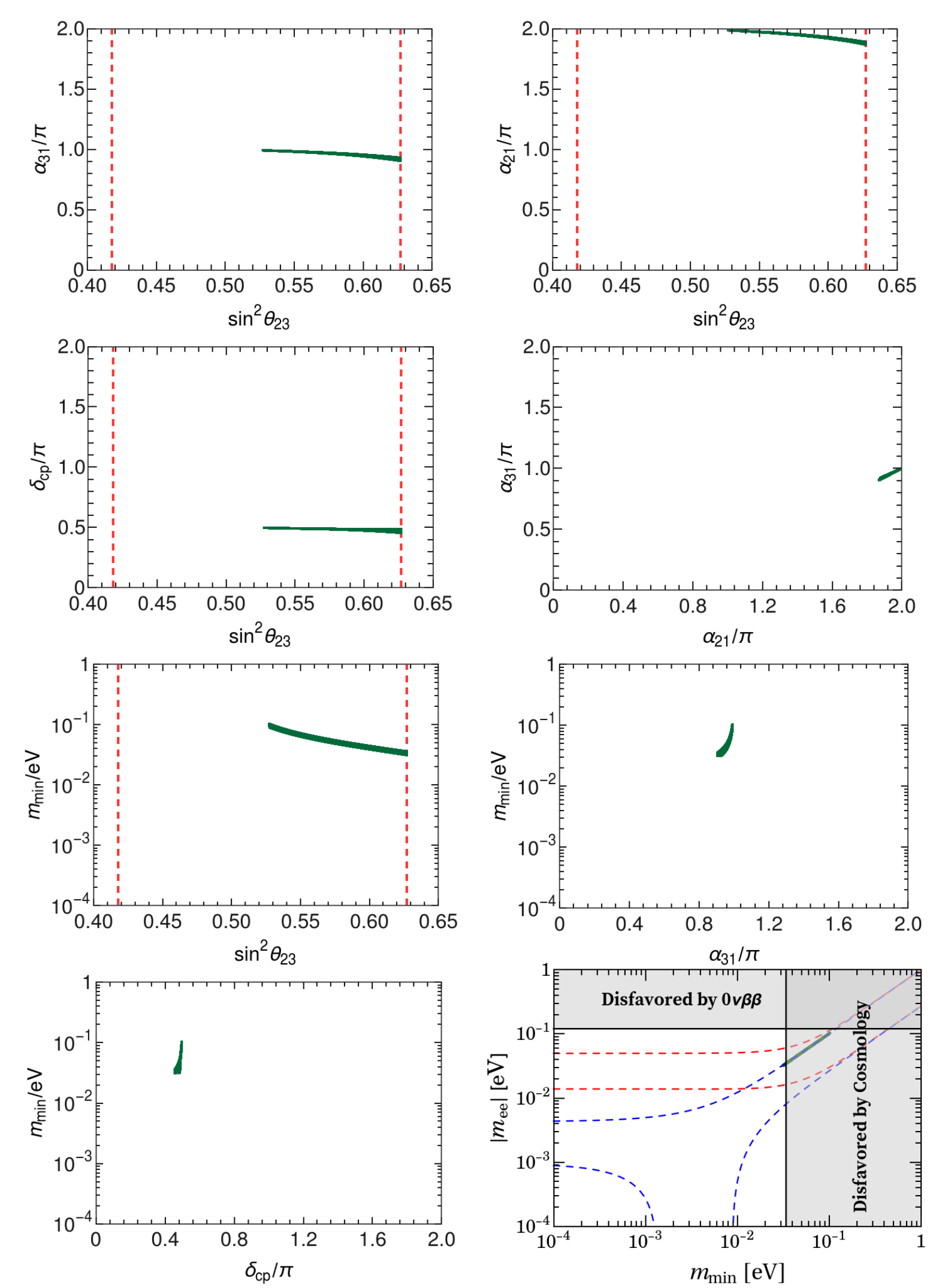}
\caption{The predictions for the correlations among the
neutrino mixing angles, CP violation phases and neutrino masses in the model $\mathcal{A}7$ with NO in region$1$. Here we adopt the same conventions as figure~\ref{fig:a3IH1}. }
\label{fig:a7NH1}
\end{figure}

\begin{figure}[hptb]
\centering
\includegraphics[width=\textwidth]{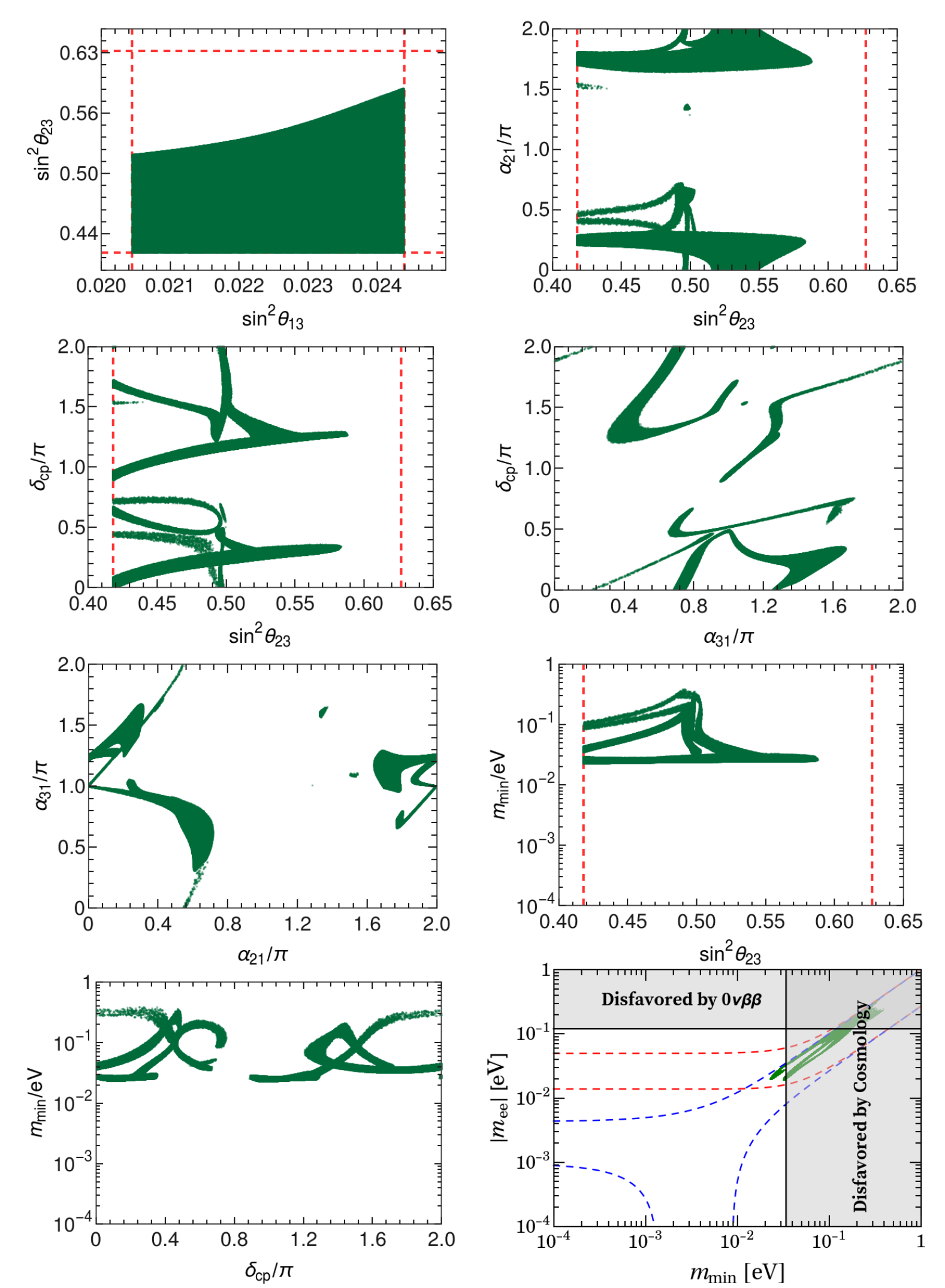}
\caption{The predictions for the correlations among the
neutrino mixing angles, CP violation phases and neutrino masses in the model $\mathcal{A}7$ with NO in region$2$. Here we adopt the same conventions as figure~\ref{fig:a3IH1}. }
\label{fig:a7NH2}
\end{figure}

\begin{figure}[hptb]
\centering
\includegraphics[width=\textwidth]{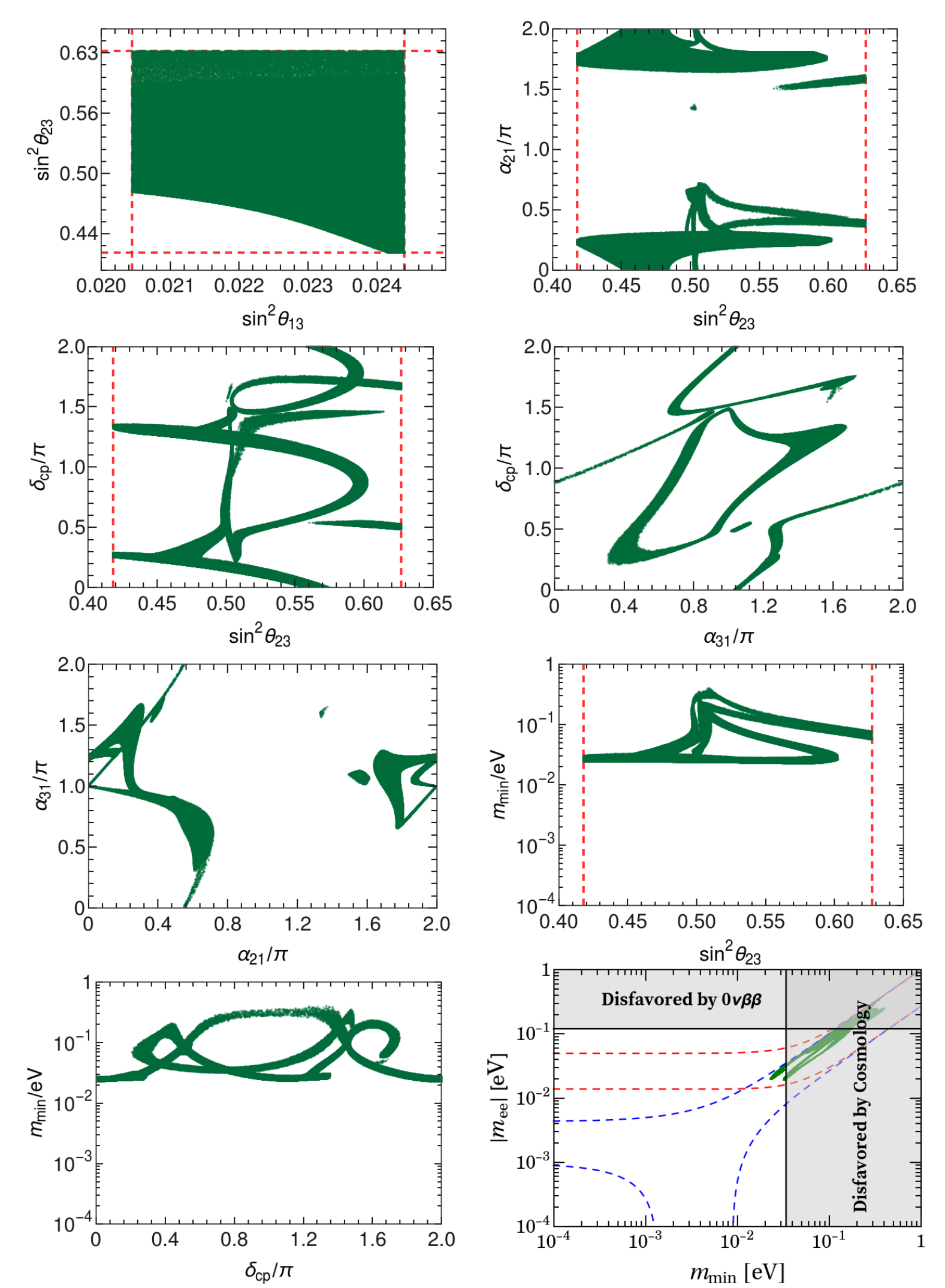}
\caption{The predictions for the correlations among the
neutrino mixing angles, CP violation phases and neutrino masses in the model $\mathcal{A}7$ with NO in region$3$. Here we adopt the same conventions as figure~\ref{fig:a3IH1}. }
\label{fig:a7NH3}
\end{figure}

Since the models depend on few free parameters, we find that the lepton mixing angles, CP violation phases, the lightest neutrino mass $m_{\rm min}$ and the effective Majorana mass $|m_{ee}|$ are generally correlated with each other. We display the different correlations in figures~\ref{fig:a3IH1}-\ref{fig:a7NH3}. We see that the Dirac CP violation phase $\delta_{CP}$ is predicted to lie in narrow regions in most cases. The forthcoming long-baseline neutrino oscillation experiments will considerably improve the sensitivity to $\delta_{CP}$ if running in both the neutrino and the anti-neutrino modes, we expect the predictions for $\delta_{CP}$ in this paper could be tested by future neutrino facilities. Furthermore, we notice that both $m_{\text{min}}$ and $|m_{ee}|$ can only obtain values in very limited ranges. The next generation experiments searching for $0\nu\beta\beta$ decay will be able to probe almost all the IO region, up to $m_{ee}\simeq0.02$ eV,
thus allowing to testing these modular symmetry models as well.

\begin{table}
\centering
\renewcommand{\arraystretch}{1.2}
\begin{tabular}{|c|cc|cc|} \hline\hline
\texttt{Model} & \multicolumn{2}{c|}{\texttt{Region1}} &  \multicolumn{2}{c|}{\texttt{Region2} } \\ \cline{2-5}
 $\mathcal{A}3$ & \texttt{Best-fit values}  & \texttt{Allowed ranges}  & \texttt{Best-fit values}  & \texttt{Allowed ranges} \\ \hline
$\rm{Re}\,\langle \tau \rangle$   & 0.2664 & $0.0 \sim 0.5$          & 0.2872 & $0.2525 \sim 0.2902$\\
  $\rm{Im}\,\langle \tau \rangle$   & 1.6324 &  $1.6060 \sim 1.8206 $  & 0.3441 &  $0.3409 \sim 0.3949 $\\
  $|g_1/g_2|$                       & 2.0926 & $2.0568 \sim 2.1503$    & 2.1296 & $2.0854 \sim 2.1457$ \\
  $\text{Arg}(g_1/g_2)$                    & 6.2308 & $6.1825 \sim 6.3905$    & 6.2954 & $6.2028 \sim 6.3265$ \\
  $ g^2_2 v_u^2/ \Lambda$ [eV]      & 0.0173 &   ---                     & 0.0008 & --- \\ \hline
$\sin^2 \theta_{12}$              & 0.3098 & $0.2750 \sim 0.3500$    & 0.3099 & $0.2750 \sim 0.3500$ \\
  $\sin^2 \theta_{13}$              & 0.0226 & $0.02068 \sim 0.02463$  & 0.0226 & $0.02068 \sim 0.02463$\\
  $\sin^2 \theta_{23}$              & 0.5835 & $0.5521 \sim 0.5989$    & 0.5840 & $0.4230 \sim 0.6290$ \\
  $\delta_{CP}/\pi$                 & 1.5265 &  $0.4669 \sim 1.5453$   & 1.7040 &  $0.1066 \sim 1.7877$\\
  $\alpha_{21}/\pi$                 & 1.9831 &  $0.0149 \sim 1.9859$   & 1.3212 &  $0.6335 \sim 1.3604$ \\
  $\alpha_{31}/\pi$                 & 1.0485&  $0.9716 \sim 1.0709$    & 1.4525&  $0.3758 \sim 1.6190$ \\ \hline
  $r$                               & 0.02942 & $0.0260 \sim 0.0332$   & 0.0294 & $0.0260 \sim 0.0332$ \\
  $m_1$ [eV]                        & 0.0696 & $0.0593 \sim 0.0909$    & 0.0814 & $0.0650 \sim 0.0908$\\
  $m_2$ [eV]                        & 0.0702 & $0.0600 \sim 0.0913$    & 0.0819 & $0.0656 \sim 0.0913$ \\
  $m_3$ [eV]                        & 0.0491 & $0.0371 \sim 0.0741$    & 0.0647 & $0.0456 \sim 0.0741$ \\
  $\textstyle \sum_i m_i$ [eV]      & 0.1889 & $0.1566 \sim 0.2562$    & 0.2280 & $0.1762 \sim 0.2562$\\
  $|m_{ee}|$ [eV]                     & 0.0693 &  $0.0590 \sim 0.0906$   & 0.0479 &  $0.0376 \sim 0.0537$ \\ \hline
 \texttt{Ordering}                          &  \multicolumn{2}{c|}{IO}                            & \multicolumn{2}{c|}{IO}           \\ \hline
  $N\sigma$                   & 0.03   &     ---                   & 0.013   &     ---          \\ \hline\hline
\end{tabular}
\caption{The best-fit values and the allowed ranges of the model parameters and lepton mixing parameters and neutrino masses for the model $\mathcal{A}3$ with IO in region1 and region2. }
\label{tab:modela3_IH1and2}
\end{table}
\begin{table}
\centering
\renewcommand{\arraystretch}{1.2}
\begin{tabular}{|c|cc|} \hline\hline
\texttt{Model}  &  \multicolumn{2}{c|}{\texttt{Region 3} } \\ \cline{2-3}
   $\mathcal{A}3$       ~&~ \texttt{ Best-fit values}  &  \texttt{Allowed ranges} \\ \hline
  $\rm{Re}\,\langle \tau \rangle$   & 0.4937 & $0.4829 \sim 0.5000$\\
  $\rm{Im}\,\langle \tau \rangle$   & 0.4830 &  $0.4680 \sim 0.5285 $\\
  $|g_1/g_2|$                       & 1.7244 & $1.7207 \sim 1.7365$ \\
  $\text{Arg}(g_1/g_2)$                    & 6.4672 & $6.0579 \sim 6.5125$ \\
  $ g^2_2 v_u^2/ \Lambda$ [eV]      & 0.0022  &  --- \\ \hline
  $\sin^2 \theta_{12}$              & 0.3099 & $0.2750 \sim 0.3500$ \\
  $\sin^2 \theta_{13}$              & 0.0226 & $0.02068 \sim 0.02463$\\
  $\sin^2 \theta_{23}$              & 0.5840 & $0.4230 \sim 0.6290$ \\
  $\delta_{CP}/\pi$                 & 0.5300 &  $0 \sim 2$\\
  $\alpha_{21}/\pi$                 & 1.2400 &  $0.6920 \sim 1.3041$ \\
  $\alpha_{31}/\pi$                 & 0.2211 &  $0 \sim 2$ \\ \hline
  $r$                               & 0.0294 & $0.0260 \sim 0.0332$ \\
  $m_1$ [eV]                        & 0.0494 & $0.0464 \sim 0.0526$\\
  $m_2$ [eV]                        & 0.0501 & $0.0472 \sim 0.0533$ \\
  $m_3$ [eV]                        & 0.00035 & $0 \sim 0.0011$ \\
  $\textstyle \sum_i m_i$ [eV]      & 0.0998 & $0.0935 \sim 0.1071$\\
  $|m_{ee}|$ [eV]                     & 0.0245 &  $0.0226 \sim 0.0272$ \\ \hline
\texttt{Ordering} & \multicolumn{2}{c|}{IO} \\ \hline
  $N\sigma$  & 0.012  &  ---  \\ \hline\hline
\end{tabular}
\caption{The best-fit values and the allowed ranges of the model parameters and lepton mixing parameters and neutrino masses for the model $\mathcal{A}3$ with IO in region3. Notice that the Dirac CP phase $\delta_{CP}\simeq1.47 \pi$ at the conjugate of the best fit point. }
\label{tab:modela3_IH3}
\end{table}
\begin{table}
\centering
\renewcommand{\arraystretch}{1.2}
\begin{tabular}{|c|cc|cc|} \hline\hline
\texttt{Model} & \multicolumn{2}{c|}{\texttt{ Region$1$} } & \multicolumn{2}{c|}{\texttt{Region$2$ }} \\ \cline{2-5}
 $\mathcal{A}7$ & \texttt{ Best-fit values}  &  \texttt{ Allowed ranges} & \texttt{ Best-fit values}  &  \texttt{ Allowed ranges} \\ \hline
  $\rm{Re}\,\langle \tau \rangle$   & 0.0285 & $0.0085 \sim 0.0490$   & 0.1541 & $0 \sim 0.2154$        \\
  $\rm{Im}\,\langle \tau \rangle$   & 2.7418 &  $2.2323 \sim 2.7776 $ & 0.3319 &  $0.1800 \sim 0.4049 $  \\
  $|g_1/g_2|$                       & 2.1867 & $1.6078 \sim 2.1889$   & 1.7603 & $1.4208\sim 2.2314$     \\
  $\text{Arg}(g_1/g_2)$                    & 3.1478 & $1.8976 \sim 4.3823$   & 0.7635 & $0 \sim 6.2832$         \\
  $ g^2_2 v_u^2/ \Lambda$ [eV]      & 0.0266 &   ---   & 0.0003 &  ---                         \\ \hline
  $\sin^2 \theta_{12}$              & 0.3106 & $0.2750 \sim 0.3500$   & 0.3099 & $0.2750 \sim 0.3500$      \\
  $\sin^2 \theta_{13}$              & 0.0226 & $0.02068 \sim 0.02463$ & 0.0226 & $0.02068 \sim 0.02463$   \\
  $\sin^2 \theta_{23}$              & 0.4714 & $0.4230 \sim 0.5039$   & 0.5840 & $0.4966 \sim 0.6290$     \\
  $\delta_{CP}/\pi$                 & 0.5006 &  $0.4965 \sim 0.5039$  & 0.7096 &  $0.0024 \sim 2$         \\
  $\alpha_{21}/\pi$                 & 0.0119 & $0.0106 \sim 0.0275$   & 1.6433 & $0 \sim 2$               \\
  $\alpha_{31}/\pi$                 & 1.0084 & $1.0071 \sim 1.0208$   & 1.0794 & $0.0015 \sim 2$           \\ \hline
  $r$                               & 0.02940 & $0.0260 \sim 0.0332$  & 0.02942 & $0.02601 \sim 0.03321$  \\
  $m_1$ [eV]                        & 0.1066 & $0.0669 \sim 0.1156$   & 0.0652 & $0.0522 \sim 0.3826$     \\
  $m_2$ [eV]                        & 0.1070 & $0.0675 \sim 0.1159$   & 0.0658 & $0.0529 \sim 0.3827$      \\
  $m_3$ [eV]                        & 0.0945 & $0.0482 \sim 0.1030$   & 0.0426 & $0.0238 \sim 0.3791$      \\
  $\textstyle \sum_i m_i$ [eV]      & 0.3081 & $0.1826 \sim 0.3345$   & 0.1736 & $0.1288 \sim 1.1444$     \\
  $|m_{ee}|$ [eV]                     & 0.1064 &  $0.0666 \sim 0.1154$  & 0.0564 &  $0.0329 \sim 0.3262$     \\ \hline
  \texttt{Ordering}    & \multicolumn{2}{c|}{IO}            & \multicolumn{2}{c|}{IO}    \\ \hline
  $N\sigma$      & 5.63   &     ---     & 0.005  &    ---                        \\ \hline\hline
\end{tabular}
\caption{The best-fit values and the allowed ranges of the model parameters and lepton mixing parameters and neutrino masses for the model $\mathcal{A}7$ with IO in region$1$ and region$2$. Notice that the Dirac CP phase is $\delta_{CP}\simeq1.499\pi$ in region$1$ and $\delta_{CP}\simeq1.29\pi$ in region$2$ at the conjugate best-fit points. }
\label{tab:modela7_IH1and2}
\end{table}
\begin{table}
\centering
\renewcommand{\arraystretch}{1.2}
\begin{tabular}{|c|cc|cc|} \hline\hline
\texttt{ Model} & \multicolumn{2}{c}{\texttt{ Region$3$} } & \multicolumn{2}{c|}{\texttt{Region$4$ }} \\ \cline{2-5}
 $\mathcal{A}7$& \texttt{Best-fit values}  &  \texttt{ Allowed ranges}  & \texttt{ Best-fit values}  &  \texttt{ Allowed ranges}\\ \hline
  $\rm{Re}\,\langle \tau \rangle$   & 0.3808 & $0.3734 \sim 0.4211$       & 0.4686 & $0.4411 \sim 0.4930$   \\
  $\rm{Im}\,\langle \tau \rangle$   & 0.1836 &  $0.1741 \sim 0.2165 $     & 0.5535 &  $0.5185 \sim 0.5600$  \\
  $|g_1/g_2|$                       & 4.4862 & $4.2475 \sim 4.8215$       & 8.0775 & $6.5062 \sim 17.3445$   \\
  $\text{Arg}(g_1/g_2)$                    & 1.5227 & $1.4285 \sim 1.6106$       & 4.7888 & $4.6699 \sim 4.8307$    \\
  $ g^2_2 v_u^2/ \Lambda$ [eV]      & 0.00005 &   ---  & 0.00015 &  ---                      \\ \hline
  $\sin^2 \theta_{12}$              & 0.3100 & $0.2750 \sim 0.3450$       & 0.3086 & $0.2750 \sim 0.3500$      \\
  $\sin^2 \theta_{13}$              & 0.0226 & $0.02068 \sim 0.02463$     & 0.0235 & $0.02068 \sim 0.02463$    \\
  $\sin^2 \theta_{23}$              & 0.5840 & $0.4230 \sim 0.6290$       & 0.4497 & $0.4230 \sim 0.4593$     \\
  $\delta_{CP}/\pi$                 & 1.4709 &  $0.4552 \sim 1.4944$      & 1.1137 &  $0.016 \sim 1.3366$     \\
  $\alpha_{21}/\pi$                 & 1.9181 & $0 \sim 2$                 & 1.8428 & $0.1181 \sim 1.9226$      \\
  $\alpha_{31}/\pi$                 & 0.2260 & $0 \sim 2$                 & 1.6057 & $1.3481 \sim 1.7857$      \\ \hline
  $r$                               & 0.02942 & $0.02601 \sim 0.03321$    & 0.02973 & $0.02601 \sim 0.03321$   \\
  $m_1$ [eV]                        & 0.0494 & $0.0464 \sim 0.0527$       & 0.0491 & $0.0464 \sim 0.0526$      \\
  $m_2$ [eV]                        & 0.0501 & $0.0472 \sim 0.0534$       & 0.0498 & $0.0472 \sim 0.0533$     \\
  $m_3$ [eV]                        & 0.0015 & $0 \sim 0.0035$            & 0.0005 & $0 \sim 0.0006$         \\
  $\textstyle \sum_i m_i$ [eV]      & 0.1010 & $0.0935 \sim 0.1096$       & 0.0995 & $0.0936 \sim 0.1066$     \\
  $|m_{ee}|$ [eV]                     & 0.0481 &  $0.0448 \sim 0.0517$      & 0.0469 &  $0.0440 \sim 0.0512$    \\ \hline
 \texttt{Ordering}        & \multicolumn{2}{c|}{IO}   &  \multicolumn{2}{c|}{IO}      \\ \hline
  $N\sigma$   & 0.006  &   ---   & 6.8373  &  ---  \\ \hline\hline
\end{tabular}
\caption{The best-fit values and the allowed ranges of the model parameters and lepton mixing parameters and neutrino masses for the model $\mathcal{A}7$ with IO in region$3$ and region$4$. }
\label{tab:modela7_IH3and4}
\end{table}
\begin{table}
\centering
\renewcommand{\arraystretch}{1.2}
\begin{tabular}{|c|cc|cc|} \hline\hline
\texttt{ Model} & \multicolumn{2}{c|}{\texttt{ Region$1$ }} & \multicolumn{2}{c|}{\texttt{ Region$2$}} \\ \cline{2-5}
 $\mathcal{A}7$ & \texttt{ Best-fit values}  &  \texttt{ Allowed ranges} & \texttt{ Best-fit values}  &  \texttt{ Allowed ranges} \\ \hline
 $\rm{Re}\,\langle \tau \rangle$   & 0.4691 & $0.4493 \sim 0.4866$      & 0.3921  & $0 \sim 0.4410$          \\
  $\rm{Im}\,\langle \tau \rangle$   & 2.1224 &  $1.7957 \sim 2.7549 $    & 0.3267  &  $0.1704 \sim 0.4765 $   \\
  $|g_1/g_2|$                       & 1.9604 & $1.9494 \sim 2.2923$      & 2.0702  & $1.7430 \sim 2.4915$     \\
  $\text{Arg}(g_1/g_2)$                    & 4.4413 & $1.3466 \sim 4.9197$      & 4.3707  & $0 \sim 6.2832$          \\
$ g^2_2 v_u^2/ \Lambda$ [eV]      & 0.0125 & ---  & 0.0004  &  ---                   \\ \hline
$\sin^2 \theta_{12}$              & 0.3099 & $0.2750 \sim 0.3500$      & 0.3099  & $0.2750 \sim 0.3500$    \\
  $\sin^2 \theta_{13}$              & 0.0224 & $0.02045 \sim 0.02439$    & 0.0230  & $0.02045 \sim 0.02439$   \\
  $\sin^2 \theta_{23}$              & 0.5800 & $0.5274 \sim 0.6270$      & 0.5519  & $0.4180 \sim 0.5873$     \\
  $\delta_{CP}/\pi$                 & 0.4847 &  $0.4542 \sim 0.4966$     & 1.2307  &  $0 \sim 2$              \\
  $\alpha_{21}/\pi$                 & 1.9486 & $1.8688 \sim 1.9881$      & 1.7959  & $0 \sim 2$               \\
  $\alpha_{31}/\pi$                 & 0.9647 & $0.9038 \sim 0.9924$      & 1.2150  & $0.0018 \sim 1.9655$    \\ \hline
$r$             & 0.02927 & $0.02587 \sim 0.03300$   & 0.02928 & $0.02587 \sim 0.03300$ \\
  $m_1$ [eV]                        & 0.04930 & $0.0314 \sim 0.1029$     & 0.0260  & $0.0228 \sim 0.3784$    \\
  $m_2$ [eV]                        & 0.05004 & $0.0325 \sim 0.1032$     & 0.0274  & $0.0244 \sim 0.3785$    \\
  $m_3$ [eV]                        & 0.07039 & $0.0568 \sim 0.1159$     & 0.0566  & $0.0525 \sim 0.3821$    \\
  $\textstyle \sum_i m_i$ [eV]      & 0.1697  & $0.1207 \sim 0.3220$     & 0.1100  & $0.0997 \sim 1.1390$    \\
  $|m_{ee}|$ [eV]                     & 0.0499  &  $0.0318 \sim 0.1033$    & 0.0236  &  $0.0189 \sim 0.3355$    \\ \hline
\texttt{Ordering}       &  \multicolumn{2}{c|}{NO}    &  \multicolumn{2}{c|}{NO}  \\ \hline
$N\sigma$     & 0.005   &   ---     & 1.622   & ---     \\ \hline\hline
\end{tabular}
\caption{The best-fit values and the allowed ranges of the model parameters and lepton mixing parameters and neutrino masses for the model $\mathcal{A}7$ with NO in region$1$ and region$2$. Notice that the Dirac CP phase $\delta_{CP}\simeq1.515\pi$ at the conjugate best fit point in region1. }
\label{tab:modela7_NH1}
\end{table}
\begin{table}
\centering
\renewcommand{\arraystretch}{1.2}
\begin{tabular}{|c|cc|} \hline\hline
\texttt{Model} & \multicolumn{2}{c|}{\texttt{Region$3$}}   \\ \cline{2-3}
  $\mathcal{A}7$ & \texttt{Best-fit values}  & \texttt{Allowed ranges}  \\ \hline
  $\rm{Re}\,\langle \tau \rangle$   & 0.4186 & $0.3836 \sim 0.500$\\
  $\rm{Im}\,\langle \tau \rangle$    & 0.1171 &  $0.0101 \sim 0.1250 $\\
  $|g_1/g_2|$   & 2.0955 & $1.7422 \sim 2.4913$ \\
  $\text{Arg}(g_1/g_2)$  & 4.7360 & $0 \sim 6.2832$ \\
  $ g^2_2 v_u^2/ \Lambda$ [eV] & 0.00004 & ---  \\ \hline
  $\sin^2 \theta_{12}$ & 0.3100 & $0.2750 \sim 0.3500$ \\
  $\sin^2 \theta_{13}$ & 0.0224 & $0.02045 \sim 0.02439$\\
  $\sin^2 \theta_{23}$ & 0.5800 & $0.4180 \sim 0.6270$ \\
  $\delta_{CP}/\pi$ & 1.0437 &  $0 \sim 2$\\
  $\alpha_{21}/\pi$ & 0.2403 & $0 \sim 2$\\
  $\alpha_{31}/\pi$ & 1.3293 & $0.0016 \sim 1.9925$\\ \hline
  $r$ & 0.02927 & $0.02587 \sim 0.03300$ \\
  $m_1$ [eV] & 0.0259 & $0.02278 \sim 0.37556$\\
  $m_2$ [eV] & 0.0273 & $0.02435 \sim 0.37566$ \\
  $m_3$ [eV] & 0.0565 & $0.05252 \sim 0.37934$ \\
  $\textstyle \sum_i m_i$ [eV] & 0.1098 & $0.0997 \sim 1.13057$\\
  $|m_{ee}|$ [eV] & 0.0231&  $0.0190 \sim 0.3343$ \\ \hline
\texttt{Ordering} &  \multicolumn{2}{c|}{NO} \\ \hline
  $N\sigma$  & 0.003 &  --- \\ \hline\hline
\end{tabular}
\caption{The best-fit values and the allowed ranges of the model parameters and lepton mixing parameters and neutrino masses for the model $\mathcal{A}7$ with NO in region$3$. }
\label{tab:modela7_NH3}
\end{table}


\subsection{\label{subsec:numerical_WO_flavons}Numerical results of the models without flavons}

Taking into account the freedom of field redefinition, the parameters $\alpha$, $\beta$ and $\gamma_1$ in the charged lepton mass matrix can be set to be real. However, the coupling constant $\gamma_2$ is generally complex. The neutrino mass matrix $m_{\nu}$ only depends on the modular parameter $\tau$ up to an overall factor in the models $\mathcal{B}1$, $\mathcal{B}2$, $\mathcal{B}4$, $\mathcal{B}5$, $\mathcal{B}6$ and $\mathcal{B}8$, and an additional complex parameter $g_1/g_2$ is involved in the models $\mathcal{B}3$ and $\mathcal{B}7$. The vacuum expectation value of the complex modulus $\tau$ is the unique source of flavor symmetry breaking in these models. In the same fashion as section~\ref{subsec:numerical_W_flavons}, we use the package MINUIT to search for the minimum of the $\chi^2$ function. We find that only the models $\mathcal{B}3$ and $\mathcal{B}7$ can accommodate the experimental results, and the neutrino mass spectrum can be either NO or IO. Since the number of the input parameters is larger than that of the models with flavons, it takes a lot of time to comprehensively scan the whole parameter space of the models $\mathcal{B}3$ and $\mathcal{B}7$. We report the numerical results of the $\chi^2$ analysis in table~\ref{tab:modelb3b7}, and the values of the lepton mixing parameters and neutrino masses and charged lepton masses at the best fitting points are displayed. It is remarkable that all the observables are quite close to their central values, and the Dirac CP phase is approximately maximal $\delta_{CP}\simeq1.599\pi$ in the $\mathcal{B}3$ model for NO neutrino masses.

\begin{table}[hptb]
\centering
\renewcommand{\arraystretch}{1.2}
\begin{tabular}{|c|cc|cc|} \hline\hline
\texttt{Models } & \multicolumn{2}{c|}{$\mathcal{B}3$} & \multicolumn{2}{c|}{$\mathcal{B}7$} \\ \hline
  & \multicolumn{4}{c|}{\texttt{Best-fit values}}  \\ \hline
$\rm{Re}\,\langle \tau \rangle$   & 0.4972 & 0.2105 & 0.0355 & 0.1686\\
$\rm{Im}\,\langle \tau \rangle$    & 0.6833 & 1.6860 & 1.2192 & 1.2519 \\
$\beta/\alpha$ & 172.531 & 76.55 & 632.301 & 349.182 \\
$\gamma_1/\alpha$ & 1003.07  & 160.105 & 86.55 & 49.182\\
$|\gamma_2/\alpha|$ & 744.12  & 188.914 &1.854 & 0.6863\\
$\text{Arg}(\gamma_2/\alpha)$  & 3.340 & 2.146 & 5.979 & 2.342\\
$|g_1/g_2|$   & 0.2185 & 1.180 & 38.741 & 1.198 \\
$\text{Arg}(g_1/g_2)$  & 6.241 & 3.521 & 4.216 & 3.086 \\
$\alpha v_d $ [MeV]  & 0.162 & 3.049 & 1.33 & 2.412\\
$ g^2_2 v_u^2/ \Lambda$ [eV] & 0.0065 & 0.0066 & 0.00003 & 0.0107 \\ \hline
$m_e/m_{\mu}$ & 0.0048& 0.0048 & 0.0048 & 0.0048 \\
$m_{\mu}/m_{\tau}$ & 0.0565 & 0.0565 & 0.0565 & 0.0565 \\ \hline
$\sin^2 \theta_{12}$ & 0.31009 & 0.30999 & 0.31000 & 0.30999\\
$\sin^2 \theta_{13}$ & 0.02264 & 0.02241 & 0.02264 & 0.02241\\
$\sin^2 \theta_{23}$ & 0.58389 & 0.57999 & 0.58398 & 0.58000\\
$\delta_{CP}/\pi$ & 0.808  & 1.599 & 1.164 & 1.933\\
$\alpha_{21}/\pi$ & 1.814  & 0.981 & 0.126 & 1.088\\
$\alpha_{31}/\pi$ & 0.823  & 0.101 & 1.735 & 1.710\\ \hline
$r$ & 0.02942 & 0.02927 & 0.02942 & 0.02927\\
$m_1$ [eV] & 0.0504 & 0.0023 & 0.0494 & 0.0020\\
$m_2$ [eV] & 0.0511 & 0.0089 & 0.0501 & 0.0088\\
$m_3$ [eV] & 0.0102  & 0.0503 & 0.0001 & 0.0503\\
$\textstyle \sum_i m_i$ [eV] & 0.1117 & 0.0615 & 0.0995 & 0.0611\\
$|m_{ee}|$ [eV] & 0.0475 & 0.00226 & 0.04767 & 0.0013\\ \hline
\texttt{Ordering} & IO & NO & IO & NO\\ \hline
$N\sigma$  & 0.009 & 0.001 & 0.002 & 0.001\\ \hline\hline
\end{tabular}
\caption{The best-fit values of the model parameters and lepton mixing parameters and neutrino masses for the models $\mathcal{B}3$ and $\mathcal{B}7$. }
\label{tab:modelb3b7}
\end{table}

\section{\label{sec:conclusion}Conclusion}
\indent
In this paper we have provided a comprehensive analysis of neutrino mass and lepton mixing in theories with $\Gamma_5\cong A_5$ modular symmetry. We have constructed the weight 2, weight 4 and weight 6 modular forms of  level 5 in terms of Dedekind eta-function and Klein forms, and their decomposition into irreducible representation of $A_5$. We have provided the necessary mathematical tools which may be useful for future studies of model building based on $A_5$ modular symmetry.

We have constructed all the simplest models based on $A_5$ modular symmetry, including models with and without flavons in the charged lepton sectors. For each case, the neutrino masses are considered
both using the Weinberg operator and the type I seesaw mechanism.
This extends the scope of the previous analyses for $A_5$ modular symmetry in the literature. The hope is that the compendium of models presented here may be a useful guide to future model building directions.

We have performed an exhaustive numerical analysis for each model, presenting our results in the form of extensive sets of figures and tables. Surveying the comprehensive results in this paper, we observe that the predictions for lepton mixing strongly depend on the modular parameter $\tau$, in other words the assumed vacuum expectation value of the modulus field. The modular symmetry models are therefore
not as predictive as might have been expected. If the value of $\tau$
is not fixed by any mechanism, the lepton mixing parameters and neutrino masses can vary widely in different regions of $\tau$ parameter space. This observation is even true for benchmark modular symmetry models which only depend on $\tau$ and overall couplings. For this reason we divided the modular parameter $\tau$ into several regions when performing the numerical analysis. This motivates models in which the modulus $\tau$ is fixed, as recently discussed in~\cite{deAnda:2018ecu} or in the context of string compactifications and supergravity~\cite{Ferrara:1990ei,Font:1990nt,Cvetic:1991qm}.

\subsection*{Acknowledgements}
G.-J.\, D. and X.-G.\, L. acknowledges the support of the National Natural Science Foundation of China under Grant Nos
11522546 and 11835013. S.\,F.\,K. acknowledges the STFC Consolidated Grant ST/L000296/1 and the European Union's Horizon 2020 research and innovation programme under the Marie Sk\l{}odowska-Curie grant agreements Elusives ITN No.\ 674896 and InvisiblesPlus RISE No.\ 690575. G.-J.\, D. and X.-G.\, L. are grateful to Jun-Nan Lu, Ping-Tao Chen and Chang-Yuan Yao for their help on numerical analysis. The numerical calculations are performed on the cluster of the experimental particle physics group of USTC.

\section*{Appendix}

\setcounter{equation}{0}
\renewcommand{\theequation}{\thesection.\arabic{equation}}

\begin{appendix}

\section{\label{sec:App_CG}Group Theory of $A_{5}$}
$A_{5}$ is the group of even permutations of five objects, and it has $5!/2=60$ elements. Geometrically it is the symmetry group of a regular icosahedron. $A_{5}$ group can be generated by two generators $S$ and $T$ which obey the multiplication rules~\cite{Ding:2011cm,Li:2015jxa}:
\begin{equation}
  S^{2}=T^{5}=(ST)^{3}=1\,.
\end{equation}
The 60 element of $A_{5}$ group are divided into 5 conjugacy classes,
\begin{eqnarray}
\nonumber 1C_{1} :&&1\\
\nonumber15C_{2} :&& ST^2ST^3S, TST^4, T^4(ST^2)^2, T^2ST^3, (T^2S)^2T^3S, ST^2ST, S, T^3ST^2ST^3,\\
\nonumber && ~T^3ST^2ST^3S, T^3ST^2, T^4ST^2ST^3S, TST^2S, ST^3ST^2S, T^4ST, (T^2S)^2T^4\\
\nonumber20C_{3} : && ST, TS, ST^4, T^4S, TST^3, T^2ST^2, T^2ST^4, T^3ST, T^3ST^3, T^4ST^2, TST^3S, T^2ST^3S, \\
\nonumber&&~T^3ST^2S, ST^2ST^3, ST^3ST, ST^3ST^2, (T^2S)^2T^2, T^2(T^2S)^2, (ST^2)^2S, (ST^2)^2T^2\\
\nonumber12C_{5}: && T, T^4, ST^2, T^2S, ST^3, T^3S, STS, TST, TST^2, T^2ST, T^3ST^4, T^4ST^3\\
\nonumber12C^{\prime}_5: &&T^2, T^3, ST^2S, ST^3S,(ST^2)^2, (T^2S)^2, (ST^3)^2, (T^3S)^2, (T^2S)^2T^3,\\
&&~T^3(ST^2)^2, T^3ST^2ST^4, T^4ST^2ST^3\,,
\end{eqnarray}
where $nC_k$ denotes a class with $n$ elements which are of order $k$.
The group structure of $A_5$ has been elaborately analyzed in Refs.~\cite{Ding:2011cm,Li:2015jxa}. The $A_5$ group has five irreducible representations: one singlet representation $\bf{1}$, two three-dimensional representations $\bf{3}$ and $\bf{3^\prime}$, one four-dimensional representation $\mathbf{4}$ and one five-dimensional representation $\mathbf{5}$. In the present work, we choose the same basis as that of Refs.~\cite{Ding:2011cm,Li:2015jxa}. The explicit forms of the generators $S$ and $T$ in the five irreducible representations are as follows
\begin{eqnarray}
\label{eq:irr_reps} \begin{array}{ccc}
    \bf{1:} &   ~S=1\,,~ &  T=1 \,, ~\\[-14pt] \\[4pt]
    \bf{3:} &   ~S=\frac{1}{\sqrt{5}}
\begin{pmatrix}
 1 &~ -\sqrt{2} &~ -\sqrt{2} \\
 -\sqrt{2} &~ -\phi_g  &~ \phi_g-1 \\
 -\sqrt{2} &~ \phi_g-1 &~ -\phi_g
\end{pmatrix}\,,~
& T=
\begin{pmatrix}
 1 &~ 0 &~ 0 \\
 0 &~ \omega_{5}  &~ 0 \\
 0 &~ 0 &~ \omega_{5} ^4
\end{pmatrix}
\,,~~\\[-14pt] \\[4pt]
\bf{3^{\prime}:} &  ~S=\frac{1}{\sqrt{5}}
\begin{pmatrix}
 -1 &~ \sqrt{2} &~ \sqrt{2} \\
 \sqrt{2} &~ 1-\phi_g &~ \phi_g  \\
 \sqrt{2} &~ \phi_g  &~ 1-\phi_g
\end{pmatrix}\,, ~
&T=
\begin{pmatrix}
 1 &~ 0 &~ 0 \\
 0 &~ \omega_{5} ^2 &~ 0 \\
 0 &~ 0 &~ \omega_{5} ^3
\end{pmatrix}
\,,~~\\[-14pt] \\[4pt]
\bf{4:} &  ~S=\frac{1}{\sqrt{5}}
\begin{pmatrix}
 1 &~ \phi_g-1 &~ \phi_g  &~ -1 \\
\phi_g-1 &~ -1 &~ 1 &~ \phi_g  \\
 \phi_g  &~ 1 &~ -1 &~ \phi_g-1 \\
 -1 &~ \phi_g  &~ \phi_g-1 &~ 1
\end{pmatrix}\,,~
&T=
\begin{pmatrix}
 \omega_{5}  &~ 0 &~ 0 &~ 0 \\
 0 &~ \omega_{5} ^2 &~ 0 &~ 0 \\
 0 &~ 0 &~ \omega_{5}^3 &~ 0 \\
 0 &~ 0 &~ 0 &~ \omega_{5}^4
\end{pmatrix}\,,~~\\[-14pt] \\[4pt]
\bf{5:} &  ~S=\frac{1}{5}
\begin{pmatrix}
 -1 &~ \sqrt{6} &~ \sqrt{6} &~ \sqrt{6} &~ \sqrt{6} \\
 \sqrt{6} &~ (\phi_g-1)^{2} &~ -2 \phi_g  &~ 2(\phi_g-1) &~ \phi_g ^2 \\
 \sqrt{6} &~ -2\phi_g  &~ \phi_g^2&~ (\phi_g-1)^{2} &~ 2(\phi_g-1) \\
 \sqrt{6} &~ 2(\phi_g-1) &~ (\phi_g-1)^{2} &~ \phi_g ^2 &~ -2 \phi_g  \\
 \sqrt{6} &~ \phi_g^2 &~ 2(\phi_g-1) &~ -2 \phi_g  &~ (\phi_g-1)^{2}
\end{pmatrix}\,,~
&T=
\begin{pmatrix}
 1 &~ 0 &~ 0 &~ 0 &~ 0 \\
 0 &~ \omega_{5}  &~ 0 &~ 0 &~ 0 \\
 0 &~ 0 &~ \omega_{5} ^2 &~ 0 &~ 0 \\
 0 &~ 0 &~ 0 &~ \omega_{5} ^3 &~ 0 \\
 0 &~ 0 &~ 0 &~ 0 &~ \omega_{5} ^4
\end{pmatrix}\,,~~
\end{array}
\end{eqnarray}
where $\omega_{5}=e^{\frac{2\pi i}{5}}$. The Kronecker products between various irreducible representations are as
follows,
\begin{eqnarray}
\nonumber&&\mathbf{1}\otimes\mathbf{R}=\mathbf{R}\otimes\mathbf{1}=\mathbf{R},\quad \mathbf{3}\otimes\mathbf{3}=\mathbf{1}\oplus\mathbf{3}\oplus\mathbf{5},\quad\mathbf{3}'\otimes\mathbf{3}'=\mathbf{1}\oplus\mathbf{3}'\oplus\mathbf{5}, \quad\mathbf{3}\times\mathbf{3}'=\mathbf{4}\oplus\mathbf{5}\,,\\
\nonumber&&\mathbf{3}\otimes\mathbf{4}=\mathbf{3}'\oplus\mathbf{4}\oplus\mathbf{5},\quad\mathbf{3}'\otimes\mathbf{4}=\mathbf{3}\oplus\mathbf{4}\oplus\mathbf{5},\quad\mathbf{3}\otimes\mathbf{5}
=\mathbf{3}\oplus\mathbf{3}'\oplus\mathbf{4}\oplus\mathbf{5}\,,\\
\nonumber&&\mathbf{3}'\otimes\mathbf{5}=\mathbf{3}\oplus\mathbf{3}'\oplus\mathbf{4}\oplus\mathbf{5},\quad\mathbf{4}\otimes\mathbf{4}=\mathbf{1}\oplus\mathbf{3}\oplus\mathbf{3}'\oplus\mathbf{4}\oplus\mathbf{5},
\quad\mathbf{4}\otimes\mathbf{5}=\mathbf{3}\oplus\mathbf{3}'\oplus\mathbf{4}\oplus\mathbf{5_{1}}\oplus\mathbf{5_{2}}\,,\\ \label{eq:Kronecker}&&\mathbf{5}\otimes\mathbf{5}=\mathbf{1}\oplus\mathbf{3}\oplus\mathbf{3}'\oplus\mathbf{4_{1}}\oplus\mathbf{4_{2}}\oplus\mathbf{5_{1}}\oplus\mathbf{5_{2}}\,.
\end{eqnarray}
where $\bf{R}$ represents any irreducible representation of $A_{5}$, and $\bf{4_{1}}$, $\bf{4_{2}}$, $\bf{5_{1}}$ and $\bf{5_{2}}$ denote the two $\bf{4}$ and two $\bf{5}$ representations which appear in the Kronecker products.

Subsequently we list the Clebsch-Gordan coefficients in the chosen basis. We use $\alpha_{i}$ to denote the elements of the first representation, $\beta_{i}$ to indicate these of the second representation of the product.
The subscripts "$\mathbf{S}$" and "$\mathbf{A}$" indicate a combination which is symmetric or antisymmetric respectively.
\begin{itemize}
\item $\mathbf{3}\otimes\mathbf{3}=\mathbf{1}_S\oplus\mathbf{3}_A \oplus\mathbf{5}_S$

\begin{equation}
\mathbf{1}_S\sim\alpha_1\beta_1+\alpha_2\beta_3+\alpha_3\beta_2\,,
\end{equation}

\begin{equation}
\mathbf{3}_A\sim\begin{pmatrix}
\alpha_2\beta_3-\alpha_3\beta_2 \\
\alpha_1\beta_2-\alpha_2\beta_1 \\
\alpha_3\beta_1-\alpha_1\beta_3
\end{pmatrix}\,,
\end{equation}

\begin{equation}
\mathbf{5}_S\sim\begin{pmatrix}
2\alpha_1\beta_1-\alpha_2\beta_3-\alpha_3\beta_2 \\
-\sqrt{3}\alpha_1\beta_2-\sqrt{3}\alpha_2\beta_1 \\
\sqrt{6}\alpha_2\beta_2 \\
\sqrt{6}\alpha_3\beta_3 \\
-\sqrt{3}\alpha_1\beta_3-\sqrt{3}\alpha_3\beta_1
\end{pmatrix}\,.
\end{equation}
\end{itemize}
\begin{itemize}

\item $\mathbf{3}'\otimes\mathbf{3}'=\mathbf{1}_S\oplus\mathbf{3}'_A\oplus\mathbf{5}_S$

\begin{equation}
\mathbf{1}_S\sim\alpha_1\beta_1+\alpha_2\beta_3+\alpha_3\beta_2\,,
\end{equation}

\begin{equation}
\mathbf{3}'_A\sim\begin{pmatrix}
\alpha_2\beta_3-\alpha_3\beta_2 \\
\alpha_1\beta_2-\alpha_2\beta_1 \\
\alpha_3\beta_1-\alpha_1\beta_3
\end{pmatrix}\,,
\end{equation}

\begin{equation}
\mathbf{5}_S\sim\begin{pmatrix}
2\alpha_1\beta_1-\alpha_2\beta_3 -\alpha_3\beta_2\\
\sqrt{6}\alpha_3\beta_3 \\
-\sqrt{3}\alpha_1\beta_2 -\sqrt{3}\alpha_2\beta_1\\
-\sqrt{3}\alpha_1\beta_3-\sqrt{3}\alpha_3\beta_1 \\
\sqrt{6}\alpha_2\beta_2
\end{pmatrix}\,.
\end{equation}
\end{itemize}

\begin{itemize}
\item $\mathbf{3}\otimes\mathbf{3}'=\mathbf{4}\oplus\mathbf{5}$

\begin{equation}
\mathbf{4}\sim\begin{pmatrix}
\sqrt{2}\alpha_2\beta_1+\alpha_3\beta_2 \\
-\sqrt{2}\alpha_1\beta_2-\alpha_3\beta_3 \\
-\sqrt{2}\alpha_1\beta_3-\alpha_2\beta_2 \\
\sqrt{2}\alpha_3\beta_1+\alpha_2\beta_3
\end{pmatrix}\,,
\end{equation}

\begin{equation}
\mathbf{5}\sim\begin{pmatrix}
\sqrt{3}\alpha_1\beta_1 \\
\alpha_2\beta_1-\sqrt{2}\alpha_3\beta_2 \\
\alpha_1\beta_2-\sqrt{2}\alpha_3\beta_3 \\
\alpha_1\beta_3-\sqrt{2}\alpha_2\beta_2 \\
\alpha_3\beta_1-\sqrt{2}\alpha_2\beta_3
\end{pmatrix}\,.
\end{equation}
\end{itemize}

\begin{itemize}
\item $\mathbf{3}\otimes\mathbf{4}=\mathbf{3}'\oplus\mathbf{4}\oplus\mathbf{5}$

\begin{equation}
\mathbf{3}'\sim\begin{pmatrix}
-\sqrt{2}\alpha_2\beta_4-\sqrt{2}\alpha_3\beta_1 \\
\sqrt{2}\alpha_1\beta_2-\alpha_2\beta_1+\alpha_3\beta_3 \\
\sqrt{2}\alpha_1\beta_3+\alpha_2\beta_2-\alpha_3\beta_4
\end{pmatrix}\,,
\end{equation}

\begin{equation}
\mathbf{4}\sim\begin{pmatrix}
\alpha_1\beta_1-\sqrt{2}\alpha_3\beta_2 \\
-\alpha_1\beta_2-\sqrt{2}\alpha_2\beta_1 \\
\alpha_1\beta_3+\sqrt{2}\alpha_3\beta_4 \\
 -\alpha_1\beta_4+\sqrt{2}\alpha_2\beta_3
\end{pmatrix}\,,
\end{equation}

\begin{equation}
\mathbf{5}\sim\begin{pmatrix}
\sqrt{6}\alpha_2
\beta_4-\sqrt{6}\alpha_3\beta_1 \\
2\sqrt{2}\alpha_1\beta_1+2\alpha_3 \beta_2\\
-\sqrt{2}\alpha_1 \beta_2+\alpha_2\beta_1+3\alpha_3\beta_3 \\
\sqrt{2}\alpha_1 \beta_3-3\alpha_2\beta_2-\alpha_3\beta_4\\
 -2\sqrt{2}\alpha_1\beta_4-2\alpha_2\beta_3
\end{pmatrix}\,.
\end{equation}
\end{itemize}

\begin{itemize}
\item $\mathbf{3}'\otimes\mathbf{4}=\mathbf{3}\oplus\mathbf{4}\oplus\mathbf{5}$

\begin{equation}
\mathbf{3}\sim\begin{pmatrix}
-\sqrt{2}\alpha_2\beta_3-\sqrt{2}\alpha_3\beta_2 \\
\sqrt{2}\alpha_1\beta_1+\alpha_2\beta_4-\alpha_3\beta_3 \\
\sqrt{2}\alpha_1\beta_4 -\alpha_2\beta_2+\alpha_3\beta_1
\end{pmatrix}\,,
\end{equation}

\begin{equation}
\mathbf{4}\sim\begin{pmatrix}
\alpha_1\beta_1+\sqrt{2}\alpha_3\beta_3 \\
\alpha_1\beta_2-\sqrt{2}\alpha_3\beta_4 \\
-\alpha_1\beta_3+\sqrt{2}\alpha_2\beta_1 \\
-\alpha_1\beta_4-\sqrt{2}\alpha_2\beta_2
\end{pmatrix}\,,
\end{equation}

\begin{equation}
\mathbf{5}\sim\begin{pmatrix}
\sqrt{6}\alpha_2\beta_3-\sqrt{6}\alpha_3
\beta_2 \\
\sqrt{2}\alpha_1\beta_1-3\alpha_2\beta_4-\alpha_3 \beta_3 \\
2\sqrt{2}\alpha_1\beta_2+2\alpha_3\beta_4\\
-2\sqrt{2}\alpha_1 \beta_3-2\alpha_2\beta_1 \\
-\sqrt{2}\alpha_1\beta_4+\alpha_2\beta_2+3\alpha_3\beta_1
\end{pmatrix}\,.
\end{equation}
\end{itemize}

\begin{itemize}
\item $\mathbf{3}\otimes\mathbf{5}=\mathbf{3}\oplus\mathbf{3}'\oplus\mathbf{4}\oplus\mathbf{5}$\,,

\begin{equation}
\mathbf{3}\sim\begin{pmatrix}
-2\alpha_1\beta_1+\sqrt{3}\alpha_2\beta_5+\sqrt{3}\alpha_3\beta_2 \\
\sqrt{3}\alpha_1\beta_2+\alpha_2\beta_1-\sqrt{6}\alpha_3\beta_3 \\
\sqrt{3}\alpha_1\beta_5-\sqrt{6}\alpha_2\beta_4+\alpha_3\beta_1
\end{pmatrix}\,,
\end{equation}

\begin{equation}
\mathbf{3}'\sim\begin{pmatrix}
\sqrt{3}\alpha_1\beta_1+\alpha_2\beta_5+\alpha_3\beta_2 \\
\alpha_1\beta_3-\sqrt{2}\alpha_2\beta_2-\sqrt{2}\alpha_3\beta_4 \\
\alpha_1\beta_4-\sqrt{2}\alpha_2\beta_3-\sqrt{2}\alpha_3\beta_5
\end{pmatrix}\,,
\end{equation}

\begin{equation}
\mathbf{4}\sim\begin{pmatrix}
2\sqrt{2}\alpha_1\beta_2-\sqrt{6}\alpha_2\beta_1+\alpha_3\beta_3 \\
-\sqrt{2}\alpha_1\beta_3+2\alpha_2\beta_2-3\alpha_3\beta_4 \\
\sqrt{2}\alpha_1\beta_4+3\alpha_2\beta_3-2\alpha_3\beta_5 \\
-2\sqrt{2}\alpha_1\beta_5-\alpha_2\beta_4+\sqrt{6}\alpha_3\beta_1
\end{pmatrix}\,,
\end{equation}

\begin{equation}
\mathbf{5}\sim\begin{pmatrix}
\sqrt{3}\alpha_2\beta_5-\sqrt{3}\alpha_3\beta_2 \\
-\alpha_1 \beta_2-\sqrt{3}\alpha_2\beta_1-\sqrt{2}\alpha_3\beta_3 \\
-2\alpha_1\beta_3-\sqrt{2}\alpha_2\beta_2 \\
2\alpha_1\beta_4+\sqrt{2}\alpha_3\beta_5 \\
\alpha_1\beta_5+\sqrt{2}\alpha_2\beta_4+ \sqrt{3}\alpha_3\beta_1
\end{pmatrix}\,.
\end{equation}
\end{itemize}

\begin{itemize}
\item $\mathbf{3}'\otimes\mathbf{5}=\mathbf{3}\oplus\mathbf{3}'\oplus\mathbf{4}\oplus\mathbf{5}$\,,

\begin{equation}
\mathbf{3}\sim\begin{pmatrix}
\sqrt{3}\alpha_1\beta_1+\alpha_2\beta_4+\alpha_3\beta_3 \\
\alpha_1\beta_2-\sqrt{2}\alpha_2\beta_5 -\sqrt{2}
\alpha_3\beta_4\\
\alpha_1\beta_5-\sqrt{2}\alpha_2 \beta_3-\sqrt{2}\alpha_3\beta_2
\end{pmatrix}\,,
\end{equation}

\begin{equation}
\mathbf{3}'\sim\begin{pmatrix}
-2\alpha_1\beta_1+\sqrt{3}\alpha_2\beta_4 +\sqrt{3}\alpha_3\beta_3\\
\sqrt{3}\alpha_1\beta_3+\alpha_2\beta_1-\sqrt{6}\alpha_3\beta_5 \\
\sqrt{3}\alpha_1\beta_4-\sqrt{6}\alpha_2\beta_2+\alpha_3\beta_1
\end{pmatrix}\,,
\end{equation}

\begin{equation}
\mathbf{4}\sim\begin{pmatrix}
\sqrt{2}\alpha_1\beta_2+3\alpha_2\beta_5-2\alpha_3\beta_4 \\
2\sqrt{2}\alpha_1 \beta_3-\sqrt{6}\alpha_2\beta_1+\alpha_3\beta_5 \\
-2\sqrt{2}\alpha_1\beta_4-\alpha_2\beta_2 +\sqrt{6}\alpha_3\beta_1\\
-\sqrt{2}\alpha_1\beta _5+2\alpha_2\beta_3-3\alpha_3\beta_2
\end{pmatrix}\,,
\end{equation}

\begin{equation}
\mathbf{5}\sim\begin{pmatrix}
\sqrt{3}\alpha_2\beta_4-\sqrt{3}\alpha_3\beta_3   \\
2\alpha_1\beta_2+\sqrt{2}\alpha_3\beta_4 \\
-\alpha_1\beta_3-\sqrt{3}\alpha_2\beta_1-\sqrt{2}\alpha_3\beta_5 \\
\alpha_1\beta_4+\sqrt{2}\alpha_2 \beta_2 + \sqrt{3}\alpha_3\beta_1\\
-2\alpha_1\beta_5-\sqrt{2}\alpha_2\beta_3
\end{pmatrix}\,.
\end{equation}
\end{itemize}

\begin{itemize}
\item $\mathbf{4}\otimes\mathbf{4}=\mathbf{1}_S\oplus\mathbf{3}_A\oplus\mathbf{3}'_A\oplus\mathbf{4}_S\oplus\mathbf{5}_S$

\begin{equation}
\mathbf{1}_S\sim\alpha_1\beta _4+\alpha_2\beta_3+\alpha_3 \beta_2+\alpha_4\beta_1\,,
\end{equation}

\begin{equation}
\mathbf{3}_A\sim\begin{pmatrix}
-\alpha_1\beta_4+\alpha_2\beta_3-\alpha_3\beta_2+\alpha_4\beta_1\\
\sqrt{2}\alpha_2\beta_4-\sqrt{2}\alpha_4\beta_2\\
\sqrt{2}\alpha_1 \beta_3-\sqrt{2}\alpha_3\beta_1
\end{pmatrix}\,,
\end{equation}

\begin{equation}
\mathbf{3}'_A\sim\begin{pmatrix}
\alpha_1\beta_4 +\alpha_2\beta_3-\alpha_3\beta_2 -\alpha_4\beta_1\\
\sqrt{2}\alpha_3\beta_4-\sqrt{2}\alpha_4\beta_3   \\
\sqrt{2}\alpha_1\beta_2-\sqrt{2}\alpha_2\beta_1
\end{pmatrix}\,,
\end{equation}

\begin{equation}
\mathbf{4}_S\sim\begin{pmatrix}
\alpha_2\beta_4+\alpha_3\beta _3+\alpha_4\beta_2 \\
\alpha_1\beta_1+\alpha_3\beta_4 +\alpha_4\beta_3\\
\alpha_1\beta _2+\alpha_2\beta_1+\alpha_4\beta_4 \\
\alpha_1\beta_3+\alpha_2\beta _2+\alpha_3\beta_1
\end{pmatrix}\,,
\end{equation}

\begin{equation}
\mathbf{5}_S\sim\begin{pmatrix}
\sqrt{3}\alpha_1\beta_4-\sqrt{3}\alpha_2\beta_3-\sqrt{3}\alpha_3 \beta_2+\sqrt{3}\alpha_4 \beta_1 \\
-\sqrt{2}\alpha_2\beta_4+2\sqrt{2}\alpha_3\beta_3-\sqrt{2}\alpha_4\beta_2\\
-2\sqrt{2}\alpha_1\beta_1+\sqrt{2}\alpha_3\beta_4+\sqrt{2}\alpha_4\beta_3\\
\sqrt{2}\alpha_1\beta_2+\sqrt{2}\alpha_2\beta_1-2 \sqrt{2}\alpha_4\beta_4 \\
-\sqrt{2}\alpha_1\beta_3+2\sqrt{2}\alpha_2\beta_2-\sqrt{2}\alpha_3\beta_1
\end{pmatrix}\,.
\end{equation}
\end{itemize}

\begin{itemize}
\item $\mathbf{4}\otimes\mathbf{5}=\mathbf{3}\oplus\mathbf{3}'\oplus\mathbf{4}\oplus\mathbf{5}_1\oplus\mathbf{5}_2$

\begin{equation}
\mathbf{3}\sim\begin{pmatrix}
2 \sqrt{2}\alpha_1\beta_5-\sqrt{2}\alpha_2 \beta_4+\sqrt{2}\alpha_3\beta_3-2 \sqrt{2}\alpha_4 \beta_2\\
-\sqrt{6}\alpha_1 \beta_1+2\alpha_2\beta_5+3\alpha_3\beta_4-\alpha_4\beta_3 \\
\alpha_1\beta_4-3\alpha_2\beta_3-2\alpha_3\beta_2+\sqrt{6}\alpha_4\beta_1
\end{pmatrix}\,,
\end{equation}

\begin{equation}
\mathbf{3}'\sim\begin{pmatrix}
\sqrt{2}\alpha_1\beta_5+2\sqrt{2}\alpha_2\beta_4-2\sqrt{2}\alpha_3\beta_3-\sqrt{2}\alpha_4\beta_2\\
3\alpha_1\beta_2-\sqrt{6}\alpha_2\beta_1-\alpha_3\beta_5+2\alpha_4\beta_4\\
-2\alpha_1\beta_3+\alpha_2\beta_2+\sqrt{6}\alpha_3 \beta_1-3\alpha_4\beta_5
\end{pmatrix}\,,
\end{equation}

\begin{equation}
\mathbf{4}\sim\begin{pmatrix}
\sqrt{3}\alpha_1\beta_1-\sqrt{2}\alpha_2\beta_5+\sqrt{2}\alpha_3 \beta_4-2\sqrt{2}\alpha_4 \beta_3\\
-\sqrt{2}\alpha_1 \beta_2-\sqrt{3}\alpha_2 \beta_1+2\sqrt{2}\alpha_3 \beta_5+\sqrt{2}\alpha_4 \beta_4 \\
\sqrt{2}\alpha_1 \beta_3+2\sqrt{2}\alpha_2\beta_2-\sqrt{3}\alpha_3\beta_1-\sqrt{2}\alpha_4\beta_5\\
-2\sqrt{2}\alpha_1 \beta_4+\sqrt{2}\alpha_2 \beta_3-\sqrt{2}\alpha_3 \beta_2+\sqrt{3}\alpha_4 \beta_1
\end{pmatrix}\,,
\end{equation}

\begin{equation}
\mathbf{5}_1\sim\begin{pmatrix}
\sqrt{2}\alpha_1\beta_5-\sqrt{2}\alpha_2 \beta_4-\sqrt{2}\alpha_3\beta_3+\sqrt{2}\alpha_4 \beta_2\\
-\sqrt{2}\alpha_1 \beta_1-\sqrt{3}\alpha_3\beta_4 -\sqrt{3}\alpha_4\beta_3\\
\sqrt{3}\alpha_1 \beta_2+\sqrt{2}\alpha_2\beta_1+\sqrt{3}\alpha_3\beta_5  \\
\sqrt{3}\alpha_2 \beta_2+\sqrt{2}\alpha_3\beta_1+\sqrt{3}\alpha_4\beta_5\\
-\sqrt{3}\alpha_1\beta_4-\sqrt{3}\alpha_2 \beta_3-\sqrt{2}\alpha_4\beta_1
\end{pmatrix}\,,
\end{equation}

\begin{equation}
\mathbf{5}_2\sim\begin{pmatrix}
2\alpha_1\beta _5+4\alpha_2\beta_4+4\alpha_3 \beta_3 +2\alpha_4\beta_2\\
4\alpha_1\beta_1+2 \sqrt{6}\alpha_2\beta_5 \\
-\sqrt{6}\alpha_1\beta _2+2\alpha_2\beta_1-\sqrt{6}\alpha_3\beta_5 +2\sqrt{6}\alpha_4 \beta_4\\
2\sqrt{6}\alpha_1 \beta_3-\sqrt{6}\alpha_2\beta_2+2\alpha_3\beta_1-\sqrt{6}\alpha_4\beta_5\\
2\sqrt{6}\alpha_3\beta_2+4\alpha_4\beta_1
\end{pmatrix}\,.
\end{equation}
\end{itemize}

\begin{itemize}
\item $\mathbf{5}\otimes\mathbf{5}=\mathbf{1}_S\oplus\mathbf{3}_A\oplus\mathbf{3}'_A\oplus\mathbf{4}_{S}\oplus\mathbf{4}_{A}\oplus\mathbf{5}_{S,1}\oplus\mathbf{5}_{S,2}$

\begin{equation}
\mathbf{1}_S\sim\alpha_1\beta_1+\alpha_2\beta_5+\alpha_3\beta_4+\alpha_4\beta_3+\alpha_5\beta_2\,,
\end{equation}

\begin{equation}
\mathbf{3}_A\sim\begin{pmatrix}
\alpha_2\beta_5+2\alpha_3\beta_4-2\alpha_4\beta_3-\alpha_5\beta_2 \\
-\sqrt{3}\alpha_1\beta_2+\sqrt{3}\alpha_2\beta_1+\sqrt{2}\alpha_3\beta_5-\sqrt{2}\alpha_5\beta_3 \\
\sqrt{3}\alpha_1\beta_5+\sqrt{2}\alpha_2\beta_4-\sqrt{2}\alpha_4\beta_2-\sqrt{3}\alpha_5\beta_1
\end{pmatrix}\,,
\end{equation}

\begin{equation}
\mathbf{3}'_A\sim\begin{pmatrix}
2\alpha_2\beta_5-\alpha_3\beta_4+\alpha_4\beta_3-2\alpha_5\beta_2\\
\sqrt{3}\alpha_1\beta_3-\sqrt{3}\alpha_3\beta_1+\sqrt{2}\alpha_4\beta_5-\sqrt{2}\alpha_5\beta_4\\
-\sqrt{3}\alpha_1\beta_4+\sqrt{2}\alpha_2\beta_3-\sqrt{2}\alpha_3\beta_2+\sqrt{3}\alpha_4\beta_1
\end{pmatrix}\,,
\end{equation}

\begin{equation}
\mathbf{4}_S\sim\begin{pmatrix}
3\sqrt{2}\alpha_1\beta_2+3\sqrt{2}\alpha_2\beta_1-\sqrt{3}\alpha_3\beta_5+4 \sqrt{3}\alpha_4\beta_4-\sqrt{3}\alpha_5\beta_3 \\
3\sqrt{2}\alpha_1\beta_3+4 \sqrt{3}\alpha_2\beta_2+3 \sqrt{2}\alpha_3\beta_1-\sqrt{3}\alpha_4\beta_5-\sqrt{3}\alpha_5\beta_4\\
3\sqrt{2}\alpha_1\beta_4-\sqrt{3}\alpha_2\beta_3-\sqrt{3}\alpha_3\beta_2+3 \sqrt{2}\alpha_4\beta_1+4 \sqrt{3}\alpha_5\beta _5 \\
3\sqrt{2}\alpha_1\beta_5-\sqrt{3}\alpha_2\beta_4+4\sqrt{3}\alpha_3\beta_3-\sqrt{3}\alpha_4\beta_2+3 \sqrt{2}\alpha_5\beta_1
\end{pmatrix}\,,
\end{equation}

\begin{equation}
\mathbf{4}_A\sim\begin{pmatrix}
\sqrt{2}\alpha_1\beta_2-\sqrt{2}\alpha_2\beta_1+\sqrt{3}\alpha_3\beta_5-\sqrt{3}\alpha_5\beta_3\\
-\sqrt{2}\alpha_1\beta_3+\sqrt{2}\alpha_3\beta_1+\sqrt{3}\alpha_4\beta_5-\sqrt{3}\alpha_5\beta_4\\
-\sqrt{2}\alpha_1\beta_4-\sqrt{3}\alpha_2\beta_3+\sqrt{3}\alpha_3\beta_2+\sqrt{2}\alpha_4\beta_1\\
\sqrt{2}\alpha_1\beta_5-\sqrt{3}\alpha_2\beta_4+\sqrt{3}\alpha_4\beta_2-\sqrt{2}\alpha_5\beta_1
\end{pmatrix}\,,
\end{equation}

\begin{equation}
\mathbf{5}_{S,1}\sim\begin{pmatrix}
2\alpha_1\beta_1+\alpha_2\beta_5-2\alpha_3\beta_4-2\alpha_4\beta_3+\alpha_5\beta_2\\
\alpha_1\beta_2+\alpha_2\beta_1+\sqrt{6}\alpha_3\beta_5+\sqrt{6}\alpha_5\beta_3 \\
-2\alpha_1\beta_3+\sqrt{6}\alpha_2\beta_2-2\alpha_3\beta_1 \\
-2\alpha_1\beta_4-2\alpha_4\beta_1+\sqrt{6}\alpha_5\beta_5 \\
\alpha_1\beta_5+\sqrt{6}\alpha_2\beta_4+\sqrt{6}\alpha_4\beta_2+\alpha_5\beta_1
\end{pmatrix}\,,
\end{equation}

\begin{equation}
\mathbf{5}_{S,2}\sim\begin{pmatrix}
2\alpha_1\beta_1-2\alpha_2\beta_5+\alpha_3\beta_4+\alpha_4\beta_3-2\alpha_5\beta_2\\
-2\alpha_1\beta_2-2\alpha_2\beta_1+\sqrt{6}\alpha_4\beta_4 \\
\alpha_1\beta_3+\alpha_3\beta_1+\sqrt{6}\alpha_4\beta_5+\sqrt{6}\alpha_5\beta_4\\
\alpha_1\beta_4+\sqrt{6}\alpha_2\beta_3+\sqrt{6}\alpha_3\beta_2+\alpha_4\beta_1\\
-2\alpha_1\beta_5+\sqrt{6}\alpha_3\beta_3-2\alpha_5\beta_1
\end{pmatrix}\,.
\end{equation}
\end{itemize}

\section{\label{sec:App_charged_lepton} Charged lepton sector with flavons}
\cleqn

In this appendix, we shall show that a diagonal charged lepton mass matrix can be achieved if the VEVs of the flavon fields are aligned in appropriate directions. The left-handed lepton doublets $L$ are assigned to a triplet $\mathbf{3}$ of $A_5$, and the right-handed charged leptons $E^{c}_i$ transform as $\mathbf{1}$. In order to achieve the desired vacuum alignment and to reproduce the observed charged lepton mass hierarchies, the auxiliary symmetry $Z_5$ is invoked. We use the supersymmetric driving field mechanism to determine the vacuum alignment of the flavons~\cite{Altarelli:2005yx}, and a continuous $U(1)_R$ symmetry related to the usual $R-$parity is assumed.
The $R-$charge of the matter superfields is $+1$, the Higgs and flavon fields are uncharged and the so-called driving fields, indicated with the superscript ``0'' carry two units of $R-$charge. The field content and the symmetry assignments are shown in table~\ref{tab:model2_fields}. The superpotential of the charged lepton sector can be written as
\begin{table}[th!]
\renewcommand{\tabcolsep}{1.05mm}
\centering
\begin{tabular}{|c||c|c|c|c|c||c|c|c||c|c|c|c|c|}
\hline \hline
\texttt{Fields} &   $L$    &  $E^c_1$     &   $E^c_2$    &    $E^c_3$  &  $H_d$ & $\varphi$ &  $\phi$  &  $\psi$ & $\sigma^{0}$ & $\phi^{0}$  & $\psi^{0}$  \\ \hline

$\Gamma_5\cong A_5$  &   $\mathbf{3}$  &    $\mathbf{1}$  & $\mathbf{1}$   &  $\mathbf{1}$   & $\mathbf{1}$  & $\mathbf{3}$  & $\mathbf{3}^\prime$ &   $\mathbf{5}$ & $\mathbf{1}$   & $\mathbf{4}$ & $\mathbf{5}$ \\ \hline

$k_I$  & $k_L$ & $k_{E_1}$ & $k_{E_2}$ & $k_{E_3}$ & $k_d$ & $k_{\varphi}$ & $k_{\phi}$ & $k_{\psi}$ & $k_{\sigma^0}$ & $k_{\phi^0}$ & $k_{\psi^0}$ \\ \hline

$Z_5$ & $1$ & $1$ & $\omega_5$ &  $\omega^4_5$ & $1$ & $\omega_5$ & $\omega^2_5$ & $\omega^2_5$ & $\omega^3_5$  & $\omega^2_5$ & $\omega^3_5$  \\ \hline

$U(1)_R$  &  $1$  &  $1$  &   $1$  & $1$  &  $0$ &  $0$ &  $0$ & $0$  & $2$ & $2$ & $2$  \\ \hline \hline
\end{tabular}
\caption{\label{tab:model2_fields} The charge assignment of the matter fields, flavon fields and driving fields under the $\Gamma_5$ modular symmetry, $Z_5$ auxiliary symmetry and $U(1)_R$ in the charged lepton sector.}
\end{table}
\begin{equation}
w=w_{l}+w_d\,,
\end{equation}
with
\begin{eqnarray}
w_{d}=&&f_{1}\sigma^{0}(\varphi\varphi)_{\bf{1}}+f_{2}(\phi^{0}(\varphi\phi)_{\bf{4}})_{\bf{1}}+f_{3}(\phi^{0}(\varphi\psi)_{\bf{4}})_{\bf{1}}
+M_{\psi}(\psi^{0}\psi)_{\bf{1}}+f_{4}(\psi^{0}(\varphi\varphi)_{\bf{5}})_{\bf{1}}\,,\\
\nonumber w_{l}=&&\frac{y_{\tau}}{\Lambda}E^c_3(L\varphi)_{\bf1}H_d+\frac{y_{\mu_{1}}}{\Lambda^{2}}E^c_2(L(\phi\psi)_{\bf3})_{\bf1}H_d+\frac{y_{\mu_{2}}}{\Lambda^{2}}E^c_2
(L(\psi\psi)_{\bf3})_{\bf1}H_d+\frac{y_{e_{1}}}{\Lambda^{3}}E^c_1(L\varphi)_{\bf{1}}(\phi\phi)_{\bf1}H_d\\
\nonumber&&+\frac{y_{e_{2}}}{\Lambda^{3}}E^c_1((L\varphi)_{\bf{5}}
(\phi\phi)_{\bf5})_{\bf{1}}H_d+\frac{y_{e_{3}}}{\Lambda^{3}}E^c_1((L\varphi)_{\bf{3}}(\phi\psi)_{\bf3})_{\bf{1}}H_d+\frac{y_{e_{4}}}{\Lambda^{3}}E^c_1((L\varphi)_{\bf{5}}
(\phi\psi)_{\bf5})_{\bf{1}}H_d\\
\nonumber&&+\frac{y_{e_{5}}}{\Lambda^{3}}E^c_1(L\varphi)_{\bf{1}}(\psi\psi)_{\bf1}H_d+\frac{y_{e_{6}}}{\Lambda^{3}}E^c_1((L\varphi)_{\bf{3}}
(\psi\psi)_{\bf3})_{\bf{1}}H_d+\frac{y_{e_{7}}}{\Lambda^{3}}E^c_1((L\varphi)_{\bf{5}}(\psi\psi)_{\bf{5_{1}}})_{\bf{1}}H_d \\
&&+\frac{y_{e_{8}}}{\Lambda^{3}}E^c_1((L\varphi)_{\bf{5}}(\psi\psi)_{\bf{5_{2}}})_{\bf{1}}H_d\,.
\end{eqnarray}
The modular weight of each term should be zero such that the following constraints should be satisfied,
\begin{subequations}
\begin{eqnarray}
\label{eq:weight_ch1}&&-\frac{2}{5}\left(k_{E_1}+k_L+k_d\right)=-\frac{1}{2}\left(k_{E_2}+k_L+k_d\right)=-2\left(k_{E_3}+k_L+k_d\right)=k_{\phi}\,,\\
\label{eq:weight_ch2}&&k_{\phi}=k_{\psi}=2k_{\varphi}=-k_{\sigma^0}=-k_{\psi^0}=-\frac{2}{3}k_{\phi^0}\,.
\end{eqnarray}
\end{subequations}
In the limit of supersymmetry, the vacuum configuration is determined by the vanishing of the derivative of the driving superpotential $w_d$ with respect to each component of the driving fields,
\begin{eqnarray}
\nonumber&&\frac{\partial w_{d}}{\partial\sigma^{0}}=f_{1}(\varphi^{2}_{1}+2\varphi_{2}\varphi_{3})=0,\\
\nonumber&&\frac{\partial w_{d}}{\partial\phi^{0}_{1}}=f_{2}(\varphi_{2}\phi_{3}+\sqrt{2}\varphi_{3}\phi_{1})-f_{3}(2\sqrt{2}\varphi_{1}\psi_{5}+\varphi_{2}\psi_{4}
-\sqrt{6}\varphi_{3}\psi_{1})=0,\\
\nonumber&&\frac{\partial w_{d}}{\partial\phi^{0}_{2}}=-f_{2}(\sqrt{2}\varphi_{1}\phi_{3}+\varphi_{2}\phi_{2})+f_{3}(\sqrt{2}\varphi_{1}\psi_{4}+3\varphi_{2}\psi_{3}
-2\varphi_{3}\psi_{5})=0,\\
\nonumber&&\frac{\partial w_{d}}{\partial\phi^{0}_{3}}=-f_{2}(\sqrt{2}\varphi_{1}\phi_{2}+\varphi_{3}\phi_{3})-f_{3}(\sqrt{2}\varphi_{1}\psi_{3}-2\varphi_{2}\psi_{2}
+3\varphi_{3}\psi_{4})=0,\\
\nonumber&&\frac{\partial w_{d}}{\partial\phi^{0}_{4}}=f_{2}(\sqrt{2}\varphi_{2}\phi_{1}+\varphi_{3}\phi_{2})+f_{3}(2\sqrt{2}\varphi_{1}\psi_{2}-\sqrt{6}\varphi_{2}\psi_{1}
+\varphi_{3}\psi_{3})=0,\\
\nonumber&&\frac{\partial w_{d}}{\partial\psi^{0}_{1}}=M_{\psi}\psi_{1}+2f_{4}(\varphi^{2}_{1}-\varphi_{2}\varphi_{3})=0,\\
\nonumber&&\frac{\partial w_{d}}{\partial\psi^{0}_{2}}=M_{\psi}\psi_{5}-2\sqrt{3}f_{4}\varphi_{1}\varphi_{3}=0,\\
\nonumber&&\frac{\partial w_{d}}{\partial\psi^{0}_{3}}=M_{\psi}\psi_{4}+\sqrt{6}f_{4}\varphi^{2}_{3}=0,\\
\nonumber&&\frac{\partial w_{d}}{\partial\psi^{0}_{4}}=M_{\psi}\psi_{3}+\sqrt{6}f_{4}\varphi^{2}_{2}=0,\\
&&\frac{\partial w_{d}}{\partial\psi^{0}_{5}}=M_{\psi}\psi_{2}-2\sqrt{3}f_{4}\varphi_{1}\varphi_{2}=0\,.
\end{eqnarray}
These equations are satisfied by the alignment
\begin{equation}\label{eq:vacuum2}
\langle\varphi\rangle=v_{\varphi}(0,\,1,\,0),\qquad \langle\phi\rangle=v_{\phi}(0,\,1,\,0),\qquad \langle\psi\rangle=v_{\psi}(0,\,0,\,1,\,0,\,0)\,,
\end{equation}
up to symmetry transformations of $A_5$, where the VEVs $v_{\varphi}$, $v_{\phi}$ and $v_{\psi}$ are related through
\begin{equation}\label{eq:ch_LO_VEV_relation}
v_{\phi}=-\frac{3\sqrt{6}f_{3}f_{4}}{M_{\psi}f_{2}}v^{2}_{\varphi}, \qquad v_{\psi}=-\frac{\sqrt{6}f_{4}}{M_{\psi}}v^{2}_{\varphi},\quad v_{\varphi}~{\rm undetermined}\,.
\end{equation}
With the vacuum configuration in Eq.~\eqref{eq:vacuum2}, we find the charged lepton mass matrix is diagonal and the three charged lepton masses are given by,
\begin{eqnarray}
\nonumber&&m_{e}=\sqrt{2}\left|3y_{e_2}\frac{v^2_{\phi}v_{\varphi}}{\Lambda^3}+(y_{e_3}-\sqrt{3}y_{e_4})\frac{v_{\phi}v_{\varphi}v_{\psi}}{\Lambda^3}+3y_{e_8}\frac{v_{\varphi}v^2_{\psi}}{\Lambda^3}\right|v_{d}\\
\nonumber
&&m_{\mu}=\sqrt{2}\left|y_{\mu_1}\frac{v_{\phi}v_{\psi}}{\Lambda^2}\right|v_{d}\\
&&m_{\tau}=\left|y_{\tau}\frac{v_{\varphi}}{\Lambda}\right|v_{d}\,.
\end{eqnarray}
We see that the electron, muon and tau masses involve three flavons, two flavons and one flavon respectively because of the $Z_5$ auxiliary symmetry and modular weight assignments. The observed charged lepton mass hierarchies $m_{e}:m_{\mu}:m_{\tau}\simeq\lambda^{4}_c: \lambda^{2}_c: 1$ can be reproduced if all the VEVs are of the same order of magnitude
\begin{equation}
\label{eq:ch_VEV}\frac{v_{\varphi}}{\Lambda}\sim\frac{v_{\phi}}{\Lambda}\sim\frac{v_{\psi}}{\Lambda}\sim\mathcal{O}(\lambda^{2}_c)\,,
\end{equation}
where $\lambda_c\simeq0.23$ is the Cabibbo angle. It is easy to see that the model for $\rho_{L}=\mathbf{3}'$ can be constructed in a similar way.

\end{appendix}

\providecommand{\href}[2]{#2}\begingroup\raggedright\endgroup

\end{document}